\documentclass[prx,aps,amssymb,onecolumn,noshowpacs,hyperref]{revtex4-1}
\usepackage{color}
\usepackage{graphicx}
\usepackage{hyperref}
\usepackage{amsmath}
\usepackage{latexsym}
\usepackage{amssymb}
\usepackage{mathrsfs}
\usepackage{mathtools}
\usepackage{layout}
\usepackage{verbatim}
\usepackage{multirow}
\usepackage{bm}
\usepackage{amsfonts,epsfig}
\usepackage{lineno}
\usepackage{textcomp}
\usepackage[dvipsnames]{xcolor}
\definecolor{myblue}{RGB}{0, 100, 200}

\begin{document}

\title{Fast measurement-based generation of large-scale Greenberger-Horne-Zeilinger state with atomic nuclear-spin qubits}
\date{\today}
\author{Yan Lu}
\affiliation{Center for Theoretical Physics and School of Physics and Optoelectronic Engineering, Hainan University, Haikou 570228, China}
\author{Xiao-Feng Shi}
\email[Corresponding author: ]{xshi@hainanu.edu.cn}
\affiliation{Center for Theoretical Physics and School of Physics and Optoelectronic Engineering, Hainan University, Haikou 570228, China}
\begin{abstract}
Large-scale Greenberger–Horne–Zeilinger~(GHZ) state is useful for quantum technologies but difficult to be prepared. Here, we propose fast measurement-based preparation of large-scale GHZ states by a four-qubit quantum phase gate with nuclear-spin qubits of alkaline-earth-like atoms, which is named as quantum ferromagnetic gate~(QFG) due to its analogy to the alignment of molecular magnetic moments in a classical magnet. A high-fidelity Rydberg-mediated QFG can be realized in a time of $6\pi/\Omega_{\text{m}}$ with $\Omega_{\text{m}}$ the maximal Rydberg Rabi frequency. From a product state of three data atom and one ancilla atom, a gluing circuit with one QFG, two single-qubit gates, and a projective measurement of the ancilla can generate a 3-qubit GHZ state, and repetition of this gluing circuit can lead to 9, 27, 81, 243, $\cdots$-qubit GHZ states. Analyses based on currently available techniques show that a 243-qubit GHZ state is realizable, and more qubits can be entangled with higher detection fidelity.

\end{abstract}
\maketitle

\section{Introduction}
The n-qubit Greenberger–Horne–Zeilinger~(GHZ) state~\cite{GHZ1989} in the form of
\begin{eqnarray}
\lvert \text{GHZ}_{\pm}^{(\text{n})}\rangle \equiv (\lvert 0_10_2\cdots0_{n}\rangle \pm \lvert 1_11_2\cdots1_{n}\rangle)/\sqrt{2}
\end{eqnarray}
or its variant is practically helpful to realize quantum secret sharing~\cite{Hillery1999} and quantum key distribution~\cite{proietti2021}, where $\lvert 0_k(1_k)\rangle$ denotes the state of the $k$th qubit being in the qubit state 0~(1). Though GHZ states with $n$ equal to 32~\cite{Moses2023} or 60~\cite{Bao2024} were experimentally demonstrated, it is difficult to scale to large $n$  
because GHZ states are globally entangled, and, hence, the preparation often requires interaction nearly between any two qubits, which is why theoretical protocols for generating large atomic GHZ states often require cavity assisted atomic interactions~\cite{zhao2021}.
Nonetheless, it is nontrivial to prepare and manipulate large-scale atomic array inside cavities as in free space, rendering extreme difficulty for generating large-scale GHZ states.
In neutral Rydberg atoms as an example~\cite{Gallagh2005}, 
the largest experimentally realized GHZ state is limited to $n=20$ either via optimal control~\cite{Omran2019} or with Rydberg entangling gates~\cite{senoo2025}.

In this work, we propose an approach to fast generation of GHZ states in free space by using optically trapped neutral atoms via Rydberg mediated interactions~\cite{Saffman2010}.
Our approach relies on an n-qubit quantum gate that adds a $\pi$ phase simultaneously to the following two components
\begin{eqnarray}
\{\lvert 0_10_2\cdots0_{n}\rangle ,\lvert 1_11_2\cdots1_{n}\rangle\} \label{2state}
\end{eqnarray}
in a general input superposition of a nuclear-spin state but does nothing to any other state components. For want of a better term, the gate is called a quantum ferromagnetic gate~(QFG) because it imprints a $\pi$ phase only if all the qubits are either in $|0\rangle$ or in $|1\rangle$, which is analogous to that in a classical ferromagnet all the molecular magnetic moments are nearly parallel to each other. The extra thing we shall bare in mind is that QFG takes action on a superposition of the two state components in Eq.~(\ref{2state}) while in a classical ferromagnet we can't have superposition of parallel and anti-parallel alignment of magnetic moments. 

Two general steps can generate large-scale GHZ states. (i) Preparation of small GHZ states: using a quantum circuit with two single-qubit rotations, one (n+1)-qubit QFG on n data qubits and one ancillary qubit, and one projective state measurement of the ancillary qubit, the state $\lvert \text{GHZ}^{(\text{n})}\rangle$ can be generated with a probability $p_n=2^{1-n}$ starting from an (n+1)-qubit product state. This circuit has the result of gluing smaller GHZ states to form a larger one, and hence is termed a gluing circuit. Unless otherwise specified, we use $\lvert \text{GHZ}^{(\text{n})}\rangle$ to denote $\lvert \text{GHZ}_{\pm}^{(\text{n})}\rangle$ for brevity. (ii) Repetitive gluing of GHZ states: in an $nm+1$ atom system where one is an ancillary atom and the rest $nm$ atoms are initialized in $n$ $m$-qubit GHZ states $\lvert \text{GHZ}^{(\text{m})}\rangle$, we can use two single-qubit rotations, one $(n+1)$-qubit QFG, and one projective measurement of the ancillary qubit for generating the state $\lvert \text{GHZ}^{\text{(mn)}}\rangle$ with a success probability $2^{1-n}$. Via this strategy, by using the 4-qubit QFG, stage (i) can realize $\lvert \text{GHZ}^{(3)}\rangle$, and subsequently stage (ii) can realize $\lvert \text{GHZ}^{(9)}\rangle$; starting from $\lvert \text{GHZ}^{(9)}\rangle$, a further execution of stage (ii) can realize $\lvert \text{GHZ}^{(27)}\rangle$; repetition of this cycle can lead to $\lvert \text{GHZ}^{(81)}\rangle$, $\lvert \text{GHZ}^{(243 )}\rangle$, $\lvert \text{GHZ}^{(729)}\rangle$, and so on. Later, we will show that by using the basic gluing circuit, one can deterministically generate a large-scale GHZ state starting from product states.

The theory requires fast generation of QFG, for which we resort to coding qubits in the clock state of an alkaline-earth-like atom like $^{171}$Yb. The required technique for rapidly exciting the nuclear spin qubits to Rydberg state has been recently realized in, e.g.,  Ref.~\cite{Ma2023}, and a nuclear-spin gate was realized with a fidelity 0.9978~\cite{senoo2025}. The theory here also requires the feasibility to assembly large-scale atomic array. Recently, Ref.~\cite{6100atoms} realized tweezer array with over 6000 atoms, and Ref.~\cite{Chiu2025} demonstrated continuous operation of a coherent 3,000-qubit system. These point to the possibility to carry out the large-scale GHZ-state generation by measurement.

The remainder of this article is organized as follows. In Sec.~\ref{sec02}, we introduce the generation of QFG with nuclear-spin qubits and study its fidelity. In Sec.~\ref{sec03}, we study the gluing circuits required to create large-scale GHZ state based on QFG and destructive measurement of atomic states. In Sec.~\ref{sec04}, we study strategy to suppress cross blockade during the parallel execution of gluing circuits in the atomic array. In Sec.~\ref{sec05}, we study the fidelity for generating a 243-qubit GHZ state. In Sec.~\ref{sec06}, we discuss the limitation of the protocol, and Sec.~\ref{sec07} gives a brief summary.

\begin{figure}
\includegraphics[width=3.0in]
{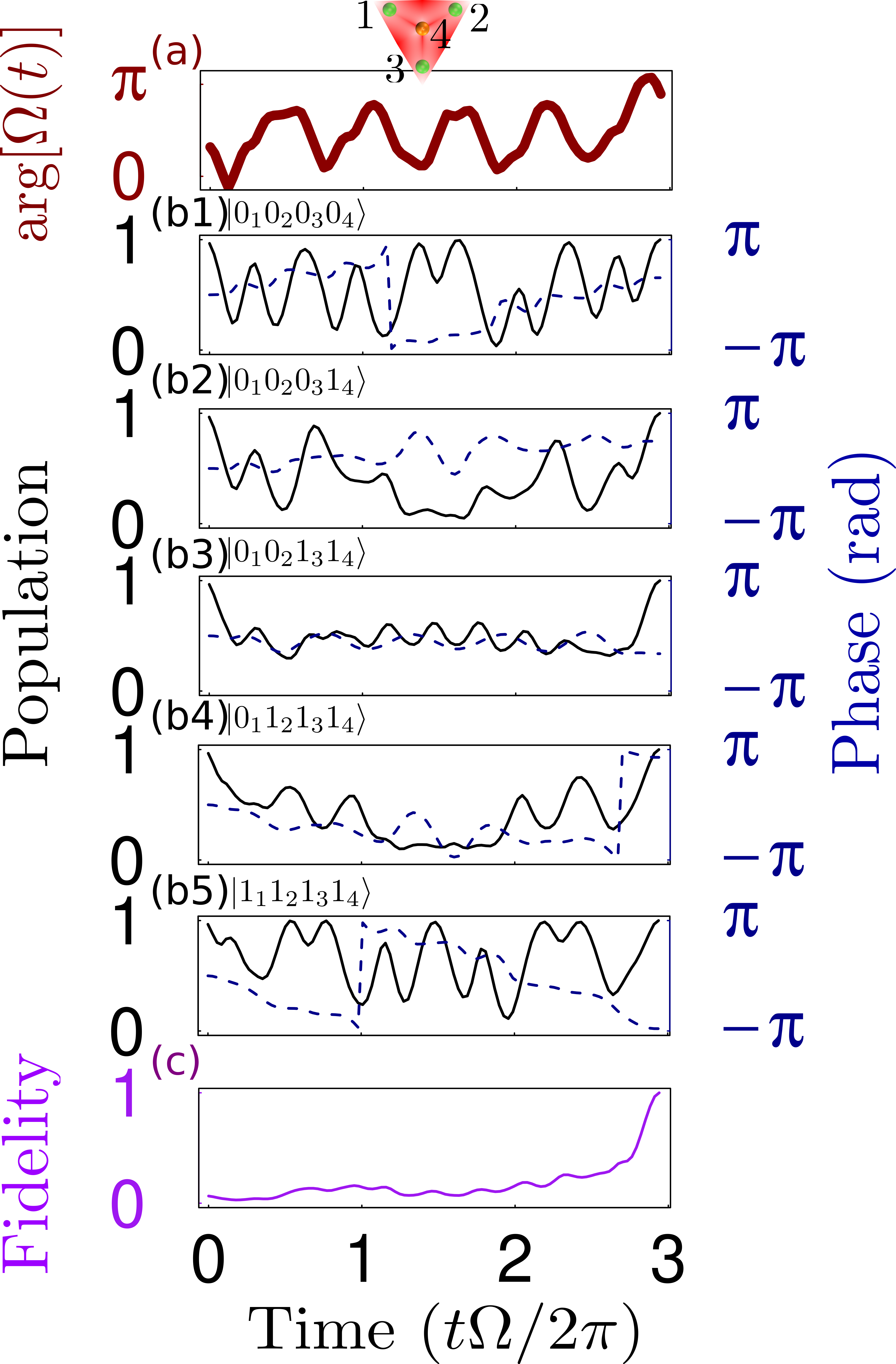}
\caption{(a) Phase of Rydberg Rabi frequency $\Omega(t)$ for realizing the 4-qubit QFG with $t_{\text{\tiny g}}=2.97\times\frac{2\pi}{\Omega}$ and $\Delta=2\Omega$, where the magnitude of the Rabi frequency is equal to $\Omega$ and the phase profile was found numerically~\cite{ShiLu2024}. The four atoms have equal distances, i.e., they are trapped at the vortexes of a tetrahedron. The red triangle represents the effect of looking through the tetrahedron. (b1-b5) The solid and dashed curves show the population and phase of the ground-state component of the wavefunction with the input state $\lvert0_10_20_30_4\rangle,\lvert0_10_20_31_4\rangle,\lvert0_10_21_31_4\rangle,\lvert0_11_21_31_4\rangle$, and $\lvert1_11_21_31_4\rangle$, respectively. The state dynamics for other input states are not shown because they are similar to those shown here: the state dynamics for the input states $\lvert 0_10_21_30_4\rangle,\lvert 0_11_20_30_4\rangle,\lvert 1_10_20_30_4\rangle$ are identical to that of $\lvert 0_10_20_31_4\rangle$, and similarly for other input states. (c) Fidelity of the QFG with Rydberg-state decay ignored. With the data in (b1-b5) we get the average Rydberg superposition time $t_{\text{Ryd}}\approx \frac{3.18\pi}{\Omega}$, leading to a Rydberg-state decay $3.18\pi/(\Omega\tau)$, where $\tau$ is the lifetime of the Rydberg state.   \label{figure-gate} }
\end{figure}

\section{Fast quantum ferromagnetic gate}\label{sec02}
\subsection{Benefit of using nuclear-spin qubits}\label{sec02A}
Before launching to the detail of the nuclear-spin-based QFG, we would like to first point out why it is beneficial to realize QFG with nuclear-spin qubits of alkaline-earth like atoms. The realization of the QFG needs to create a $\pi$ phase to an arbitrary superposition of the two state components in Eq.~(\ref{2state}), which means that in general we need to excite both $\lvert 0\rangle$ and $\lvert 1\rangle$ of the atoms to Rydberg states, where $\lvert 0\rangle$ and $\lvert 1\rangle$ are the two states for defining a qubit. If one tries to realize the QFG by alkali-metal atoms, which should be possible as discussed in Appendix~\ref{app-A}, the two hyperfine levels of the ground states in frequently used heavy alkali-metal atoms, cesium and rubidium, are separated by several GHz. Then, two sets of Rydberg laser fields are required if hyperfine qubits of cesium and rubidium atoms are used, with one set for exciting $\lvert 0\rangle$ to Rydberg state, and the other for exciting $\lvert 1\rangle$ to Rydberg state.

In contrast, only one Rydberg laser field is required for exciting both $\lvert 0\rangle$ and $\lvert 1\rangle$ to Rydberg state. This is because for nuclear-spin qubits of alkaline-earth like atoms, two nearby nuclear-spin Zeeman substates $\lvert 0\rangle$ and $\lvert 1\rangle$ can be nearly degenerate due to that the nuclear-spin g-factor is three orders of magnitude smaller than that of electrons~\cite{Shi2021}. The near degeneracy of nuclear spin states in a Gauss-scale magnetic field is a special strength of nuclear-spin-possessing alkaline-earth-like atoms, and it can enable hyperentanglement~\cite{Shi2021pra}, electrons-nuclei entanglement~\cite{Shi2024}, and fast nuclear-spin gates compatible with large-scale atomic arrays~\cite{ShiLu2024}. By the fact that one set of laser fields can simultaneously excite both $\lvert 0\rangle$ and $\lvert 1\rangle$ to Rydberg states, the experimental overhead for realizing the QFG is substantially reduced.

\subsection{QFG by nuclear-spin qubits}\label{sec02B}
This article considers QFG with four atoms because the four two-atom separations can be equal in a tetrahedron so as to have strong Rydberg blockade. For the purpose of creating GHZ state, one among of the four atoms is an ancilla, i.e., an atom helping to entangle the other three data atoms. For better blockade among the four atoms, we consider them located in the corners of a tetrahedron as shown in Fig.~\ref{figure-product-GHZ}(a), where the length of the side  of the tetrahedron is $\mathbb{L}$. The three data atoms, labeled 1, 2, and 3, lie in the x-y plane, and the ancilla atom, labeled a, lies in a plane away by $\sqrt{6}\mathbb{L}/3$. A qubit is defined by two nuclear-spin Zeeman substates $\lvert 0\rangle$ and $\lvert 1\rangle$ in the ground or clock states of an alkaline-earth like atom. When $\lvert 0(1)\rangle$ is excited to Rydberg states, the state $\lvert 1(0)\rangle$ can also be excited to Rydberg states due to that the two Rydberg excitations of both qubits have a frequency difference which is on MHz scale in a Gauss-scale magnetic field~\cite{Chen2022}. To have larger Rydberg Rabi frequency of tens of MHz, we suppose that the qubit is defined in the clock state. In experiment, $\Omega$ over $2\pi\times15$~GHz~\cite{Muniz2025} is realizable with qubits in the clock state of $^{171}$Yb. One Rydberg laser sent to them can induce the two ground-Rydberg transitions
\begin{eqnarray}
\lvert 0_\alpha\rangle  & \xrightarrow{\Omega(t),\Delta}  & \lvert r_\alpha\rangle ,\nonumber\\
\lvert 1_\alpha\rangle   & \xrightarrow{\Omega(t),-\Delta} &\lvert R_\alpha\rangle ,\label{twoTran}
\end{eqnarray}
for each atom $\alpha\in\{1,2,3,\text{a}\}$. Here, the two transitions above have opposite detunings $\Delta$ and $-\Delta$ when the central frequency of the laser is tuned to the middle of the Zeeman gap between the two Rydberg states $\lvert r\rangle$ and $\lvert R\rangle$. $\lvert r\rangle$ and $\lvert R\rangle$ are two different Zeeman substates of a principal Rydberg level~\cite{ShiLu2024}, and $2\Delta$ is the Zeeman splitting between them. With dipole and rotating-wave approximation, the Hamiltonian in the rotating frame is
\begin{eqnarray}
\hat{H}&=&  \sum_{\alpha\in\{1,2,3,\text{a}\}} \left\{  \left[ \left(\frac{\Omega(t)}{2} (\lvert r_\alpha \rangle   \langle 0_\alpha\rvert+\lvert R_\alpha \rangle   \langle1_\alpha\rvert)+\text{H.c.} \right) +  \Delta   (\lvert  r_\alpha  \rangle\langle r_\alpha\rvert-\lvert  R_\alpha  \rangle\langle R_\alpha\rvert)  \right]\right.\nonumber\\
&& \left. \otimes\left[ \bigotimes_{\beta\in\{1,2,3,\text{a}\},\beta\neq \alpha} (\lvert  0_\beta  \rangle\langle 0_\beta\rvert+\lvert  1_\beta  \rangle  \langle 1_\beta\rvert+ \lvert  r_\beta  \rangle\langle r_\beta\rvert+\lvert  R_\beta  \rangle\langle R_\beta\rvert)\right]\right\}+\hat{H}_{\text{Ryd}},\label{Ryd-H-0}
\end{eqnarray}
where H.c. denotes Hermitian conjugate, and $\hat{H}_{\text{Ryd}}$ is the Hamiltonian involving Rydberg interaction,
\begin{eqnarray}
 \hat{H}_{\text{Ryd}} &=& \sum_{\alpha\in\{1,2,3,\text{a}\}}\sum_{\beta\in\{1,2,3,\text{a}\},\beta\neq \alpha} V_{\alpha\beta} \left\{  \left (\lvert  r_\alpha r_\beta   \rangle\langle r_\alpha r_\beta\rvert+\lvert  r_\alpha R_\beta   \rangle\langle r_\alpha R_\beta\rvert+\lvert  R_\alpha r_\beta   \rangle\langle R_\alpha r_\beta\rvert+\lvert  R_\alpha R_\beta   \rangle\langle R_\alpha R_\beta\rvert   \right)\right.\nonumber\\
&& \left. \otimes\left[ \bigotimes_{\zeta\in\{1,2,3,\text{a}\},\zeta\neq \alpha,\beta} (\lvert  0_\zeta  \rangle\langle 0_\zeta\rvert+\lvert  1_\zeta  \rangle  \langle 1_\zeta\rvert+ \lvert  r_\zeta  \rangle\langle r_\zeta\rvert+\lvert  R_\zeta  \rangle\langle R_\zeta\rvert)\right]\right\}, \label{Ryd-H}
\end{eqnarray}
where we have assumed that the blockade interaction between two atoms are equal when each of them is either in $\lvert r\rangle$ or in $\lvert R\rangle$.

Without Rydberg blockade, the two qubit states will accumulate opposite phases linear in time, which can be equal to $\pi$ and $-\pi$ respectively at a certain time~\cite{Shi2017}. Since both the phases $\pm \pi$ yield $e^{\pm i\pi}=-1$, it is possible to have useful phase accumulation in a multi-qubit system. With Rydberg blockade, the phase accumulation in different two-qubit input states has a nonlinear dependence on $\Delta$~\cite{Shi2024}, so that it becomes useful to use numerical optimization to find a pulse profile for realizing a Rydberg gate~\cite{ShiLu2024}. Nonetheless, the numerical method employed in Ref.~\cite{ShiLu2024}, which was originally proposed in Ref.~\cite{Khaneja2005}, needs to calculate the time-evolution operator for a tiny time step throughout the total duration $t_{\text{\tiny g}}$ of the gate for each iteration of optimization. When we include Eq.~(\ref{Ryd-H}) in the optimization, the Hamiltonian matrix that should be propagated has a dimension of 35, which is difficult to handle even in a HPC center. But if we assume that the blockade is so strong that states with two or more Rydberg excitations are not populated, the dimension of the Hamiltonian matrix that shall be propagated in the optimization is only thirteen, a case that can be handled in a HPC center within days of simulation. In the strong blockade limit, this approximation can yield useful pulse profiles as experimentally verified in recent high-fidelity gate demonstrations with fidelity up to 0.995 in Ref.~\cite{Evered2023}, 0.9962 in Ref.~\cite{Finkelstein2024}, 0.9971 in Ref.~\cite{tsai2024fid}, 0.9972 in Ref.~\cite{Muniz2025}, and 0.9978 in Ref.~\cite{senoo2025}. 

By using the numerical method employed in Ref.~\cite{ShiLu2024} as detailed in Appendix~\ref{app-B}, we find that the gate can be realized within a time $t_{\text{\tiny g}}$ less than $3\times\frac{2\pi}{\Omega}$, where we assume constant magnitude $\Omega$ in the Rydberg Rabi frequency~\cite{Jandura2022}. The phase profile is shown in Fig.~\ref{figure-gate}(a), and the state evolution for typical input eigenstates are shown in Fig.~\ref{figure-gate}(b1-b5). Following Ref.~\cite{Jandura2022}, the Rydberg-state decay is not included in the numerical optimization and the decay error is analyzed later. The gate realized by the pulse of Fig.~\ref{figure-gate}(a) is
\begin{eqnarray}
 \lvert 0_10_20_30_4\rangle&\mapsto& e^{ix_1}\lvert 0_10_20_30_4\rangle,\nonumber\\
 \lvert 0_10_20_31_4\rangle&\mapsto& e^{ix_2}\lvert 0_10_20_31_4\rangle,\nonumber\\
 \lvert 0_10_21_31_4\rangle&\mapsto& e^{ix_3}\lvert 0_10_21_31_4\rangle,\nonumber\\
 \lvert 0_11_21_31_4\rangle&\mapsto& e^{ix_4}\lvert 0_11_21_31_4\rangle,\nonumber\\
 \lvert 1_11_21_31_4\rangle&\mapsto& e^{ix_5}\lvert 1_11_21_31_4\rangle, \label{QFG01}
\end{eqnarray}
where the final populations are all 1, and their final phases $x_1,x_2,x_3,x_4$, and $x_5$ are 0.9749152,1.549656, -1.017333,2.699127, and -3.009320, respectively. In the equation above, the phase accumulation for
$\lvert 0_10_21_30_4\rangle,\lvert 0_11_20_30_4\rangle,\lvert 1_10_20_30_4\rangle$ are not shown for it is identical to that of $\lvert 0_10_20_31_4\rangle$, and similar for others. If we introduce $(y_0,y_1)=(-2.112464,1.603853)$ and $y=y_0-y_1$, then by using the single-qubit phase gate
\begin{eqnarray}
 \lvert 1\rangle&\mapsto& e^{iy}\lvert 1\rangle,\label{phase-gate0}
\end{eqnarray}
the gate in Fig.~\ref{figure-gate} transforms to
\begin{eqnarray}
 \lvert 0_10_20_30_4\rangle\mapsto& -e^{4iy_0}&\lvert 0_10_20_30_4\rangle,\nonumber\\
 \lvert 0_10_20_31_4\rangle\mapsto& e^{4iy_0}&\lvert 0_10_20_31_4\rangle,\nonumber\\
 \lvert 0_10_21_31_4\rangle\mapsto& e^{4iy_0}&\lvert 0_10_21_31_4\rangle,\nonumber\\
 \lvert 0_11_21_31_4\rangle\mapsto& e^{4iy_0}&\lvert 0_11_21_31_4\rangle,\nonumber\\
 \lvert 1_11_21_31_4\rangle\mapsto& -e^{4iy_0}&\lvert 1_11_21_31_4\rangle.\label{QFG02}
\end{eqnarray}
So, there will be an overall phase $4y_0$ to each input state. The overall phase to the QFG is trivial and has no physical consequence for the content in this article. The single-qubit phase gate is used for clarity, while in the appendix we will show that there is no need to carry out Eq.~(\ref{phase-gate0}) when using the QFG for GHZ-state generation if we use appropriate phase in the external field in the gluing cycle.

One can see that the QFG in Fig.~\ref{figure-gate} is perfect because we have excluded the Rydberg-state decay and the blockade error. Below, we estimate the fidelity the gate with fundamental and technical issues. 

\subsection{Error due to Rydberg-state decay and the blockade error}\label{sec02C}
There are two fundamental errors that can not be removed, the Rydberg-state decay and the blockade error.
\subsubsection{Rydberg-state decay}\label{sec02C1}
The infidelity due to Rydberg-state decay can be estimated by averaging over the sixteen input states. For the gate pulse of Fig.~\ref{figure-gate}, we find that the average Rydberg superposition time $t_{\text{Ryd}}\approx\frac{3.18\pi}{\Omega}$, so the Rydberg-state decay induced gate error is $t_{\text{Ryd}}/\tau$, where $\tau$ is the lifetime of the Rydberg state. $\tau$ was measured to be $65~\mu$s for the $6sns~^3S_1$ state of $^{171}$Yb with a principal quantum number $n=59$~\cite{Ma2023}.
The radiative lifetime is limited by two processes, the spontaneous emission and black-body radiation, where the former approximately scales as $n^{-3}$ for alkali-metal atoms with $n$ the effective principal quantum number, and the latter is proportional to $n^{-2}$~\cite{Saffman2010}. At room temperature, the radiative decay of Rydberg atoms is dominated by the black-body radiation. Take rubidium which is an alkali-metal atom as an example, $\tau$ are 330~$\mu$s and $1.2$~ms at room temperature and 4.2~K, respectively~\cite{Beterov2009}. Due to lack of quantum defect numbers for triplet s,~p,~d Rydberg states that the ytterbium $~^3S_1$ Rydberg state can transit to in a thermal photon bath, we assume an $n^{-2}$ scaling for $\tau$. So, we take the lifetime of the $6s100s~^3S_1$ state to be around $187~\mu$s according to the measured lifetime for $n=59$ in Ref.~\cite{Ma2023}. By assuming a Rabi frequency $\Omega/(2\pi)=10$~MHz, the Rydberg-state decay error is 
\begin{eqnarray}
 E_{\text{decay}}=8.5\times10^{-4}. \label{decay-e1}
\end{eqnarray}

\begin{figure}
\includegraphics[width=3.0in]
{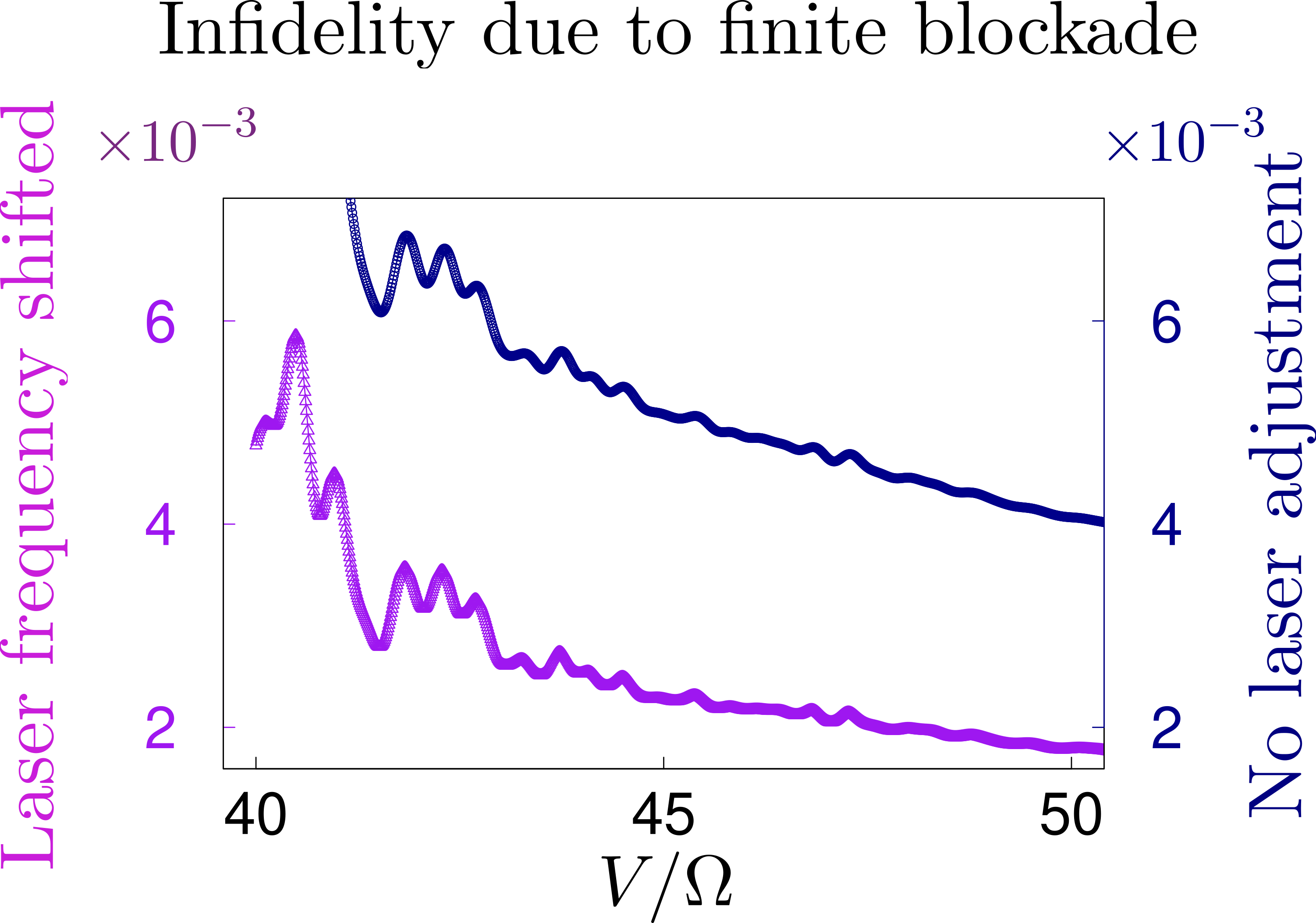}
\caption{The blue curve shows $10^3(1-F)$ as a function of $V/\Omega$ by using the pulse profile of Fig.~\ref{figure-gate}, where $F$ is defined in Eq.~(\ref{F0-definition}), and the simulation is with Hamiltonians listed in Appendix~\ref{app-C}. The purple curve shows $10^3(1-F)$ with $\pm\Delta$ shifted by $\Omega^2/V$ according to the method of Ref.~\cite{Levine2019} as detailed at the end of Appendix~\ref{app-C}. Because the detuning is given by laser frequency deducted by the transition frequency, a frequency shift $\Omega^2/V$ of the Rydberg laser means change of $\pm\Delta$ to $\Omega^2/V\pm\Delta$.     \label{figure-Verror} }
\end{figure}

The decay error can be reduced by using high-power laser fields. According to Ref.~\cite{Madjarov2020}, a 1266~nm seed laser at 2~W was used via two rounds of frequency doubling to generate the 317~nm UV laser with a power up to 0.1~W. Reference~\cite{Evered2023} realized high-fidelity Rydberg gates with 1013~nm laser of power up to $100$~W, and Ref.~\cite{6100atoms} realized a tweezer array with 6100 highly coherent atomic qubits thanks to near-infrared wavelengths provided by commercial fiber amplifiers which can yield continuous-wave laser powers that exceed 100~W. On the experimental side, however, though the clock-Rydberg excitation of two-valence atoms can be achieved with a UV Rabi frequency over $2\pi\times15$~MHz~\cite{Muniz2025}, the blockade mechanism requires that $\Omega$ shall be much smaller than the Rydberg blockade strength~\cite{Shi2021qst}. Current experiments usually choose Rydberg states with $n$ around 60, and the interaction between Rydberg atoms with frequently employed atomic separations is of order $2\pi\times100$~MHz~\cite{Muniz2025}. This makes it necessary to use not too large $\Omega$, which is why though Ref.~\cite{Muniz2025} showed that $\Omega/2\pi>15~$MHz was readily available for exciting the $6s65s~^3S_1$ state of $^{117}$Yb, they used a small $\Omega$ to optimize their CZ gate fidelity to a value of 0.9972. But when we choose high-n Rydberg states and place atoms with separations of several $\mu$m, the interaction is nearer to the resonant dipole-dipole interaction~\cite{Shi2025pra}, which can be quite strong~\cite{Saffman2010}. These mean that it is feasible to assume large $\Omega$ if high-n Rydberg states are used.  According to laser powers over 100~W as experimentally demonstrated in Rydberg-atom technologies~\cite{Madjarov2020,Evered2023,6100atoms}, it is possible to assume $\Omega/(2\pi)=30$~MHz, which leads to $E_{\text{decay}}=2.8\times10^{-4}$.

It is also possible to use low-lying Rydberg state and small Rydberg Rabi frequency with Rydberg decay detection~\cite{senoo2025}. The above details show that if a ytterbium Rydberg state with $n\approx58$ excited by a $\Omega=2\pi\times3$~MHz without Rydberg decay detection as in Ref.~\cite{senoo2025}, the error would be about 0.009. But Rydberg decay detection can enhance the lifetime to a much longer effective Rydberg lifetime $1.2$~ms~\cite{senoo2025}, which leads to $E_{\text{decay}}=4.4\times10^{-4}$ even with a small $\Omega=2\pi\times3$~MHz. If we assume a similar factor of enhancement~\cite{senoo2025} of the Rydberg state lifetime for $n=100$, one can estimate $\tau$ to be 3.5~ms. This would lead to 
\begin{eqnarray}
 E_{\text{decay}}=1.5\times10^{-4}, \label{decay-e2}
\end{eqnarray}
even with a small $\Omega=2\pi\times3$~MHz.

\subsubsection{Blockade leakage}\label{sec02C2}
For the blockade error, it is in principle possible to numerically search a pulse profile to have a perfect gate with a certain finite Rydberg interaction~\cite{ShiLu2024}. However, we found it difficult due to the large Hilbert space with a finite blockade. But it was shown in Ref.~\cite{ShiLu2024} that for the two-qubit gate with nuclear spin qubits, the blockade induced gate error with reasonable position fluctuation induced variation of blockade strength is below $10^{-4}$. Because the gate here and that of Ref.~\cite{ShiLu2024} have a similar mechanism of exciting the two transitions of Eq.~(\ref{twoTran}) with opposite detunings for each atom, the blockade leakage occurs to each atom in a similar manner. The blockade leakage denotes unwanted population of multi-Rydberg states. Because a state with $m$ Rydberg atoms should be excited from a state with $m-1$ Rydberg atoms, where $m=2,3$, or 4, the probability to have a three-atom Rydberg state is much smaller than that to have a two-atom Rydberg state. Similarly, the probability to have a two-atom Rydberg state is much smaller than that to have a one-atom Rydberg state. So, we ignore states with more than three Rydberg atoms, as shown in Appendix~\ref{app-C}. In this case, the probability to have a blockade leakage would be larger with a larger Rydberg superposition time $t_{\text{\tiny Ryd}}$. The two-qubit gate of Ref.~\cite{ShiLu2024} with $V=50\Omega$ and circular lasers has $t_{\text{\tiny Ryd}}=2.2\pi/\Omega$ and the error due to fluctuating blockade would be smaller than $E_{\text{blockade}}=10^{-4}$ with $|\Delta V/V|\leq 0.22$, where $V=50\Omega$ is the desired Rydberg interaction and $\Delta V$ is the deviation of the interaction. The QFG in Fig.~\ref{figure-gate} has $t_{\text{Ryd}}\approx \frac{3.18\pi}{\Omega}$, based on which we estimate that the blockade error here is 
\begin{eqnarray}
 E_{\text{blockade}}=1.4\times10^{-4}. \label{blockade-e1}
\end{eqnarray}

The analysis above is based on the assumption that the optimization can yield a pulse profile that can yield a perfect gate when $V$ does not fluctuate. Even with the pulse obtained with infinite Rydberg interaction $V$ as used in Fig.~\ref{figure-gate}, it is possible to suppress the blockade leakage. For example, it was shown in Ref.~\cite{Levine2019} that a finite Rydberg blockade can be treated as a renormalization of the energy of the Rydberg state. This means that in principle we can adjust the frequency of the Rydberg laser to partially compensate the finiteness of the blockade. As shown in Appendix~\ref{app-C}, we can shift the frequency of the Rydberg laser by $\Omega/V$. By using the Hamiltonians derived in Appendix~\ref{app-C}, we find that a finite $V=50\Omega$ causes an infidelity $1-F$ about $4.1\times10^{-3}$. But if we shift the laser frequency, the infidelity drops to $E_{\text{blockade}}=1.8\times10^{-3}$. When $V$ decreases, the infidelity increases as shown in Fig.~\ref{figure-Verror} where data with $V/\Omega>40$ are shown. If $V$ is very weak, for example, with $V=10\Omega$, $1-F$ would be 0.13 if we do not shift the laser frequency, and even with the frequency of the Rydberg laser shifted by $\Omega/V$, $1-F$ is as large as 0.078. But if $V$ is strong, the infidelity would be smaller. For instance, if we assume $V=90\Omega$, $1-F$ is $1.3\times10^{-3}$ without shifting the laser frequency, and is 
\begin{eqnarray}
 E_{\text{blockade}}=6.5\times10^{-4}, \label{blockade-e2}
\end{eqnarray}
with the laser frequency shifted. Though not easy, strong blockade is possible by placing atoms close enough~\cite{Saffman2010}. In theory, $V$ over $2\pi\times3$~GHz was assumed in designing Rydberg gate~\cite{Saffman2020}; in experiments, Ref.~\cite{Graham2022} employed blockade strengths over $2\pi\times 1$~GHz when studying quantum algorithms on a programmable gate-model neutral-atom quantum computer.

Large enough $V$ is possible if we consider using $^{171}$Yb for the QFG. Experimental aided multichannel quantum defect theory on the Rydberg states of $^{171}$Yb in Ref.~\cite{Peper2025} showed that for a large range of effective principal quantum number $n$, the $C_6$ van der Waals coefficient of the $^3S_1$ Rydberg state has a $\nu^{11}$ scaling as expected when no F\"{o}rster resonance occurs~\cite{Saffman2010}. The value of $C_6$ with $\nu\approx 54$ is $2\pi\times 34$~GHz$\mu$m$^6$ with two atoms perpendicular to the quantization axis, and Fig.~6 of Ref.~\cite{Peper2025} shows that the interaction is nearly isotropic. If we consider states with $n\sim100$, then $C_6$ would be $2\pi\times 30$~THz$\mu$m$^6$. With $\mathbb{L}=5.5~\mu$m, we expect $V\approx2\pi\times1$~GHz between two Rydberg atoms within one QFG as highlighted by the red triangle in Fig.~\ref{figure-product-GHZ}. This means that it is feasible to employ large $V$ for the QFG. When it comes to such a close qubit separation, the position fluctuation of qubits should be considered. If we include position fluctuation of the atomic qubit with typical variations of qubit positions~\cite{Graham2022}, Appendix~\ref{app-D} shows that the blockade error will increase by about 3\% when $V/\Omega=50$ if we take $\mathbb{L}=5.5~\mu$m.

Using high-n Rydberg state and strong laser leads to increased sensitivity to stray DC electric fields. The energy of the Rydberg atom becomes dependent on the unknown stray DC electric field around the atom because high-n Rydberg electron is too sensitive to stray DC electric fields. With an $n^7$ scaling for the polarizability~\cite{Booth2017}, the sensitivity of an $n=100$ Rydberg state would be 40 times of that of an $n=59$ Rydberg state. This means the $n=100$ Rydberg state would be 40 times more sensitive compared to that of the $n=59$ Rydberg state as employed in Ref.~\cite{Ma2023}. To tackle this, Ref.~\cite{Booth2017} proposed microwave-induced dressing of Rydberg states for reducing the sensitivity of atomic ground-to-Rydberg transitions to stray DC electric fields, which was tested in Ref.~\cite{Bohorquez_2023}. With a cesium Rydberg state of $n\sim 90$, Ref.~\cite{Booth2017} showed that the sensitivity to stray DC electric field can be reduced by 1600 times with two microwave fields for dressing the targeted Rydberg states. This means that it is possible to reduce the sensitivity to stray DC electric fields if we use a high-n Rydberg state.


\begin{figure}
\includegraphics[width=6.0in]
{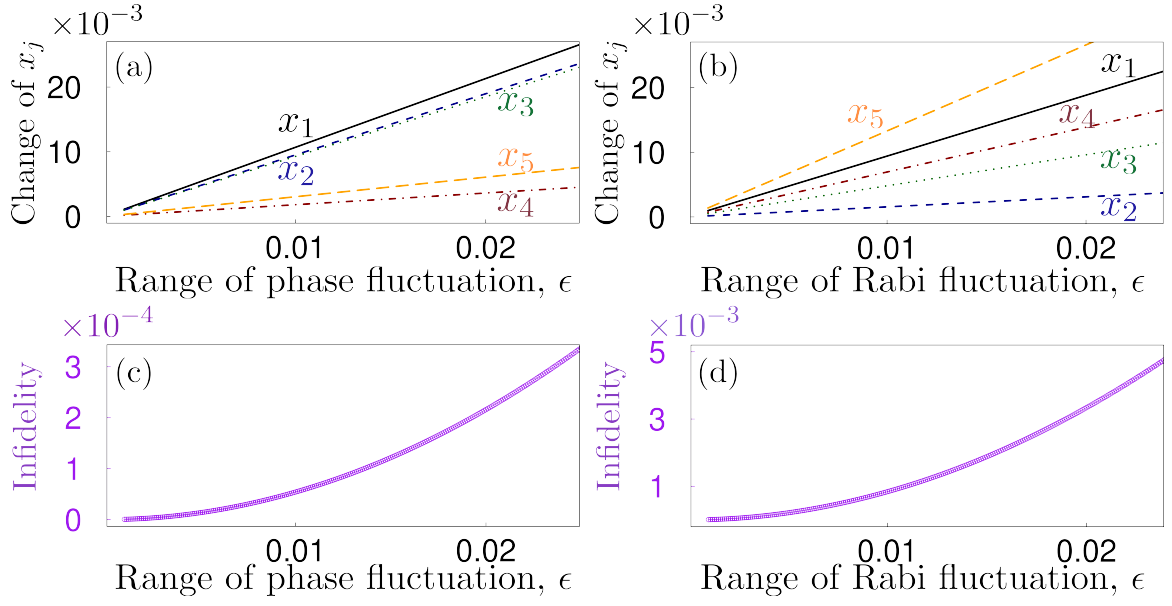}
\caption{Sensitivity of the final accumulated phases $x_j$ and the gate fidelity to small
phase and amplitude fluctuations in $\Omega(t)$. The solid, short-dashed, dotted, dash-dotted, and long dashed curves in (a) and (b) show the fractional error of the angle $|(x_j'-x_j)/x_j|$ with $j=1,~2,~3,~4,$ and 5, respectively. A fractional variation $\delta_\phi$ in the phase and variation $\delta_{\Omega}$ in the amplitude of $\Omega(t)$ are used in (a) and (b), respectively, and the data is an average over results with $\delta_\phi$  and $\delta_{\Omega}$ uniformly sampled in $[-\epsilon,~\epsilon]$, respectively. The infidelity due to the fluctuation of phase and amplitude of $\Omega(t)$ is shown in (c) and (d), respectively.   \label{figure-Omega-change} }
\end{figure}

\begin{table*}[ht]
  \centering
  \begin{tabular}{|c|c|c|c|c|c| }
    \hline
  Rydberg decay  & Blockade leakage & Error in arg$[\Omega(t)]$ & Error in $|\Omega(t)|$  &   Doppler & $F_0$   \\\hline
  $ 8.5\times10^{-4} $~[Eq.~(\ref{decay-e1})] &  $ 6.5\times10^{-4} $~[Eq.~(\ref{blockade-e2} )] &  $ 2\times10^{-4} $~[Eq.~(\ref{phase-e})]&  $ 8.4\times10^{-4} $~[Eq.~(\ref{amplitude-e})]& $ 2.9\times10^{-4} $~[Eq.~(\ref{Doppler-e})]&$ 0.9972 $ \\
   \hline
     $ 1.5\times10^{-4} $~[Eq.~(\ref{decay-e2})]  &  $ 1.4\times10^{-4} $~[Eq.~(\ref{blockade-e1})]  &   $ 0.66\times10^{-4} $~[Eq.~(\ref{phase-e})]& $ 2.1\times10^{-4} $~[Eq.~(\ref{amplitude-e})] &  $ 0.32\times10^{-4} $~[Eq.~(\ref{Doppler-e})]&  $ 0.9994 $  \\
   \hline
  \end{tabular}
  \caption{Estimated errors due to Rydberg-state decay, imperfect blockade, noise in laser phase and amplitude, and Doppler effect. The second~(third) row of the table shows the larger~(smaller) error estimated in the text, with the reference equations shown here. Details for the analyses that lead to the value shown here can be found around the quoted equations.     }\label{table-error}
\end{table*}

\subsection{Error due to noise}\label{sec02D}
Noise from laser and atomic motion can also hamper the gate. It was pointed out in Ref.~\cite{Jiang_2023} that for Rydberg excitations, the noise from lasers predominantly occurs in the phase and amplitude variables. Detailed experimental examinations revealed that the noise can be well captured by a combination of white noise and a Gaussian servo bump~\cite{Jiang_2023}. The white noise is from the laser source, while the the gate error is dominated by the white-noise background, and  servo bump is from the frequency locking system. Reference~\cite{Jiang_2023} found that the gate error is dominated by the white-noise background when the servo bump peak is well separated from the Rabi frequency. In the experiment of Ref.~\cite{Graham2022}, a very narrow linewidth for the Rydberg lasers was demonstrated with servo resonance peaks below -50 dBC for carrier frequencies over 20 kHz. For servo bump peak below $1$~MHz~\cite{Jiang_2023}, the infidelity from the phase noise will be dominated by the white noise when we consider $\Omega$ over $2\pi\times10$~MHz. For a quantum gate of duration $\frac{\pi (2N+1)}{\Omega}$ with $N$ an nonnegative integer, Ref.~\cite{Jiang_2023} showed that the error due to the phase noise is $E_{\text{\tiny phase}} = \frac{2\pi^3 (2N+1)S_f}{3\Omega},$
where $S_f$ is the spectral density of frequency noise, which is constant in the case of white noise. For the QFG of Fig.~\ref{figure-gate}, we have $N \approx 2.5$. With $S_f=100$Hz$^2/$Hz from Ref.~\cite{Jiang_2023}, the QFG has an error from the laser phase noise 
\begin{eqnarray}
 E_{\text{phase}}\approx (20,6.6)\times10^{-5} \label{phase-e}
\end{eqnarray}
with $\Omega = 2\pi\times(10,30)$~MHz, respectively.

For a numerical analysis, we would like to examine what occurs if arg$[\Omega(t)]$ has a fractional error $\delta_\phi$, i.e., instead of the desired arg$\Omega(t)$ shown in Fig.~\ref{figure-gate}(a), we have $\text{arg}[\Omega(t)](1+\delta_\phi)$
as the phase of the Rabi frequency. Further, we examine the fractional change of the angles $|x_j'-x_j|/x_j$ averaged over the $\phi$ in the interval $[-\epsilon,~\epsilon]$, where $x_j$ is the desired angles given below Eq.~(\ref{QFG01}). As shown in Fig.~\ref{figure-Omega-change}(a), the average fractional changes of the angles $x_j$ are small, being $\{2.1,~1.9,~1.9,~0.036,~0.061\}\%$ with $j=\{1,~2,~3,~4,~5\}$ even with $\epsilon=0.02$. Accordingly, the infidelity of the gate due to the phase fluctuation is tiny, less than $2.2\times10^{-4}$ when $\epsilon<0.02$ as shown in Fig.~\ref{figure-Omega-change}(c). Meanwhile, Fig.~\ref{figure-Omega-change}(b) shows the fractional change of the angles when there is a fractional error $\delta_{\Omega}$ in the amplitude of $\Omega(t)$, where the data is averaged over $\delta_{\Omega}\in[-\epsilon,~\epsilon]$. Compared to Fig.~\ref{figure-Omega-change}(a), the error in $x_4$ and $x_5$ shoots up in Fig.~\ref{figure-Omega-change}(b). Figure~\ref{figure-Omega-change}(d) shows that the gate error due to the amplitude fluctuation of $\Omega(t)$ is an order of magnitude larger than those in Fig.~\ref{figure-Omega-change}(c), leading to 
\begin{eqnarray}
 E_{\text{ampli}}\approx (2.1,~8.4)\times10^{-4} \label{amplitude-e}
\end{eqnarray}
with $\epsilon = \{0.005,~0.01\} $, respectively. This shows that the gate is prone to amplitude fluctuation in the laser field.

Doppler broadening will arise with atoms in free flight. To avoid differential light shift in the dipole trap, the trap is usually switched off during the Rydberg excitation. When we assume Rydberg Rabi frequency over $2\pi\times10$~MHz, the velocity change of the atom due to photon recoil can be ignored for effective atomic temperature $T_{\text{eff}}$ on the order $10~\mu$K. Then, we assume a constant velocity of the atom during the Rydberg excitation. As an estimate, we can examine the error by averaging data with each of the four atoms in a QFG flying with a speed $\pm v_{\text{rms}}$ where $v_{\text{rms}}=\sqrt{k_{\text{\tiny B}}T_{\text{eff}}/m}$, where $k_{\text{\tiny B}}$ is the Boltzmann constant and $m$ is the mass of the atom. In this case, the error with more random velocities in the four atoms will be larger, as shown in Fig.~9 of Ref.~\cite{Shi2018Accuv1} with an example of two-qubit gate. This can be understood by that two atoms with opposing flying directions would experience opposite Doppler detunings, which partially suppress the undesired phase shift to the final atomic state. Then, we consider two extreme case, i.e., all the atoms with a speed $v_{\text{rms}}$ or $-v_{\text{rms}}$, and the error can be averaged over these two cases for each $T_{\text{eff}}$. With a $2\pi\times10$~MHz Rydberg Rabi frequency for the 302~nm laser excitation of the $^{171}$Yb clock state as an example, numerical simulation shows that the infidelity of the QFG due to Doppler broadening is $\{2.9,~4.4,~5.8,~7.3,~8.7\}\times10^{-4}$ at $T_{\text{eff}}=\{10,~15,~20,~25,~30\}~\mu$K. Note that the Doppler dephasing will be smaller with shorter gate. For example, with $|\Omega(t)|=2\pi\times30$~MHz, $\{3.2,~4.8,~6.5,~8.1,~9.7\}\times10^{-5}$ at $T_{\text{eff}}=\{10,~15,~20,~25,~30\}~\mu$K. Though realizing $T_{\text{eff}}<0.3~\mu$K is possible in experiments~\cite{senoo2025}, we assume a worse case with $T_{\text{eff}}=10~\mu$K, at which the Doppler error is
\begin{eqnarray}
 E_{\text{\tiny Dop}}\approx (2.9,~0.32)\times10^{-4} \label{Doppler-e}
\end{eqnarray}
with $\Omega = 2\pi\times(10,30)$~MHz, respectively.

\begin{figure*}
\includegraphics[width=7.0in]
{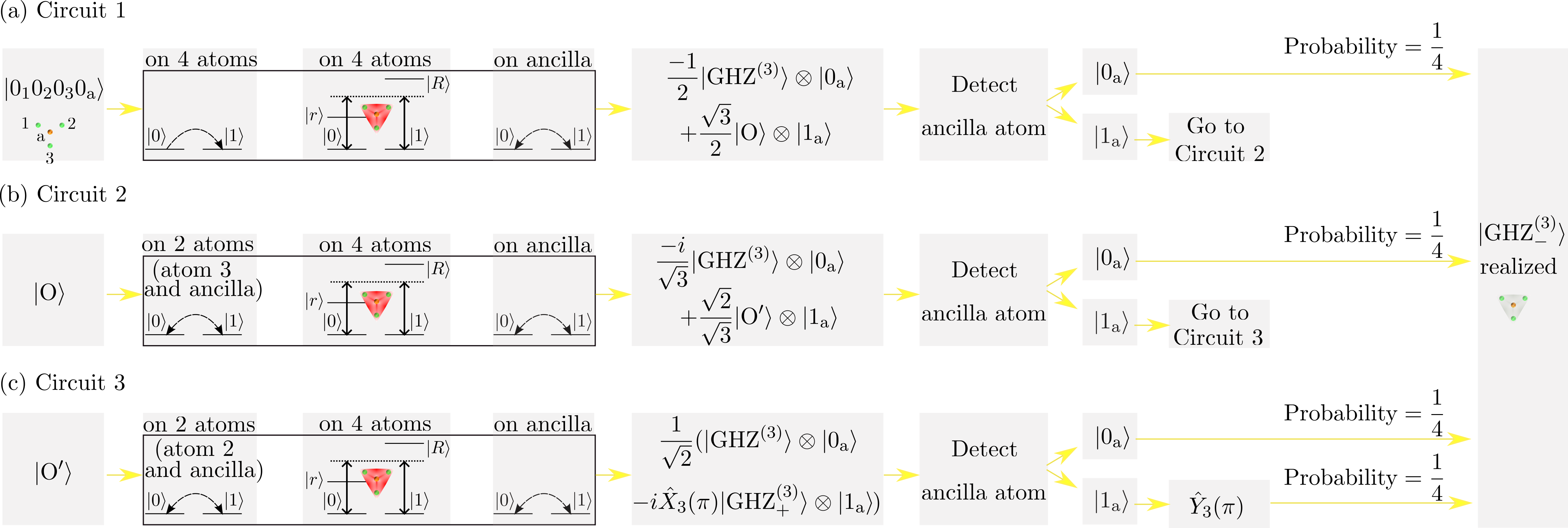}
\caption{Illustration of the elementary gluing cycle via the creation of $\lvert \text{GHZ}^{(3)}_-\rangle$ from a product state of three data atoms, labeled 1, 2, and 3, and one ancillary atom labeled a. The red triangle illustrates the effect of looking through a tetrahedron, in the four vertexes of which the four atoms are trapped. (a) The probability to have the $\lvert \text{GHZ}^{(3)}\rangle$ created in Circuit 1 is $p_{1-\text{yes}}=1/4$, and the remaining probability $p_{1-\text{no}}=3/4$ is to go to Circuit 2. (b) Within Circuit 2, the conditional probability to have the $\lvert \text{GHZ}^{(3)}\rangle$ created is $1/3$, so that the probability to succeed in Circuit 2 is $p_{2-\text{yes}}=p_{1-\text{no}}\cdot(1/3)=1/4$. The probability to go to Circuit 3 is $p_{2-\text{no}}=p_{1-\text{no}}\cdot(2/3)=1/2$. (c) In Circuit 3 without the need of a $\pi$ rotation around the y axis, the conditional probability to have $\lvert \text{GHZ}^{(3)}_-\rangle$ created is $1/2$, so the probability to end with $\lvert \text{GHZ}^{(3)}_-\rangle$ at the detection of Circuit 3 is $p_{3-\text{yes}}=p_{2-\text{no}}\cdot(1/2)=1/4$. In Circuit 3 with the need of $\hat{Y}_3(\pi)$, it is sure to end with $\lvert \text{GHZ}^{(3)}_-\rangle$.   \label{figure-product-GHZ} }
\end{figure*}

Another noise is from the light-shift inhomogeneity across the array. In Ref.~\cite{Graham2022} where a $7\times7$ array with spacing 3~$\mu$m was experimentally used, the resonance light shift is on the order of $2\pi\times100~$Hz. For the analysis of GHZ state generation, we will consider an atomic array in Fig.~\ref{figure-product-GHZ} which has an extension of $11.5\sqrt{3}\mathbb{L}$ and $14.5\mathbb{L}$ along the vertical and horizontal directions, respectively, where $\mathbb{L}$ is the length of the side of the tetrahedron. Larger system would suffer from stronger light inhomogeneity. So, we assume a large $\mathbb{L}=10$~$\mu$m. With a similar light inhomogeneity as in  Ref.~\cite{Graham2022}, the resonance light shift in the array of Fig.~\ref{figure-product-GHZ} would be less than $2\pi\times1~$kHz even with $\mathbb{L}=10$~$\mu$m. This light shift resembles a detuning, and numerical simulation with a Rydberg Rabi frequency $2\pi\times10$~MHz shows that a change $\pm2\pi\times1~$kHz in the frequency of the Rydberg transitions leads to an infidelity less than $10^{-7}$. This error is orders of magnitude smaller than other errors shown above, and hence is not included in Table~\ref{table-error}.

\section{GHZ generation by QFG}\label{sec03}
The basic step in the measurement-based GHZ state generation is a gluing cycle consisting of up to three gate-measurement circuits, where each circuit involves the preparation of an entangled state and a projection measurement of the ancilla. In an experimental perspective, the most technically demanding part is the state detection of the ancilla. However, we find that with one gluing cycle consisting of up to three gate-measurement circuits, one can create a 3-qubit GHZ state deterministically. The same gluing cycle can be used to glue three smaller GHZ states $\lvert \text{GHZ}^{(\text{m})}\rangle$ to form a larger GHZ state $\lvert \text{GHZ}^{(\text{3m})}\rangle$ when assisted by one ancillary atom. 

\subsection{A 3-qubit GHZ state by one gluing cycle}\label{sec03A}
The elementary gluing cycle is easily understood in the creation of a 3-qubit GHZ state as illustrated in Fig.~\ref{figure-product-GHZ}. We consider a product state of a 4-qubit system
\begin{eqnarray}
\lvert 0_10_2 0_30_{\text{a}}\rangle\equiv\lvert 0_10_2 0_3\rangle\otimes\lvert0_{\text{a}}\rangle ,\label{s01}
\end{eqnarray}
where the subscripts 1, 2, and 3 label the three data qubits, while the subscript ``a'' for the fourth qubit denotes the ancillary qubit. With a probability $1/4$, it is possible to use only one gate-measurement circuit to realize a 3-qubit GHZ state. The measurement can produce undesired results with a probability $3/4$, so more gate-measurement circuits may be needed. Below, we show that at most three gate-measurement circuits are needed in the worst case, where the three circuits are labeled as Circuit 1, Circuit 2, and Circuit 3, respectively, shown in Fig.~\ref{figure-product-GHZ}. Both Circuit 1 and Circuit 2 consist of three laser pulses and one measurement operation, while Circuit 3 consists of four laser pulses and one measurement operation.  \newline

{\bf Circuit 1-pulse 1}. Circuit 1 is shown in Fig.~\ref{figure-product-GHZ}(a). For the three data atoms and the ancilla, we excite the atomic transition $\lvert 0\rangle\leftrightarrow\lvert 1\rangle$ with a Rabi frequency $i\Omega_{\text{c}}$, i.e., with a Hamiltonian 
\begin{eqnarray}
\hat{H}_j(i\Omega_{\text{c}})=i\frac{\Omega_{\text{c}}}{2}\lvert 1_{\text{j}}\rangle\langle 0_{\text{j}}\rvert+\text{H.c.},\label{Hamiltonian-atom-j}
\end{eqnarray}
where j$=1,~2,~3$, or a. Then, Eq.~(\ref{s01}) becomes,
\begin{eqnarray}
\left(\prod_{i=1}^3\frac{\lvert0_{\text{i}}\rangle +\lvert1_{\text{i}}\rangle  }{\sqrt{2}}\right)\otimes \frac{\lvert0_{\text{a}}\rangle +\lvert1_{\text{a}}\rangle  }{\sqrt{2}}\label{s02}
\end{eqnarray}
with a $\frac{\pi}{2}$ pulse. The part enclosed by $(\cdots)$ in Eq.~(\ref{s02}) can be expanded, leading to
\begin{eqnarray}
\frac{1}{2}\left(\frac{\lvert0_{1}0_{2}0_{3}\rangle +\lvert1_{1}1_{2}1_{3}\rangle   }{\sqrt{2}}+ \sqrt{3}\lvert \text{O}\rangle \right)\otimes \frac{\lvert0_{\text{a}}\rangle +\lvert1_{\text{a}}\rangle  }{\sqrt{2}},\label{s04}
\end{eqnarray}
where
\begin{eqnarray}
\lvert\text{O}\rangle &=&  \frac{1}{\sqrt{6}}(\lvert0_{1}0_{2}1_{3} \rangle + \lvert0_{1}1_{2}0_{3} \rangle + \lvert1_{1}0_{2}0_{3} \rangle + \lvert0_{1}1_{2}1_{3} \rangle + \lvert1_{1}0_{2}1_{3} \rangle + \lvert1_{1}1_{2}0_{3} \rangle).\label{operatorO}
\end{eqnarray} \newline
{\bf Circuit 1-pulse 2}. To illustrate the gluing of the states by QFG, Eq.~(\ref{s04}) can be rewritten as
\begin{eqnarray}
\frac{\sqrt{3}}{2}\lvert \text{O}\rangle \otimes \frac{\lvert0_{\text{a}}\rangle +\lvert1_{\text{a}}\rangle  }{\sqrt{2}} + \frac{1}{4}\left(    \lvert0_{1}0_{2}0_{3}\rangle\otimes \lvert0_{\text{a}}\rangle +\lvert1_{1}1_{2}1_{3}\rangle\otimes \lvert0_{\text{a}}\rangle +   \lvert0_{1}0_{2}0_{3}\rangle\otimes \lvert1_{\text{a}}\rangle +\lvert1_{1}1_{2}1_{3}\rangle\otimes \lvert1_{\text{a}}\rangle    \right) .\label{s04-vary1}
\end{eqnarray}
By using the 4-qubit QFG which induces a $\pi$ phase for the following two state components
\begin{eqnarray}
\{\lvert 0_10_2 0_{3}\rangle\otimes \lvert0_{\text{a}}\rangle ,\lvert 1_11_21_{3}\rangle\otimes \lvert1_{\text{a}}\rangle\}\label{s04-vary2}
\end{eqnarray}
but does nothing to other input state components, Eq.~(\ref{s04-vary1}) becomes
\begin{eqnarray}
&&\frac{\sqrt{3}}{2}\lvert \text{O}\rangle \otimes \frac{\lvert0_{\text{a}}\rangle +\lvert1_{\text{a}}\rangle  }{\sqrt{2}} + \frac{1}{4}\left(  -  \lvert0_{1}0_{2}0_{3}\rangle\otimes \lvert0_{\text{a}}\rangle +\lvert1_{1}1_{2}1_{3}\rangle\otimes \lvert0_{\text{a}}\rangle +   \lvert0_{1}0_{2}0_{3}\rangle\otimes \lvert1_{\text{a}}\rangle -\lvert1_{1}1_{2}1_{3}\rangle\otimes \lvert1_{\text{a}}\rangle    \right)\nonumber\\
&=&\frac{\sqrt{3}}{2}\lvert \text{O}\rangle \otimes \frac{\lvert0_{\text{a}}\rangle +\lvert1_{\text{a}}\rangle  }{\sqrt{2}}-
\frac{\lvert0_{1}0_{2}0_{3}\rangle -\lvert1_{1}1_{2}1_{3}\rangle   }{2^{3/2}}\otimes \frac{\lvert0_{\text{a}}\rangle -\lvert1_{\text{a}}\rangle  }{\sqrt{2}}\nonumber\\
&=&\frac{\sqrt{3}}{2}\lvert \text{O}\rangle \otimes \frac{\lvert0_{\text{a}}\rangle +\lvert1_{\text{a}}\rangle  }{\sqrt{2}}-\frac{1}{2} \lvert \text{GHZ}^{(3)}_-\rangle
 \otimes \frac{\lvert0_{\text{a}}\rangle -\lvert1_{\text{a}}\rangle  }{\sqrt{2}}.
\label{s05}
\end{eqnarray}
In this step, the duration for realizing the QFG is less than $6\pi/\Omega$, which can be less than $1~\mu$s with a MHz-scale Rydberg Rabi frequency~\cite{Muniz2025}, 
and the time to realize the single-qubit phase gate in Eq.~(\ref{phase-gate0}) can also be of order $1~\mu$s with the method of Ref.~\cite{Jenkins2022}. We label the duration for this step by $t_{\text{g}}$ which is mainly determined by the time for realizing the single-qubit phase gate. In Appendix~\ref{app-E}, however, we show that Eq.~(\ref{phase-gate0}) is not a necessary step if we use appropriate phase in the gluing cycle. In other words, $t_{\text{g}}$ can be equal to the duration of the QFG.
\newline
{\bf Circuit 1-pulse 3}. We excite the transition $\lvert 0\rangle\leftrightarrow\lvert 1\rangle$ of the ancillary atom with a Rabi frequency $i\Omega_{\text{c}}$ for a duration $\frac{\pi}{2\Omega_{\text{c}}}$, leading to 
\begin{eqnarray}
&& \frac{\lvert0_{\text{a}}\rangle +\lvert1_{\text{a}}\rangle  }{\sqrt{2}}\xrightarrow{i\Omega_{\text{c}}, ~\frac{\pi}{2} \text{pulse}}\lvert1_{\text{a}}\rangle,\label{c1p3laser1}\\
&& \frac{\lvert0_{\text{a}}\rangle -\lvert1_{\text{a}}\rangle  }{\sqrt{2}}\xrightarrow{i\Omega_{\text{c}},  ~\frac{\pi}{2} \text{pulse}}\lvert0_{\text{a}}\rangle,\label{c1p3laser2}
\end{eqnarray}
where Eq.~(\ref{c1p3laser1}) can be understood by that the combination of Circuit 1-pulse 1 and Circuit 1-pulse 3 on the ancilla is a $\pi$ pulse with a Rabi frequency $i\Omega_{\text{c}}$, which leads to $\lvert0_{\text{a}}\rangle\rightarrow \lvert1_{\text{a}}\rangle$. For Eq.~(\ref{c1p3laser2}), we first identify
\begin{eqnarray}
&&\frac{\lvert0_{\text{a}}\rangle -\lvert1_{\text{a}}\rangle  }{\sqrt{2}}\equiv\text{exp} \left[-it_{\frac{\pi}{2}}\hat{H}_{\text{a}}(-i\Omega_{\text{c}}) \right]\lvert0_{\text{a}}\rangle
\end{eqnarray}
where $t_{\frac{\pi}{2}}=\frac{\pi}{2\Omega_{\text{c}}}$. Then, because
\begin{eqnarray}
\lvert0_{\text{a}}\rangle&\equiv&\text{exp} \left[-it_{\frac{\pi}{2}}\hat{H}_{\text{a}}(i\Omega_{\text{c}}) \right]\text{exp} \left[-it_{\frac{\pi}{2}}\hat{H}_{\text{a}}(-i\Omega_{\text{c}}) \right]\lvert0_{\text{a}}\rangle\nonumber\\
&=&\text{exp} \left[-it_{\frac{\pi}{2}}\hat{H}_{\text{a}}(i\Omega_{\text{c}}) \right] \frac{\lvert0_{\text{a}}\rangle -\lvert1_{\text{a}}\rangle  }{\sqrt{2}},
\end{eqnarray}
we get Eq.~(\ref{c1p3laser2}), where the first line in the equation above is similar to a spin echo. As a result, this pulse changes Eq.~(\ref{s05}) to
\begin{eqnarray}
|\text{M}\rangle_{1}&=&\frac{\sqrt{3}}{2}\lvert \text{O}\rangle \otimes  \lvert1_{\text{a}}\rangle -
\frac{1}{2}\frac{\lvert0_{1}0_{2}0_{3}\rangle -\lvert1_{1}1_{2}1_{3}\rangle   }{\sqrt{2}}\otimes  \lvert0_{\text{a}}\rangle\equiv \frac{\sqrt{3}}{2}\lvert \text{O}\rangle \otimes  \lvert1_{\text{a}}\rangle -
\frac{1}{2}   \lvert \text{GHZ}^{(\text{3})}_-\rangle  \otimes\lvert0_{\text{a}}\rangle .\label{s06}
\end{eqnarray}
\newline
{\bf Circuit 1-measurement}. According to Eq.~(\ref{s06}), we can measure the state of the ancillary qubit, and there are two possible outcomes with the state being in $\lvert \text{O}\rangle \otimes  \lvert1_{\text{a}}\rangle$ or $\lvert \text{GHZ}^{(\text{3})}_-\rangle  \otimes\lvert0_{\text{a}}\rangle$ with respective probability amplitudes $\frac{\sqrt{3}}{2}$ and $\frac{1}{2}$. The square of the probability amplitudes is equal to the corresponding probabilities of $3/4$ and $1/4$, i.e., if the measurement result is $\lvert0\rangle$ for the ancilla, we have got the 3-qubit GHZ state $(\lvert0_{1}0_{2}0_{3}\rangle -\lvert1_{1}1_{2}1_{3}\rangle)/\sqrt{2}$ with a success probability $1/4$. So, on average, four rounds of the above four-step circuit will yield the 3-qubit GHZ state if we always start with Eq.~(\ref{s01}). However, always starting from Eq.~(\ref{s01}) means that state re-initialization is required whenever the measurement yields failure of the state generation. This is not optimal since state re-initialization is technique-demanding.

To reduce the overhead we should avoid mid-circuit state re-initialization. To understand this, we notice that the state detection can be via the strong $^1S_0\leftrightarrow ^1P_1$ transition as in Refs.~\cite{Muniz2025}. This imaging will wipe out the nuclear-spin coherence due to the hyperfine interaction in the $^1P_1$ state~\cite{PhysRevA.107.023102}. To make sure the measurement doesn't spoil the applicability of the ancilla for further use, one can use circularly polarized field to map the qubit state $|0\rangle$ to the ground state as analyzed in Ref.~\cite{Jia2024}, where a Rabi frequency $2\pi\times0.2$~MHz is realizable~\cite{Chen2022}. This transition will have no effect if the state is in $|1\rangle$. If the imaging via the $^1S_0\leftrightarrow~^1P_1$ transition is positive, i.e., an atom is detected, the state in Eq.~(\ref{s06}) collapses to
\begin{eqnarray}
\lvert \text{GHZ}^{(\text{3})}_-\rangle \otimes  \lvert {\text{Mixed state}}\rangle,\label{s06-an02}
\end{eqnarray}
where $\lvert {\text{Mixed state}}\rangle$ denotes a state of the ancilla without any coherence in the nuclear spin due to the imaging light. But if no atom is detected, it means that the atom is still in the state $|1\rangle$. In other words, the state in Eq.~(\ref{s06}) collapses to
\begin{eqnarray}
\lvert \text{O}\rangle \otimes  \lvert1_{\text{a}}\rangle, \label{s06-an01}
\end{eqnarray}
which is still a pure state due to that no optical pumping occurs since the ancillary atom is still in the clock state.  \newline
{\bf Circuit 2-pulse 1}. Circuit 2 is shown in Fig.~\ref{figure-product-GHZ}(b). If the measurement result is $\lvert1\rangle$ in the last step, we have got Eq.~(\ref{s06-an01}). For the third data atom, we excite the transition $\lvert 0\rangle\leftrightarrow\lvert 1\rangle$ with a Rabi frequency $\Omega_{\text{c}}$ for a $\pi$ pulse that induces
\begin{eqnarray}
&&  \lvert0_{\text{3}}\rangle \xrightarrow{\Omega_{\text{c}},~ \pi \text{pulse}} -i\lvert1_{\text{3}}\rangle,\nonumber\\
&&   \lvert1_{\text{3}}\rangle \xrightarrow{\Omega_{\text{c}},~ \pi \text{pulse}} -i\lvert0_{\text{3}}\rangle,\label{rabi-pi}
\end{eqnarray}
and for the ancilla, we excite the transition $\lvert 0\rangle\leftrightarrow\lvert 1\rangle$ with a Rabi frequency $-i\Omega_{\text{c}}$ for a $\pi/2$ plus,
\begin{eqnarray}
&&  \lvert0_{\text{a}}\rangle \xrightarrow{-i\Omega_{\text{c}},~ \frac{\pi}{2} \text{pulse}}\frac{\lvert0_{\text{a}}\rangle -\lvert1_{\text{a}}\rangle  }{\sqrt{2}}\nonumber\\
&&   \lvert1_{\text{a}}\rangle \xrightarrow{-i\Omega_{\text{c}}, ~\frac{\pi}{2} \text{pulse}}\frac{\lvert0_{\text{a}}\rangle +\lvert1_{\text{a}}\rangle  }{\sqrt{2}},
\end{eqnarray}
which can be understood by identifying that the two equations above are exactly time-reversal counterparts to Eq.~(\ref{c1p3laser1}) and Eq.~(\ref{c1p3laser2}). Then,  Eq.~(\ref{s06-an01}) becomes,
\begin{eqnarray}
 &&  \frac{-i}{×\sqrt{6}} ( \lvert0_{1}0_{2}0_{3} \rangle + \lvert0_{1}1_{2}1_{3} \rangle + \lvert1_{1}0_{2}1_{3} \rangle +\lvert0_{1}1_{2}0_{3} \rangle +\lvert1_{1}0_{2}0_{3} \rangle + \lvert1_{1}1_{2}1_{3} \rangle)\otimes \frac{\lvert0_{\text{a}}\rangle +\lvert1_{\text{a}}\rangle  }{\sqrt{2}} ,\label{an03}
\end{eqnarray}
where the phase for the state transforms $\lvert 0\rangle\rightarrow\lvert 1\rangle$ and $\lvert 1\rangle\rightarrow\lvert 0\rangle$ is $-\pi/2$ as understood in the standard picture of Rabi oscillation~\cite{Shi2017}. When the excitation for the data atom and that for the ancilla are simultaneous, the above state transform needs a duration of $\pi/\Omega_{\text{c}}$. \newline
{\bf Circuit 2-pulse 2}. Before examining what will happen if we use QFG, we rewrite Eq.~(\ref{an03}) as
\begin{eqnarray}
 &&  \frac{-i}{×\sqrt{6}}\bigg [ (\lvert0_{1}1_{2}1_{3} \rangle + \lvert1_{1}0_{2}1_{3} \rangle + \lvert0_{1}1_{2}0_{3} \rangle + \lvert1_{1}0_{2}0_{3} \rangle ) \otimes \frac{\lvert0_{\text{a}}\rangle +\lvert1_{\text{a}}\rangle  }{\sqrt{2}} + ( \lvert0_{1}0_{2}0_{3} \rangle +  \lvert1_{1}1_{2}1_{3} \rangle)\otimes \frac{\lvert0_{\text{a}}\rangle + \lvert1_{\text{a}}\rangle  }{\sqrt{2}}\bigg],\label{an03-v1}
\end{eqnarray}
and with the strategy as in Eqs.~(\ref{s04-vary1}),~(\ref{s04-vary2}), and~(\ref{s05}), one can show that Eq.~(\ref{an03-v1}) becomes
\begin{eqnarray}
&& \frac{-i}{×\sqrt{6}}\bigg [ (\lvert0_{1}1_{2}1_{3} \rangle + \lvert1_{1}0_{2}1_{3} \rangle + \lvert0_{1}1_{2}0_{3} \rangle + \lvert1_{1}0_{2}0_{3} \rangle ) \otimes \frac{\lvert0_{\text{a}}\rangle +\lvert1_{\text{a}}\rangle  }{\sqrt{2}}- ( \lvert0_{1}0_{2}0_{3} \rangle -  \lvert1_{1}1_{2}1_{3} \rangle)\otimes \frac{\lvert0_{\text{a}}\rangle -\lvert1_{\text{a}}\rangle  }{\sqrt{2}}\bigg].
\label{an05}
\end{eqnarray}
if we use the 4-qubit QFG on the three data atoms and the ancilla.
\newline
{\bf Circuit 2-pulse 3}. We can use the laser field as used in Circuit 1-pulse 3, which results in the state transform of Eqs.~(\ref{c1p3laser1}) and~(\ref{c1p3laser2}). So, Eq.~(\ref{an05}) becomes
\begin{eqnarray}
|\text{M}\rangle_{2}&=& \frac{\sqrt{2}}{×\sqrt{3}}  \lvert \text{O}'\rangle \otimes \lvert1_{\text{a}}\rangle -\frac{i}{×\sqrt{6}} ( \lvert0_{1}0_{2}0_{3} \rangle -  \lvert1_{1}1_{2}1_{3} \rangle)\otimes \lvert0_{\text{a}}\rangle  \equiv \frac{\sqrt{2}}{×\sqrt{3}}  \lvert \text{O}'\rangle \otimes \lvert1_{\text{a}}\rangle -\frac{i}{×\sqrt{3}} \lvert \text{GHZ}^{(\text{3})}_-\rangle\otimes \lvert0_{\text{a}}\rangle .\label{an05-2}
\end{eqnarray}
where
\begin{eqnarray}
\lvert \text{O}'\rangle &=& \frac{-i}{2}(\lvert0_{1}1_{2}1_{3} \rangle + \lvert1_{1}0_{2}1_{3} \rangle + \lvert0_{1}1_{2}0_{3} \rangle + \lvert1_{1}0_{2}0_{3} \rangle ).
\end{eqnarray}
\newline
{\bf Circuit 2-measurement}. The state coefficients in Eq.~(\ref{an05-2}) means that when we measure the state of the ancillary qubit and the result is $\lvert0\rangle$, the state in Eq.~(\ref{an05-2}) can collapse to Eq.~(\ref{s06-an02}) with a conditional probability $1/3$, and if the measurement result is $\lvert1\rangle$, the state in Eq.~(\ref{an05-2}) collapses to
\begin{eqnarray}
\lvert \text{O}'\rangle \otimes \lvert1_{\text{a}}\rangle\label{an06}
\end{eqnarray}
with a conditional probability $2/3$ according to the coefficients in the state components in Eq.~(\ref{an05-2}). Here, the probabilities $1/3$ and $2/3$ are ``conditional'' because the probability to use Circuit 2 is $3/4$, so the actual probabilities for the two outcomes of this circuit are $3/4\cdot(1/3,~2/3)$.
\newline
{\bf Circuit 3-pulse 1}. Circuit 3 is shown in Fig.~\ref{figure-product-GHZ}(c). If the measurement result is $\lvert1\rangle$ in the last step, we shall do as in Circuit 2-pulse 1, use a $\pi$ pulse to one data atom and a $\pi/2$ pulse to the ancilla, with the change here that the second data atom should be addressed in this pulse. The $\pi$ pulse will induce a transition as in Eq.~(\ref{rabi-pi}) with the atom index changed from 3 to 2, and one can find that Eq.~(\ref{an06}) becomes,
\begin{eqnarray}
 &&    \frac{-1}{2}(\lvert0_{1}0_{2}1_{3} \rangle + \lvert1_{1}1_{2}1_{3} \rangle + \lvert0_{1}0_{2}0_{3} \rangle + \lvert1_{1}1_{2}0_{3} \rangle ) \otimes \frac{\lvert0_{\text{a}}\rangle +\lvert1_{\text{a}}\rangle  }{\sqrt{2}} .\label{an07}
\end{eqnarray} \newline
{\bf Circuit 3-pulse 2}. The 4-qubit QFG, as in Circuit 1-pulse 2, will only take action on the state components $\lvert 0_10_2 0_{3}\rangle\otimes \lvert0_{\text{a}}\rangle$ and $\lvert 1_11_21_{3}\rangle\otimes \lvert1_{\text{a}}\rangle$. So, when we use QFG in the four atoms, one can show that Eq.~(\ref{an07}) becomes
\begin{eqnarray}
&& \frac{-1}{2}\bigg [ ( \lvert0_{1}0_{2}1_{3} \rangle +   \lvert1_{1}1_{2}0_{3} \rangle )  \otimes   \frac{\lvert0_{\text{a}}\rangle +\lvert1_{\text{a}}\rangle  }{\sqrt{2}} - ( \lvert0_{1}0_{2}0_{3} \rangle -  \lvert1_{1}1_{2}1_{3} \rangle)\otimes \frac{\lvert0_{\text{a}}\rangle -\lvert1_{\text{a}}\rangle  }{\sqrt{2}}\bigg].\label{an08}
\end{eqnarray}
with analyses as used in Eqs.~(\ref{s04-vary1}),~(\ref{s04-vary2}), and~(\ref{s05}).
\newline
{\bf Circuit 3-pulse 3}. As in Circuit 1-pulse 3, we excite the transition $\lvert 0\rangle\leftrightarrow\lvert 1\rangle$ of the ancillary atom with a Rabi frequency $i\Omega_{\text{c}}$ for a $\frac{\pi}{2 }$ pulse. With the understanding shown in  Eqs.~(\ref{c1p3laser1}) and~(\ref{c1p3laser2}), one can find that Eq.~(\ref{an08}) becomes
\begin{eqnarray}
&& \frac{-1}{2}\bigg [ ( \lvert0_{1}0_{2}1_{3} \rangle +   \lvert1_{1}1_{2}0_{3} \rangle )  \otimes   \lvert1_{\text{a}}\rangle   - ( \lvert0_{1}0_{2}0_{3} \rangle -  \lvert1_{1}1_{2}1_{3} \rangle)\otimes \lvert0_{\text{a}}\rangle \bigg].\label{an08-2}
\end{eqnarray}
\newline
{\bf Circuit 3-measurement}. One can rewrite Eq.~(\ref{an08-2}) as
\begin{eqnarray}
|\text{M}\rangle_{3}&=& \frac{-1}{\sqrt{2}} \frac{\lvert0_{1}0_{2}1_{3} \rangle +   \lvert1_{1}1_{2}0_{3} \rangle }{\sqrt{2}} \otimes   \lvert1_{\text{a}}\rangle   +\frac{1}{\sqrt{2}}\lvert \text{GHZ}^{(3)}_-\rangle \otimes \lvert0_{\text{a}}\rangle .\label{an08-3}
\end{eqnarray}
We then measure the state of the ancillary qubit. If the measurement result is $\lvert0\rangle$, the state in Eq.~(\ref{an05-2}) collapses to exactly Eq.~(\ref{s06-an02}), i.e., we have got the 3-qubit GHZ state, with a conditional probability $1/2$ according to the coefficients in the state components in Eq.~(\ref{an08-3}). If the measurement result is $\lvert1\rangle$, the state in Eq.~(\ref{an08-2}) collapses to
\begin{eqnarray}
 &&-\frac{\lvert0_{1}0_{2}1_{3} \rangle +   \lvert1_{1}1_{2}0_{3} \rangle }{\sqrt{2}}
 \otimes \lvert1_{\text{a}}\rangle= -i\hat{X}_{3}(\pi)\lvert \text{GHZ}^{(3)}_+\rangle \otimes \lvert1_{\text{a}}\rangle\label{an09}
\end{eqnarray}
with a conditional probability $1/2$, where the single-qubit rotation around x and y axes for data atom $j\in\{1,~2,~3,~\text{a}\}$ with an angle $\pi$ is
\begin{eqnarray}
 \hat{X}_{j}(\pi)&=&-i\left(\begin{array}{cc}0&1\\1&0\end{array}\right),\nonumber\\
 \hat{Y}_{j}(\pi)&=&\left(\begin{array}{cc}0&-1\\1&0\end{array}\right),\label{rotation-xy}
\end{eqnarray}
where the basis of the matrix above is $\{\lvert 0_j\rangle, \lvert 1_j\rangle\}$, and $\hat{Y}_{3}(\pi)$ is to be used in Circuit 3-pulse 4.
\newline
{\bf Circuit 3-pulse 4}. If the measurement result is $\lvert1\rangle$, we get  Eq.~(\ref{an09}). Then, we use $\hat{Y}_{3}(\pi)$, which can be realized by a laser-induced transition $\lvert 0\rangle\leftrightarrow\lvert 1\rangle$ in data atom 3 with a Rabi frequency $i\Omega_{\text{c}}$ for a $\pi$ pulse, so that Eq.~(\ref{an09}) becomes
\begin{eqnarray}
 && \hat{Y}_{3}(\pi) [\text{Eq}.~(\ref{an09})]=\lvert \text{GHZ}^{(3)}_-\rangle\otimes \lvert1_{\text{a}}\rangle  ,\label{an10}
\end{eqnarray}
with a probability $1/2$. Equation~(\ref{an10}) shows that if Circuit 3-pulse 4 is required, the final state is pure in the ancilla.

\begin{table*}[ht]
  \centering
  \begin{tabular}{|c|c|c|c|c|c| }
    \hline
  & Pulse 1  & Pulse 2 & Pulse 3 & Ancilla detection & Pulse 4   \\\hline
 Circuit 1   &  $ \frac{\pi}{2\Omega_{\text{c}}} $ &  $t_{\text{g}}$  & $ \frac{\pi}{2\Omega_{\text{c}}} $& $t_{\text{detect}}$ & $\diagdown$\\
   \hline  Circuit 2 &  $ \frac{\pi}{\Omega_{\text{c}}} $ &  $t_{\text{g}} $  & $ \frac{\pi}{2\Omega_{\text{c}}} $& $t_{\text{detect}}$ & $\diagdown$
     \\
   \hline  Circuit 3~(without pulse 4)   &  $ \frac{\pi}{\Omega_{\text{c}}} $ &  $t_{\text{g}} $  & $ \frac{\pi}{2\Omega_{\text{c}}} $& $t_{\text{detect}}$ & $\diagdown$  \\
   \hline  Circuit 3~(with pulse 4)  &  $ \frac{\pi}{\Omega_{\text{c}}} $ &  $t_{\text{g}} $  & $ \frac{\pi}{2\Omega_{\text{c}}} $& $t_{\text{detect}}$ &   $ \frac{\pi}{\Omega_{\text{c}}} $ \\
   \hline
  \end{tabular}
  \caption{Required times for each step in the three circuits for one gluing cycle analyzed in Sec.~\ref{sec03A}. The three circuits are illustrated in Fig.~\ref{figure-product-GHZ}, which shows that the chance to get the GHZ state at the end of Circuit 1, Circuit 2, and Circuit 3, are $1/4$, $1/4$, and $1/2$, respectively. For Circuit 3, Fig.~\ref{figure-product-GHZ}shows that there is half chance to end at the detection stage, and another half chance to end at Circuit 3-pulse 4 with a pure state in the ancilla as shown in Eq.~(\ref{an10}). Whether we end with a pure state in the ancilla or not, the final state is a product state of $\lvert \text{GHZ}^{(3)}\rangle\otimes\lvert$the state of the ancilla$\rangle$.   }\label{table-step-time}
\end{table*}

We suppose that the time of the measurement step of Circuit 1, Circuit 2, or Circuit 3 is $t_{\text{detect}}$, and the QFG step has a time $t_{\text{g}}$ which is dominated by the single-qubit gate as discussed below Eq.~(\ref{s05}) but in practice can be equal to the duration QFG as shown in Appendix~\ref{app-E}. As shown above, before each measurement, a circularly polarized field is used to map the qubit state $|0\rangle$ to the ground state where a Rabi frequency $2\pi\times0.2$~MHz is realizable~\cite{Chen2022}. Further, $t_{\text{detect}}$ should include the exposure time, dead time, and recovery time, where the latter two can be much longer than the exposure time~\cite{Finkelstein2024}. Take Ref.~\cite{senoo2025} as an example where $t_{\text{detect}}\sim3$~ms can yield a detection fidelity over 0.998. According to Table~\ref{table-step-time}, the duration $t_{\text{circuit-x}}$ for each circuit is
\begin{eqnarray}
 \text{Circuit 1: }&t_{\text{circuit-1}}&=t_{\text{g}}+\frac{\pi}{\Omega_{\text{c}}}+t_{\text{detect}},\nonumber\\
 \text{Circuit 2: }&t_{\text{circuit-2}}&=t_{\text{g}}+\frac{3\pi}{2\Omega_{\text{c}}}+t_{\text{detect}},\nonumber\\
 \text{Circuit 3~(without pulse 4): }&t_{\text{circuit-3}}&=t_{\text{g}}+\frac{3\pi}{2\Omega_{\text{c}}}+t_{\text{detect}},\nonumber\\
 \text{Circuit 3~(with pulse 4): }&t_{\text{circuit-3}'}&=t_{\text{g}}+\frac{5\pi}{2\Omega_{\text{c}}}+t_{\text{detect}},\label{4-choice}
\end{eqnarray}
which shows that there are three possible events to yield the GHZ state by measurement, and one possible event to get the GHZ state without measurement, namely, Circuit 3 with pulse 4. According to the measurement, the conditional probabilities to have the GHZ state via measurement in Circuit x are given by
\begin{eqnarray}
p_{\text{x}} &=& \text{Tr}_{\text{a}} \langle \text{GHZ}^{(3)}_- |     \left( |\text{M}\rangle_{\text{x}} \langle \text{M}\rvert_{\text{x}}  \right)\lvert \text{GHZ}^{(3)}_-\rangle,\label{p-x-yes}
\end{eqnarray}
where $\text{Tr}_{\text{a}}$ means trace over the degrees of freedom of the ancilla, and $|\text{M}\rangle_{\text{x}}\rangle$ with x=$1,~2$, and $3$ are given in Eqs.~(\ref{s06}),~(\ref{an05-2}),~(\ref{an08-3}). Equation~(\ref{p-x-yes}) leads to $p_{\text{x}}=1/4,1/3$, and $1/2$ for x=1, 2, and 3, respectively. However, there is no need to go to Circuit 2 and 3 if the GHZ state is already created at the end of Circuit 1 and 2, respectively. The probabilities to end up with the GHZ state at the end of the four options of Eq.~(\ref{4-choice}) are all equal to $1/4$, but the time required for reaching the certain step is equal to the times of the all the preceding operations. As a result, the average number of QFG as well as the number of ancilla detection that should be used within one gluing cycle is
\begin{eqnarray}
 {\overline{\text{n}}} &=& 1+(1-p_1)\left\{1+ (1-p_2) \left[ p_3 +(1-p_3)\right]\right\}  \nonumber\\
 &= &\frac{9}{4}. \label{defineEta}
\end{eqnarray}
Similarly, the average time to finish one gluing cycle is
\begin{eqnarray}
 \overline{\mathcal{T}}&=&  t_{\text{circuit-1}}+(1-p_1)\left\{ t_{\text{circuit-2}}+ (1-p_2) \left[ p_3 t_{\text{circuit-3}}+(1-p_3)t_{\text{circuit-3}'}\right]\right\}  \nonumber\\
 &= &{\overline{\text{n}}} (t_{\text{g}}+t_{\text{detect}})+\frac{25\pi}{8\Omega_{\text{c}}}.\label{averageTime}
\end{eqnarray}

 Notably, Circuit 3 is deterministic for if the measurement yields a failure for the state generation, Circuit 3-pulse 4 can create the desired GHZ state by the ground-clock transition. So, the probability to get Eq.~(\ref{s06-an02}) is $\frac{3}{4}$, and the probability to get Eq.~(\ref{an10}) is $\frac{1}{4}$. This doesn't mean that there is entanglement between the ancilla and the data atoms because the final state for the three data atoms and the ancilla is either Eq.~(\ref{s06-an02}) or Eq.~(\ref{an10}), both of which are a product state between the data and the ancilla.

To simplify the following presentation, we redefine the qubit state $\lvert1\rangle$ for each qubit with
\begin{eqnarray}
 \lvert 1_j\rangle&\rightarrow& e^{-i\pi/3}\lvert 1_j\rangle \label{GHZ-toGHz+}
\end{eqnarray}
for each atom $j\in\{1,~2,~3\}$. Then, the GHZ state created by the above process can be written as
\begin{eqnarray}
 \lvert \text{GHZ}^{(3)}_-\rangle\rightarrow\lvert \text{GHZ}^{(3)}_+\rangle&=&\frac{ 1}{\sqrt{2}} (\lvert0_{1}0_{2}0_{3} \rangle +   \lvert1_{1}1_{2}1_{3} \rangle)\label{3GHZ-toGHz+}.\label{an11}
\end{eqnarray}

\subsection{9-qubit GHZ state}\label{sec03B}
\begin{figure*}
\includegraphics[width=7.0in]
{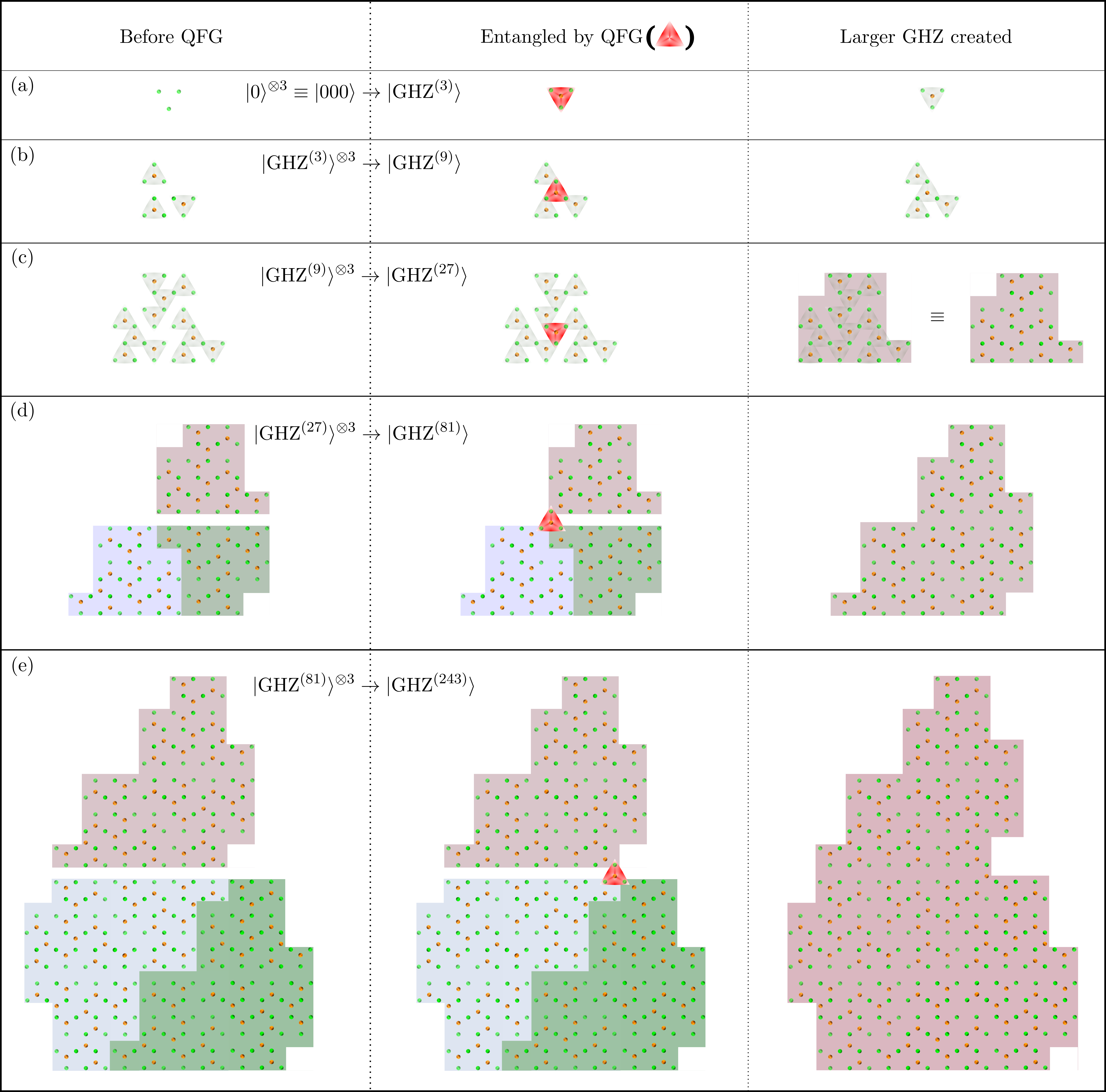}
\caption{Illustration of generating a 243-qubit GHZ state. (a,b,c,d,e) show the gluing from product state to $\lvert \text{GHZ}^{(3)}\rangle$, from $\lvert \text{GHZ}^{(3)}\rangle$ to $\lvert \text{GHZ}^{(9)}\rangle$, from $\lvert \text{GHZ}^{(9)}\rangle$ to $\lvert \text{GHZ}^{(27)}\rangle$, from $\lvert \text{GHZ}^{(27)}\rangle$ to $\lvert \text{GHZ}^{(81)}\rangle$, and from $\lvert \text{GHZ}^{(81)}\rangle$ to $\lvert \text{GHZ}^{(243)}\rangle$, respectively. The green and red dots represent the data and ancilla atoms, respectively. In each gluing, the QFG is denoted by the red triangle which represents looking through a tetrahedron where the four atoms in the QFG are the four vortexes.     \label{figure-ghz4} }
\end{figure*}

The above three-circuit gate-measurement gluing cycle can be used to glue smaller GHZ states so as to form a larger GHZ state. For example, with three GHZ states generated as described above and an ancillary atom in the state $\lvert0\rangle$, namely, starting from the following state
\begin{eqnarray}
&&\lvert \text{GHZ}^{(3)}_+\rangle\otimes\lvert \text{GHZ}^{(3)}_+\rangle\otimes\lvert \text{GHZ}^{(3)}_+\rangle\otimes\lvert0_{\text{a}}\rangle\nonumber\\
&=&2^{-3/2}(\lvert0_{1}0_{2}0_{3}\rangle +\lvert1_{1}1_{2}1_{3}\rangle    )\otimes(\lvert0_{4}0_{5}0_{6}\rangle +\lvert1_{4}1_{5}1_{6}\rangle   ) \otimes(\lvert0_{7}0_{8}0_{9}\rangle +\lvert1_{7}1_{8}1_{9}\rangle   )\otimes\lvert0_{\text{a}}\rangle, \label{9ghz-1}
\end{eqnarray}
we can use a similar three-circuit gluing cycle to generate a 9-qubit GHZ state,
\begin{eqnarray}
\lvert \text{GHZ}_{\pm}^{(9)}\rangle &=&
(\lvert 0_10_20_30_40_50_60_70_80_9\rangle\nonumber  \pm\lvert 1_11_21_31_41_51_61_71_81_9 \rangle )/\sqrt{2}. \label{9qubitGHZ}
\end{eqnarray}
Before showing the details, we rewrite Eq.~(\ref{9ghz-1}) as
\begin{eqnarray}
&&\frac{1}{2} \left(\lvert \text{GHZ}_+^{(9)}\rangle + \sqrt{3}\lvert\mathbb{O}\rangle \right)\otimes\lvert0_{\text{a}}\rangle,\label{9qubitGHZ-2}
\end{eqnarray}
where
\begin{eqnarray}
 \sqrt{6}\lvert\mathbb{O}\rangle &=&\lvert 0_10_20_30_40_50_61_71_81_9\rangle +\lvert 0_10_20_31_41_51_60_70_80_9\rangle+ \lvert 1_11_21_30_40_50_60_70_80_9\rangle + \lvert 0_10_20_31_41_51_61_71_81_9 \rangle \nonumber\\
 && +\lvert 1_11_21_30_40_50_61_71_81_9 \rangle+ \lvert 1_11_21_31_41_51_60_70_80_9 \rangle
 ,\nonumber
\end{eqnarray}
where we note that the state $\lvert\mathbb{O}\rangle$ is normalized. Note that the three consecutive data atoms $\{1,2,3\}$, or $\{4,5,6\}$, or $\{7,8,9\}$ have exactly the same states. So, we can use abbreviated symbols
\begin{eqnarray}
 \alpha_{\text{\tiny 1-3}}&\equiv& \alpha_1\alpha_2\alpha_3,\nonumber\\
 \alpha_{\text{\tiny 4-6}}&\equiv& \alpha_4\alpha_5\alpha_6,\nonumber\\
 \alpha_{\text{\tiny 7-9}}&\equiv& \alpha_7\alpha_8\alpha_9,
\end{eqnarray}
where $\alpha$ is either 0 or 1. Then, $\lvert\mathbb{O}\rangle$ can be written as
\begin{eqnarray}
 \lvert\mathbb{O}\rangle &=&\frac{1}{\sqrt{6}} (  \lvert 0_{\text{\tiny 1-3}}0_{\text{\tiny 4-6}}1_{\text{\tiny 7-9}} \rangle   +\lvert 0_{\text{\tiny 1-3}}1_{\text{\tiny 4-6}}0_{\text{\tiny 7-9}} \rangle +  \lvert 1_{\text{\tiny 1-3}}0_{\text{\tiny 4-6}}0_{\text{\tiny 7-9}} \rangle  + \lvert 0_{\text{\tiny 1-3}}1_{\text{\tiny 4-6}}1_{\text{\tiny 7-9}}\rangle +\lvert 1_{\text{\tiny 1-3}}0_{\text{\tiny 4-6}}1_{\text{\tiny 7-9}} \rangle   + \lvert 1_{\text{\tiny 1-3}}1_{\text{\tiny 4-6}}0_{\text{\tiny 7-9}} \rangle )  ,\label{9qubit-1}
\end{eqnarray}
which is of exactly the same form of Eq.~(\ref{operatorO}) if we identify the three digits in each ket of Eq.~(\ref{operatorO}) as the three digits of Eq.~(\ref{9qubit-1}). Similarly, the 9-qubit GHZ state in Eq.~(\ref{9qubitGHZ-2}) is also of identical form to the 3-qubit GHZ state in Eq.~(\ref{s04}). This means that by using a similar strategy as in Sec.~\ref{sec03A}, three $\lvert \text{GHZ}^{(3)}_+\rangle$ in three nearby blocks of Fig.~\ref{figure-ghz4} can be glued. The extra thing is that only one data atom from each $\lvert \text{GHZ}^{(3)}_+\rangle$ should be chosen to participate in the gluing. Because this gluing is similar to that in Sec.~\ref{sec03A}, we briefly describe the procedure below.
\newline
{\bf Circuit I-pulse 1}. We use the $\lvert 0\rangle\leftrightarrow\lvert 1\rangle$ transition on the ancillary qubit with a Rabi frequency $i\Omega_{\text{c}}$ for a $\pi/2$ pulse, and the state in Eq.~(\ref{9qubitGHZ-2}) becomes
\begin{eqnarray}
&&\frac{1}{2} \left(\sqrt{3}\lvert\mathbb{O}\rangle +\lvert \text{GHZ}_+^{(9)}\rangle  \right)\otimes \frac{\lvert0_{\text{a}}\rangle +\lvert1_{\text{a}}\rangle  }{\sqrt{2}}.\label{9qubitGHZ-23}
\end{eqnarray}
The state transform induced by this pulse is very similar to that discussed around Eqs.~(\ref{s02},~\ref{s04},~\ref{operatorO}).
\newline
{\bf Circuit I-pulse 2}. We use a 4-qubit QFG which induces a $\pi$ phase for the following two input states
\begin{eqnarray}
\{\lvert 0_10_40_70_{\text{a}}\rangle ,\lvert 1_11_41_71_{\text{a}}\rangle \}
\end{eqnarray}
but does nothing to other input states. Equation~(\ref{9qubitGHZ-23}) becomes
\begin{eqnarray}
&& \frac{\sqrt{3}}{2} \lvert\mathbb{O}\rangle \otimes \frac{\lvert0_{\text{a}}\rangle +\lvert1_{\text{a}}\rangle  }{\sqrt{2}}-\frac{1}{2} \lvert \text{GHZ}_-^{(9)}\rangle\otimes \frac{\lvert0_{\text{a}}\rangle -\lvert1_{\text{a}}\rangle   }{\sqrt{2}},\label{9qubitGHZ-3}
\end{eqnarray}
which is derived in a way similar to that for deriving Eq.~(\ref{s05}). Here, one should pay attention to that $\lvert \text{GHZ}_+^{(9)}\rangle$ in Eq.~(\ref{9qubitGHZ-23}) changes to $\lvert \text{GHZ}_-^{(9)}\rangle$ in Eq.~(\ref{9qubitGHZ-3}). \newline
{\bf Circuit I-pulse 3}. We repeat Circuit I-pulse 1. Then, Eq.~(\ref{9qubitGHZ-3}) becomes
\begin{eqnarray}
&&\frac{\sqrt{3}}{2} \lvert\mathbb{O}\rangle \otimes \lvert1_{\text{a}}\rangle - \frac{1}{2} \lvert \text{GHZ}_-^{(9)}\rangle\otimes \lvert0_{\text{a}}\rangle  .\label{9qubitGHZ-4}
\end{eqnarray}
which is very similar for realizing Eq.~(\ref{s06}) of Circuit 1-pulse 3 studied in Sec.~\ref{sec03A}. \newline
{\bf Circuit I-measurement}: Measure the state of the ancillary atom. If the measurement result is $\lvert0\rangle$, we have got the GHZ state $\lvert \text{GHZ}_-^{(9)}\rangle$
with a probability of $1/4$, with the other $3/4$ probability to have $\lvert\mathbb{O}\rangle \otimes \lvert1_{\text{a}}\rangle$. 

When one examines the above gate-measurement operations with those operations in Circuit 1 of Sec.~\ref{sec03A}, one can find that the operations are the same, except that here only one atom from each initial 3-qubit GHZ states is chosen to work in the QFG. By this understanding, one can immediately see that the following two gate-measurement circuits will also be similar to those in Sec.~\ref{sec03A}, except that during each gate-measurement circuit, the excitation should be for three data atoms from three nearby 3-qubit GHZ states. For example, if the measurement result is $\lvert1\rangle$ in the Circuit I-measurement, then Eq.~(\ref{9qubitGHZ-4}) collapses to
\begin{eqnarray}
&& \lvert\mathbb{O}\rangle \otimes \lvert1_{\text{a}}\rangle . \label{9qubit-pure}
\end{eqnarray}
To proceed with Eq.~(\ref{9qubit-pure}), we can use a pulse similar to Circuit 2-pulse 1 studied around Eq.~(\ref{an03}). The difference here is that now the laser shall target atoms 7, 8, 9, and the ancilla here, while for Circuit 2-pulse 1 the atom 3 and the ancilla are targeted. Then, Eq.~(\ref{9qubit-pure}) becomes
\begin{eqnarray}
 && \frac{i}{\sqrt{6}}( \lvert 0_{\text{\tiny 1-3}}0_{\text{\tiny 4-6}}0_{\text{\tiny 7-9}} \rangle   +\lvert 0_{\text{\tiny 1-3}}1_{\text{\tiny 4-6}}1_{\text{\tiny 7-9}} \rangle +  \lvert 1_{\text{\tiny 1-3}}0_{\text{\tiny 4-6}}1_{\text{\tiny 7-9}} \rangle   + \lvert 0_{\text{\tiny 1-3}}1_{\text{\tiny 4-6}}0_{\text{\tiny 7-9}}\rangle +\lvert 1_{\text{\tiny 1-3}}0_{\text{\tiny 4-6}}0_{\text{\tiny 7-9}} \rangle   + \lvert 1_{\text{\tiny 1-3}}1_{\text{\tiny 4-6}}1_{\text{\tiny 7-9}} \rangle ) \otimes \frac{\lvert0_{\text{a}}\rangle +\lvert1_{\text{a}}\rangle  }{\sqrt{2}},\label{9qubit-4}
\end{eqnarray}
which is similar to Eq.~(\ref{an03}) when we identify the three digits in each ket of Eq.~(\ref{an03}) as the three digits of Eq.~(\ref{9qubit-4}). The only difference of Eq.~(\ref{9qubit-4}) from Eq.~(\ref{an03}) is an overall $\pi$ phase which is trivial and can be absorbed in the state of the ancilla. Similarly, when the second gate-measurement circuit yields failure, then the third gate-measurement circuit shall be used, where the first pulse shall also be to three data atoms, namely, atoms 4, 5, and 6. And, if the measurement result of the third circuit is also failure, then the last step is similar to Circuit 3-pulse 4 of Sec.~\ref{sec03A}, shown around Eq.~(\ref{an10}), but with the modification here that the three atoms, atom 7,~8,~9, should be targeted. This means that by using exactly the same strategy of Sec.~\ref{sec03A}, we can perform the gluing of the three $\lvert \text{GHZ}^{(3)}\rangle$ so that the 9-qubit GHZ state forms.

One can find that the final state for the nine data qubits will be $\lvert \text{GHZ}_-^{(9)}\rangle$. But if we reuse Eq.~(\ref{GHZ-toGHz+}) to redefine the qubit state $\lvert 1_j\rangle$, then because $(e^{-i\pi/3})^9=-1$, the GHZ state created by the above process can be written as
\begin{eqnarray}
 \lvert \text{GHZ}^{(9)}_-\rangle\rightarrow\lvert \text{GHZ}^{(9)}_+\rangle.\label{9GHZ-toGHz+}
\end{eqnarray}
In the original definition of the qubit state $\lvert 1_j\rangle$ as used in Eqs.~(\ref{s01}-\ref{an10}), however, one can see that the 9-qubit GHZ state created is just $|\text{GHZ}^{(9)}_+\rangle$ without using the qubit redefinitions for Eqs.~(\ref{3GHZ-toGHz+}) and~(\ref{9GHZ-toGHz+}). This indicates that one needs no qubit redefinitions as Eqs.~(\ref{3GHZ-toGHz+}) and~(\ref{9GHZ-toGHz+}) when we use the basic gluing cycle as detailed in Sec.~\ref{sec03A}. In other words, we can use $\lvert \text{GHZ}^{(\text{n})}\rangle$ to denote $\lvert \text{GHZ}_{\pm}^{(\text{n})}\rangle$ in this article.


\subsection{Larger GHZ state}
Note that the three-circuit gate-measurement protocol shown in Sec.~\ref{sec03A} not only can be used in a similar way in Sec.~\ref{sec03B} for gluing three nearby $\lvert \text{GHZ}^{(3)}\rangle$, but can also be used for further gluing. One can easily verify that starting from three of the above 9-qubit GHZ states, a similar three-circuit gluing cycle with another ancillary atom can lead to a 27-qubit GHZ state, with a conditional probability still equal to those of Sec.~\ref{sec03A} for each gate-measurement circuit. Repetition of this can lead to $\lvert \text{GHZ}^{(81)}\rangle$, $\lvert \text{GHZ}^{(243)}\rangle,\cdots$, as illustrated in Fig.~\ref{figure-ghz4}. Importantly, each repetition requires blockade within four atoms, in which three are from the smaller GHZ states and one as an ancillary atom. The time required for each repetition is the same, and the fidelity of each repetition is the same. Though 3D atomic array can be assembled with arbitrary configurations~\cite{Barredo_2018}, the question is whether there is any possibility to have an atomic configuration so that each circuit can be processed with four atoms from the three smaller GHZ states and an ancilla.

In Fig.~\ref{figure-ghz4}, we illustrate a system with two layers of tweezer array, where the atoms in the upper and lower array are shown by green and blue. The atoms in the upper layer are data atoms to form GHZ states, while the atoms in the lower layer are ancillary atoms. Each three nearby data atoms are accompanied by an ancillary atom, forming a tetrahedron. The tetrahedron is illustrated by the red triangle in Fig.~\ref{figure-ghz4}, which is like looking through the tetrahedron. Before we analyze the achievable fidelity for creating a 243-qubit GHZ state based on the configuration of Fig.~\ref{figure-ghz4}, we first study how to avoid undesired blockade when executing the QFG in parallel.

\begin{figure*}
\includegraphics[width=7.0in]
{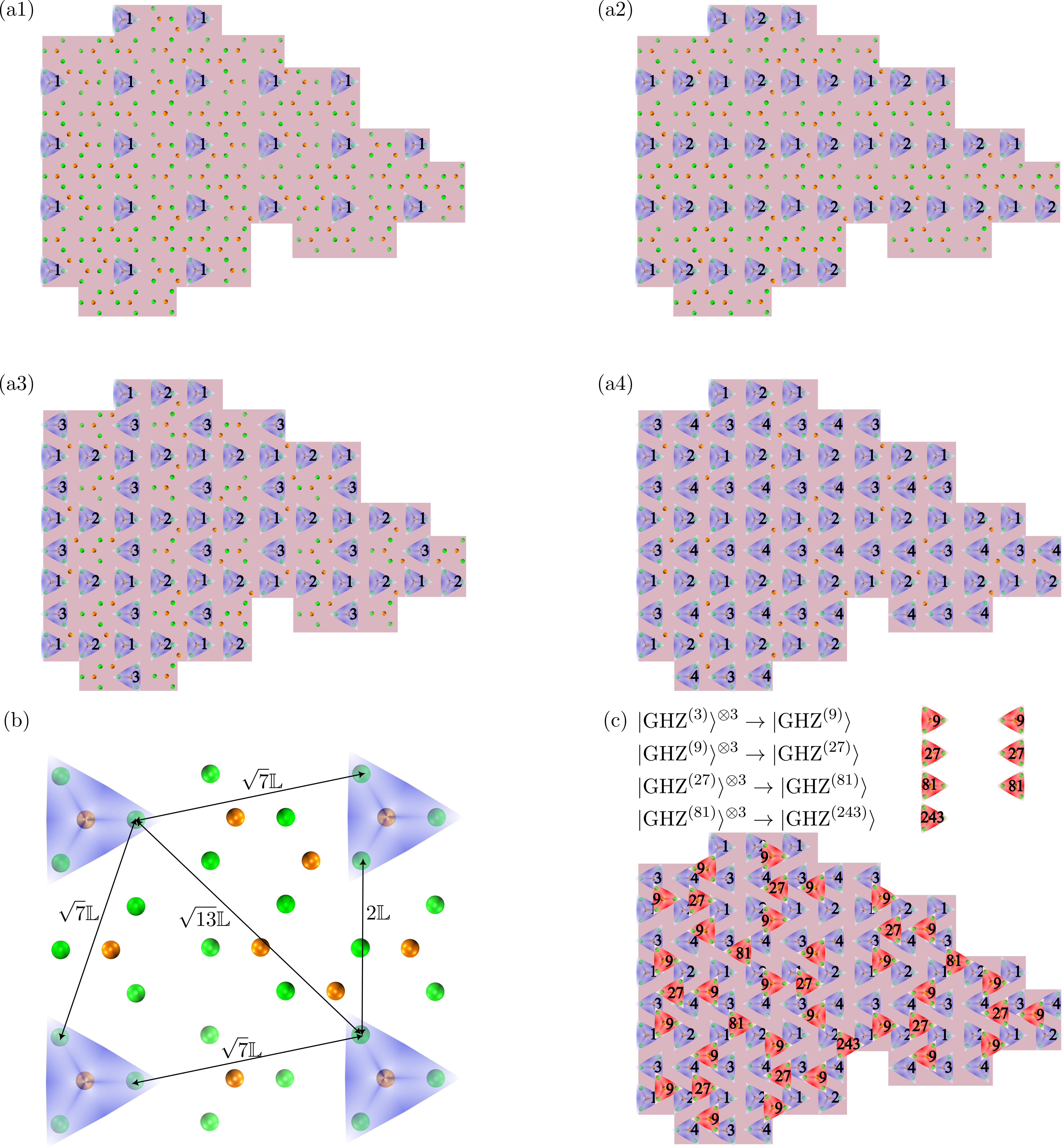}
\caption{(a1,a2,a3,a4) Show the parallel creation of 22,~19,~20,~20 sets of $\lvert \text{GHZ}^{(3)}\rangle$ with 22,~19,~20,~20 parallel elementary gluing operations, respectively. The number $1,~2,~3,~4$ are put in the corresponding triangles in the first, second, third, and fourth steps, respectively. These four steps finish the generation of 81 sets of $\lvert \text{GHZ}^{(3)}\rangle$ by four gluing cycles so as to avoid cross-blockade interference. The four atoms in each QFG execution are located in the corners of a tetrahedron. The length of the side of the tetrahedron is $\mathbb{L}$. Within the four atoms in one QFG, the Rydberg interaction is $V_{\vartriangle}=C_\alpha/\mathbb{L}^\alpha$ with $\alpha=6$ and $3$ with the van der Waals scaling and resonant dipole-dipole interactions, respectively. (b) Shows the smallest separations between two atoms belonging to two QFGs in (a1-a4). In each of the four steps shown here, the shortest distance between two Rydberg atoms belonging to two neighboring QFGs has three possible values, $\{2\mathbb{L},~\sqrt{7}\mathbb{L},~\sqrt{13}\mathbb{L}\}$. Thus, the strongest blockade interaction between two Rydberg atoms in two nearby QFGs is $V_{\vartriangle-\vartriangle}=\epsilon V_{\vartriangle}$, where $\epsilon=\{2^{-\alpha},~7^{-\alpha/2},~13^{-\alpha/2}\}$. To further shrink the undesired blockade, one can further divide each of (a1,a2,a3,a4) into four steps so that the shortest distance between two Rydberg atoms belonging to two neighboring QFGs is $5\mathbb{L}$; see texts below Eq.~(\ref{T-nu}) for more details. \label{figure-4-step} }
\end{figure*}

\begin{figure*}
\includegraphics[width=7.0in]
{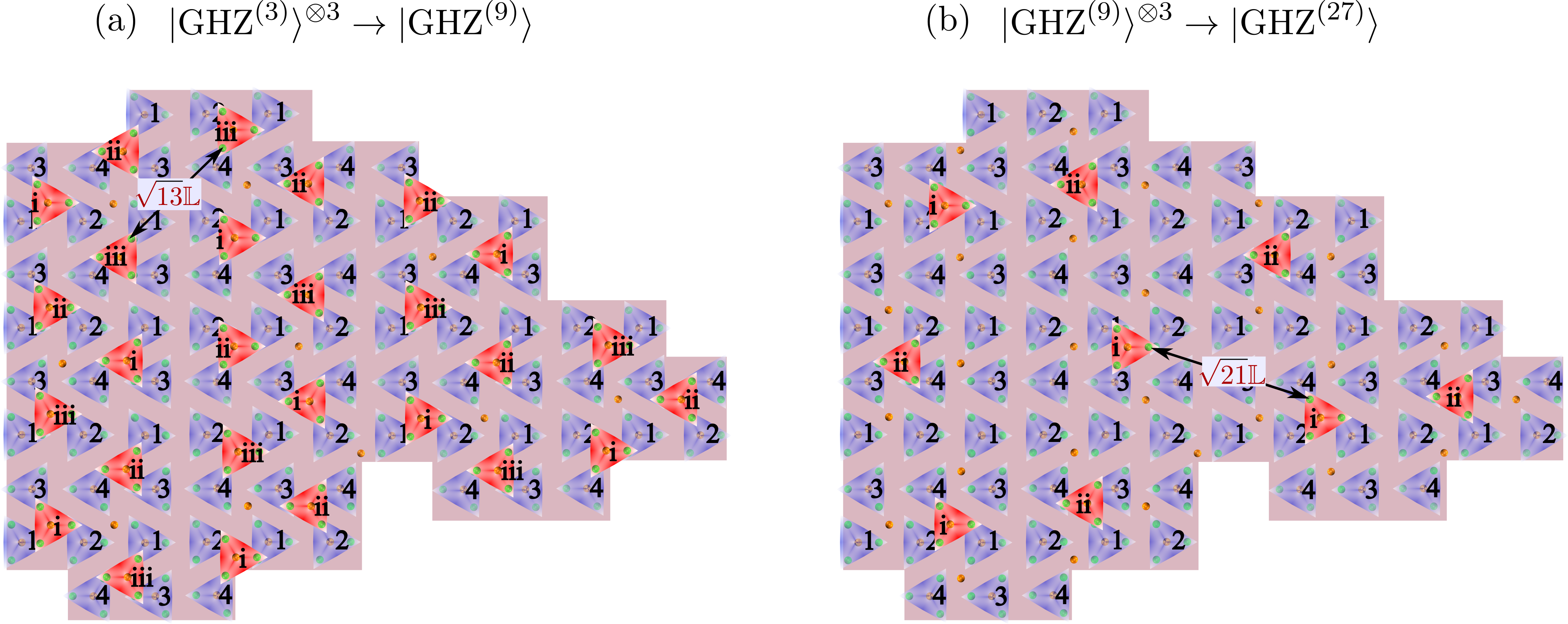}
\caption{(a) Shows the parallel creation of triple-9 sets of $\lvert \text{GHZ}^{(9)}\rangle$ with triple-9 parallel elementary gluing operations, respectively. The  numbers i, ii, iii put in the corresponding red triangles denote the order to execute the gate, where each red triangle denotes a QFG for entangling three neighboring $\lvert \text{GHZ}^{(3)}\rangle$. If not dividing the twenty-seven QFGs in this way, the nearest distance between two atoms belonging to neighboring QFGs marked by a number 9 in Fig.~\ref{figure-4-step}(c) is $\sqrt{3}\mathbb{L}$. But if we execute the twenty-seven QFGs in three steps shown here, the shortest distance between two Rydberg atoms belonging to two neighboring QFGs is $\sqrt{13}\mathbb{L}$ as denoted by the arrow here. (b) Shows the order to glue $\lvert \text{GHZ}^{(9)}\rangle$ to form $\lvert \text{GHZ}^{(27)}\rangle$ with numbers i and ii marked in the red triangles. The shortest distance between two Rydberg atoms belonging to two neighboring QFGs is $\sqrt{21}\mathbb{L}$ as denoted by the arrow.   \label{figure-9-mul} }
\end{figure*}

\section{Cross blockade}\label{sec04}
The eighty-one $\lvert \text{GHZ}^{(3)}\rangle$ states can be prepared in parallel as if entangling three qubits instead of eighty-one copies of three qubits. But there will be cross blockade between nearby QFG when QFGs are executed in parallel. To avoid undesired blockade from a nearby Rydberg atom when we execute a QFG, it is useful to use four gluing cycles instead of one to prepare the eighty-one $\lvert \text{GHZ}^{(3)}\rangle$ states, shown in Fig.~\ref{figure-4-step}(a1-a4). In this case, the nearest distance between two atoms belonging to two nearby QFGs are shown in Fig.~\ref{figure-4-step}(b). 

In order to speed up the creation of the GHZ states, one can execute Fig.~\ref{figure-4-step} with $\nu=4$ rounds of parallel execution of QFGs while the other operations in each gluing cycle are still globally parallel: (i) First, we execute Circuit 1-pulse 1 in parallel in all the 81 sets of four atoms highlighted in Fig.~\ref{figure-4-step}(a4). (ii) Then, we execute Circuit 1-pulse 2, i.e., the QFGs, according to the order of Fig.~\ref{figure-4-step}(a1-a4). By this, we avoid cross blockade. (iii) We execute Circuit 1-pulse 3 in parallel in all the 81 sets of four atoms highlighted in Fig.~\ref{figure-4-step}(a4). (iv) Do the ancilla detection, i.e., Circuit 1-measurement. In other words, only the execution of Circuit 1-pulse 2 should be divided into several steps as in Fig.~\ref{figure-4-step}(a1-a4). Similarly, in Circuit 2 and Circuit 3, it is alo necessary to execute QFGs in a way similar to the order shown in Fig.~\ref{figure-4-step}(a1-a4). As a result, the expected time for finishing one gluing cycle from Eq.~(\ref{averageTime}) should be updated by replacing $ t_{\text{g}}$ with $\nu t_{\text{g}}$, so that the average time to prepare the eighty-one $\lvert \text{GHZ}^{(3)}\rangle$ is 
\begin{eqnarray}
\overline{T}(\nu) = \overline{\mathcal{T}}+{\overline{\text{n}}}(\nu-1)t_{\text{g}}.  \label{T-nu}
\end{eqnarray}

The above analyses are based on Fig.~\ref{figure-4-step}. But if short qubit separation $\mathbb{L}$ is used, it can be useful to further increase the distance between two QFGs. For example, in Fig.~\ref{figure-4-step}(a1-a4), each step involves five rows of QFGs, which can be labeled row-1, row-2, $\cdots$, row-5. If we further divide the five rows of QFGs in each of them into two, \{row-1, row-3, row-5\}, \{row-2, row-4\}, then, the total number of rounds of parallel execution of QFGs is $\nu=8$, and the smallest distance between two atoms belonging to two neighboring QFGs is $\sqrt{7}\mathbb{L}$. Again, in Fig.~\ref{figure-4-step}(a1-a4), each step involves six columns of QFGs, and we can first execute QFGs in the odd columns in \{row-1, row-3, row-5\}, and then the even columns. Likewise, we can use two steps to execute the QFGs in \{row-2, row-4\} too, so that the smallest distance between two atoms belonging to neighboring QFGs is $5\mathbb{L}$. This means that we have $\nu=16$, and the average time to prepare the eighty-one $\lvert \text{GHZ}^{(3)}\rangle$ would be $\overline{T}(16)=\overline{\mathcal{T}}+33.75t_{\text{g}}$.  

 According to Fig.~\ref{figure-4-step}(c), the smallest distance between two atoms belonging to neighboring QFGs denoted by a number 9 in the triangle is $\sqrt{3}\mathbb{L}$. This means when we glue three $\lvert \text{GHZ}^{(3)}\rangle$, the Rydberg blockade can have crosstalk if we execute the QFGs simultaneously in all the QFGs denoted by a number 9 in the triangle of Fig.~\ref{figure-4-step}(c). So, the preparation from the eighty-one $\lvert \text{GHZ}^{(3)}\rangle$ to twenty-seven $\lvert \text{GHZ}^{(9)}\rangle$ may also need to cope with the cross blockade. Figure~\ref{figure-9-mul}(a) shows that we can use $\nu=3$ rounds of parallel execution of the QFGs labeled by the number 9 in Fig.~\ref{figure-4-step}(c), where the nine QFGs marked by i will be executed first, then the nine marked by ii, and finally the nine marked by iii. By this, the distance between two atoms belonging to two nearest neighboring QFGs is enlarged, and the shortest distance between two Rydberg atoms belonging to two neighboring QFGs is $\sqrt{13}\mathbb{L}$ denoted in Fig.~\ref{figure-9-mul}(a). This can effectively suppress the cross blockade, and the average time to prepare the twenty-seven $\lvert \text{GHZ}^{(9)}\rangle$ is $\overline{T}(3)=\overline{\mathcal{T}}+\frac{9}{2}t_{\text{g}}$.

The schedule for gluing $\lvert \text{GHZ}^{(9)}\rangle$ to form $\lvert \text{GHZ}^{(27)}\rangle$ is denoted by the red triangles labeled by the number $27$ in Fig.~\ref{figure-4-step}(c). One can see that among the twenty-seven atoms enclosed by the nine red triangles with the number $27$, there are three two-atom pairs with an atomic separation $\sqrt{7}\mathbb{L}$. To suppress the cross blockade, Fig.~\ref{figure-9-mul}(b) shows that we can use $\nu=2$ rounds of parallel execution of the QFGs, with the order denoted by i and ii, respectively. In this way, the shortest distance between two Rydberg atoms belonging to two neighboring QFGs in each round of QFG-execution is $\sqrt{21}\mathbb{L}$ as denoted in Fig.~\ref{figure-9-mul}(b), and the average time to prepare the nine $\lvert \text{GHZ}^{(27)}\rangle$ is $\overline{T}(2)=\overline{\mathcal{T}}+\frac{9}{4}t_{\text{g}}$.

One can see from Fig.~\ref{figure-4-step}(c) that for gluing $\lvert \text{GHZ}^{(27)}\rangle$ states, three QFGs are needed, with three red triangles marked by the
number 81. Among the three triangles, the two on the left are a little close. So, we can use $\nu=2$ rounds of execution of the QFGs, where the first round is to execute one QFG, e.g., the lowest one in Fig.~\ref{figure-4-step}(c). In the second round, the other two QFGs will be executed in parallel, during which the shortest distance between two Rydberg atoms belonging to the two QFGs is about $\sqrt{91}\mathbb{L}$. So the average time to prepare the three $\lvert \text{GHZ}^{(81)}\rangle$ is $\overline{T}(2)$.

The last QFG for gluing three $\lvert \text{GHZ}^{(81)}\rangle$ incurs no cross blockade. So, the average time for each gluing cycle and the total average time are
\begin{eqnarray}
|0\rangle^{\otimes3}\rightarrow\lvert \text{GHZ}^{(3)}\rangle:&\overline{T}(16)& = \overline{\mathcal{T}}+33.75t_{\text{g}}, \nonumber\\
\text{GHZ}^{(3)}\rangle^{\otimes3}\rightarrow\lvert \text{GHZ}^{(9)}\rangle:&\overline{T}(3)& = \overline{\mathcal{T}}+4.5t_{\text{g}}, \nonumber\\
\lvert\text{GHZ}^{(9)}\rangle^{\otimes3}\rightarrow\lvert \text{GHZ}^{(27)}\rangle:&\overline{T}(2)& = \overline{\mathcal{T}}+2.25t_{\text{g}}, \nonumber\\
\lvert\text{GHZ}^{(27)}\rangle^{\otimes3}\rightarrow\lvert \text{GHZ}^{(81)}\rangle :&\overline{T}(2)& = \overline{\mathcal{T}}+2.25t_{\text{g}}, \nonumber\\
\lvert\text{GHZ}^{(81)}\rangle^{\otimes3}\rightarrow\lvert \text{GHZ}^{(243)}\rangle:&\overline{T}(1)& = \overline{\mathcal{T}} , \nonumber\\
\text{Average time to create }\lvert \text{GHZ}^{(243)}\rangle:& 5\overline{\mathcal{T}}+42.75t_{\text{g}} & = 11.25t_{\text{detect}} + 54 t_{\text{g}} +\frac{125\pi}{8\Omega_{\text{c}}} , \label{time-nu}
\end{eqnarray}
where the shortest distances between two Rydberg atoms belonging to two neighboring QFGs are $5\mathbb{L},~\sqrt{13}\mathbb{L},~\sqrt{21}\mathbb{L}$ for the creation of $\text{GHZ}^{(3)}\rangle$, $\lvert \text{GHZ}^{(9)}\rangle$, $\lvert \text{GHZ}^{(27)}\rangle$, and $\lvert \text{GHZ}^{(81)}\rangle$, respectively. If we take $t_{\text{detect}}=3~$ms~\cite{senoo2025}, $\Omega=2\pi\times10$~MHz, $\Omega_{\text{c}}= 2\pi\times0.77$~MHz~\cite{Jenkins2022}, the average time to prepare
$\lvert \text{GHZ}^{(243)}\rangle$ is about 33.8~ms, dominated by the detection time. Due to the fast speed of the QFG, it is not a problem if one would like to further shrink the smallest distances between two Rydberg atoms belonging to two neighboring QFGs in each gluing cycle. By this, one can see that the cross blockade can be avoided. Compared to the detection time, this latter strategy will barely increase the average time for the state generation.   

\section{Fidelity for a 243-qubit GHZ state} \label{sec05}
As shown in Sec.~\ref{sec03A}, multiple QFG are needed for each gluing cycle. This means that the fidelity is strongly limited by the entanglement operation of QFG. Theoretically, the fundamental fidelity for generating the GHZ states is limited by the gate fidelity of the QFG because the single-qubit gate can be realized without exciting high-lying states~\cite{Shi2021qst}. The average time for each gluing cycle when generating the 243-qubit GHZ state is summarized in Eq.~(\ref{time-nu}).


We need first to prepare the eighty-one $\lvert \text{GHZ}^{(3)}\rangle$ states from product states. According to Eq.~(\ref{defineEta}), the average number of QFG needed in each gluing cycle is ${\overline{\text{n}}}=\frac{9}{4}$. When the total number of 4-atom sets is $3^{\mathbb{N}}$, the fidelity for generating the eighty-one $\lvert \text{GHZ}^{(3)}\rangle$ states is
\begin{eqnarray}
\mathcal{F}_1=(F_0F_{\text{detect}})^{3^{\mathbb{N}}{\overline{\text{n}}}}, \label{F1-define}
\end{eqnarray}
where $F_0$ is the fidelity of one FQG gate, and $F_{\text{detect}}$ is the fidelity to detect the state of the ancilla. $F_0$ is estimated in Secs.~\ref{sec02C} and~\ref{sec02D}, with typical upper and lower values listed in Table~\ref{table-error}. For the case of Fig.~\ref{figure-ghz4}, we have $\mathbb{N}=4$, then $\mathcal{F}_1=(F_0F_{\text{detect}})^{81{\overline{\text{n}}}}$.


After preparing the eighty-one $\lvert \text{GHZ}^{(3)}\rangle$, we need to glue 3-qubit GHZ states to form 9-qubit GHZ states. This is illustrated by Fig.~\ref{figure-ghz4}(b). By using the QFG, the measurement protocol can yield the 9-qubit GHZ state like the one in Eq.~(\ref{9qubitGHZ}). For a system with the total number of 4-atom sets equal to $3^{\mathbb{N}}$, the average number of QFG for gluing in this step is $3^{\mathbb{N}-1}{\overline{\text{n}}}$, so the fidelity for this step is
\begin{eqnarray}
 \mathcal{F}_2=(F_0F_{\text{detect}})^{3^{\mathbb{N}-1}{\overline{\text{n}}}}.\label{F2-define}
\end{eqnarray}
For the case of Fig.~\ref{figure-4-step}, we have $\mathbb{N}=4$ which leads to $\mathcal{F}_2=(F_0F_{\text{detect}})^{27{\overline{\text{n}}}}$.

The third step is to glue 9-qubit GHZ states to form 27-qubit GHZ states illustrated in Fig.~\ref{figure-ghz4}(c). For a system with the total number of 4-atom sets equal to $3^{\mathbb{N}}$, the average number of QFG for gluing in this step is $3^{\mathbb{N}-2}{\overline{\text{n}}}$, so the fidelity for this step is 
\begin{eqnarray}
\mathcal{F}_3=(F_0F_{\text{detect}})^{3^{\mathbb{N}-2}{\overline{\text{n}}}}.\label{F3-define}
\end{eqnarray}
For the case of Fig.~\ref{figure-ghz4}, we have $\mathbb{N}=4$ which leads to $\mathcal{F}_3=(F_0F_{\text{detect}})^{9{\overline{\text{n}}}}$.

The fourth step is to glue 27-qubit GHZ states to form 81-qubit GHZ states as illustrated in Fig.~\ref{figure-ghz4}(d). As analyzed above, by using the QFG, the measurement protocol can lead to the three 81-qubit GHZ states with a fidelity
\begin{eqnarray}
 \mathcal{F}_4=(F_0F_{\text{detect}})^{3^{\mathbb{N}-3}{\overline{\text{n}}}}.\label{F4-define}
\end{eqnarray}
  For the case of Fig.~\ref{figure-ghz4}, we have $\mathcal{F}_4=(F_0F_{\text{detect}})^{3{\overline{\text{n}}}}$.

The fifth gluing iteration is to glue the 81-qubit GHZ states. As shown in Fig.~\ref{figure-ghz4}(e), we have three $\lvert \text{GHZ}^{(81)}\rangle$, so only one QFG is required for this step. The fidelity of this step is 
\begin{eqnarray}
\mathcal{F}_5=(F_0F_{\text{detect}})^{\overline{\text{n}}}.\label{F5-define}
\end{eqnarray}
\begin{figure*}
\includegraphics[width=6.80in]
{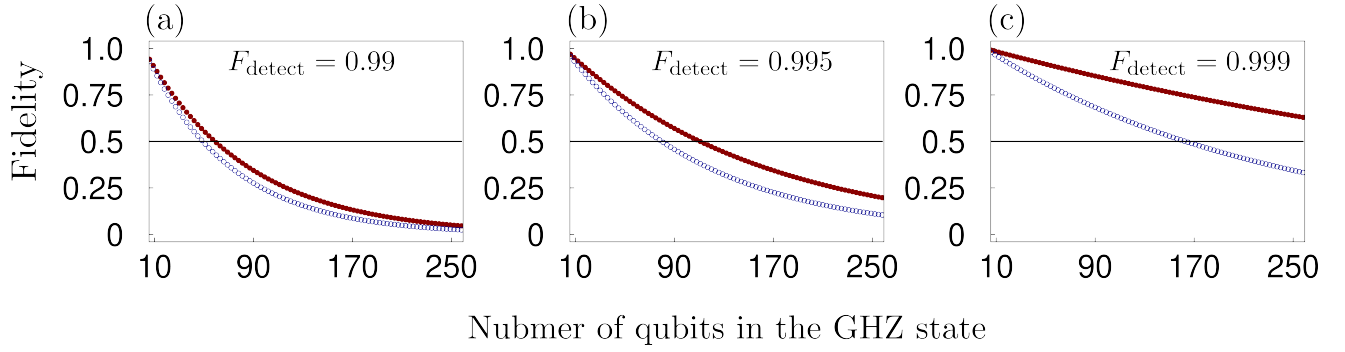}
\caption{Fidelity for creating an $\mathscr{N}$-qubit GHZ state, $\lvert \text{GHZ}^{(\mathscr{N})}\rangle$, as a function of $\mathscr{N}$ where $\mathscr{N}$ is a multiple of 3. The red~(blue) dots are with $F_0=0.9994~(0.9972)$ as from Table~\ref{table-error}. The data are evaluated with Eq.~(\ref{F-N}) with a detection fidelity equal to $0.99,~0.995$, and $0.999$ in (a), (b), and (c), respectively. In practice, $\mathcal{F}_{\text{detect}}=0.998$ was realized in Ref.~\cite{senoo2025}. The horizontal line shows the value 0.5.   \label{figure-4} }
\end{figure*}

The final fidelity for generating the $3^{\mathbb{N}+1}$-qubit GHZ state is $\mathcal{F}= \mathcal{F}_1\mathcal{F}_2\mathcal{F}_3\mathcal{F}_4\mathcal{F}_5=(F_0F_{\text{detect}})^{\sum_{k=0}^{\mathbb{N}} 3^k{\overline{\text{n}}}}$, i.e.,
\begin{eqnarray}
 \mathcal{F}
 &=&(F_0F_{\text{detect}}) ^{(3^{\mathbb{N}+1}-1){\overline{\text{n}}}/2},\label{F-N0}
\end{eqnarray}
where $ 3^{\mathbb{N}+1}$ is equal to the total number of the data atoms. If we use
$ \mathscr{N}$ to denote the number of qubits in the GHZ state, the fidelity can be written as
\begin{eqnarray}
 \mathcal{F}&=&  (F_0F_{\text{detect}}) ^{(\mathscr{N}-1){\overline{\text{n}}}/2},\label{F-N}
\end{eqnarray}
as a function of $\mathscr{N}$. In Fig.~\ref{figure-4}, we show the fidelity $\mathcal{F}$ for creating $\lvert \text{GHZ}^{(\mathscr{N})}\rangle$ with $F_{\text{detect}}=(0.99,~0.995,~0.999)$ in Fig.~\ref{figure-4}(a,~b,~c). The red~(blue) data in Fig.~\ref{figure-4} are with the upper and lower fidelities of the QFG as estimated in Secs.~\ref{sec02C} and~\ref{sec02D}, with typical errors from different sources shown in Table~\ref{table-error}. Here, Fig.~\ref{figure-4}(a) also includes $\mathscr{N}$ not equal to 3, 9, 27, $\cdots$. These are included because a GHZ state with $\mathscr{N}\neq 3^k$ can be generated by gluing three GHZ states $\lvert \text{GHZ}^{(\text{n}_1)}\rangle$, $\lvert \text{GHZ}^{(\text{n}_2)}\rangle$ and $\lvert \text{GHZ}^{(\text{n}_3)}\rangle$ with unequal $n_1,~n_2$, and $n_3$. In the example of 243-qubit GHZ state illustrated in Fig.~\ref{figure-ghz4}, we have
$\mathcal{F}= (F_0F_{\text{detect}})^{121{\overline{\text{n}}}}$, which is about 0.65 with $1-F_0=6.0~\times10^{-4}$~from Table~\ref{table-error} and $F_{\text{detect}}=0.999$. If $F_{\text{detect}}=1$, then 500-qubit GHZ state can be realized with our theory. Theoretically, cavity-assisted effective atomic interaction can help to generation GHZ state but not easy to scale to large $\mathscr{N}$~\cite{PhysRevA.96.062315}. Notably, Ref.~\cite{zhao2021} proposed an entanglement-amplification based method to generate GHZ states with thousands of atoms. Compared to Ref.~\cite{zhao2021} which showed a method to create an $\mathscr{N}$-atom GHZ state with a fidelity $0.981-2.31\ln\mathscr{N}/\mathscr{C}$, where $\mathscr{C}$ is the single-atom cooperativity, the fidelity in the present article is lower. The method of Ref.~\cite{zhao2021} needs to have equal coupling between each atom and the same cavity mode, and depends on the availability of high single-atom cooperativity. If the method of Ref.~\cite{zhao2021} is used to prepare a 243-qubit GHZ state, the fidelity would be 97\% with $\mathscr{C}=1000$, but would drop to 50\% if $\mathscr{C}=26$. Though realizing strong coupling between neutral atoms and optical cavity is challenging, hundreds of $^6$Li atoms were loaded into a high-finesse cavity and trapped by an optical lattice, leading to $\mathscr{C}=6.7$ in the experiment of Ref.~\cite{Sauerwein_2023}, and $\mathscr{C}=1.9$ was realized for trapping cesium atoms by optical tweezers inside a miniature optical cavity~\cite{Liu_2023}.

\section{Discussions}\label{sec06}
\subsection{Feasibility to prepare the required tweezer array}\label{sec06A}
The realization of the GHZ state with the theory in this article requires techniques to realize defect-free tweezer pattern of arbitrary configurations, which has been demonstrated with 1D~\cite{Endres2016}, 2D~\cite{Kim2016,Barredo_2016}, and 3D~\cite{Barredo_2018}. Moreover, AI-assisted fast assembly~\cite{Lin2024}, coherence-preserving spatial atom movement~\cite{Chiu2025} over thousands of atoms have been demonstrated. Also, Ref.~\cite{6100atoms} reported optical tweezers trapping more than 6100 neutral atoms, and Ref.~\cite{Chiu2025} demonstrated continuous operation of a coherent 3,000-qubit system. This points to the feasibility to realize the two-layer atomic tweezer arrays illustrated in Fig.~\ref{figure-ghz4}.

\subsection{Ancilla detection}\label{sec06B}
\subsubsection{Fidelity}\label{sec06B01}
The state detection can be via the strong $^1S_0\leftrightarrow ^1P_1$ transition as in Refs.~\cite{Jenkins2022,Ma2023}. This imaging, however, will wipe out the nuclear spin coherence due to the hyperfine interaction in the $^1P_1$ state. To make sure that the measurement doesn't spoil the applicability of the ancilla in a following gate-measurement circuit, one can use circularly polarized field to map the qubit state $|0\rangle$ to the ground state~\cite{Chen2022}. If the imaging via the $^1S_0\leftrightarrow~^1P_1$ transition results in a positive photon count, i.e., if the imaging tells us that the atom is in the ground state, then according to Eqs.~(\ref{s06}) and~(\ref{9qubitGHZ-4}, the state of the data atoms collapses to the desired GHZ state while the ancilla atom is in a mixed state, as in Eq.~(\ref{s06-an02}) in the example of a 3-qubit GHZ state. But when the measurement yields a negative photon count, i.e., there is no ground-state atom at the targeted tweezer site, it means that the state of the data atoms collapses to the undesired states $\lvert\text{O}\rangle$ and $\lvert\mathbb{O}\rangle$ instead of to the GHZ states, and meanwhile the ancilla atom is in a pure state, as in Eq.~(\ref{s06-an01}) and Eq.~(\ref{9qubit-pure}) in the example of creating 3- and 9-qubit GHZ states, respectively. Because the coherence of the ancillary atom is preserved in the pure states like those in Eq.~(\ref{s06-an01}) and Eq.~(\ref{9qubit-pure}), it can be used in following gate-measurement circuits. This is why up to three gate-measurement circuits are sufficient to create one larger GHZ state from three smaller GHZ state as detailed in Sec.~\ref{sec03A} in the example of 3-qubit GHZ state.

High-fidelity state imaging is required for deterministic creation of the GHZ state. In principle, by using longer exposure times, sufficiently large detection fidelity can be achieved~\cite{Jenkins2022}. But to speed up the GHZ-state generation, fast destructive detection is helpful. In Ref.~\cite{senoo2025}, a $t_{\text{imag}}=17~\mu$s exposure time realized the ground-state detection with an infidelity smaller than 0.002~(page 12 of Ref.~\cite{senoo2025}). If the fidelity $\mathcal{F}_{\text{detect}}=0.998$ for the state detection is assumed, the final fidelity for the 243-state generation would be 49.2\% when we take $F_0=0.9994$ from Table~\ref{table-error}, which means failure to generate entanglement. With $\mathcal{F}_{\text{detect}}=0.999$, the data in Fig.~\ref{figure-4}(c) shows that the fidelity is 64.7\%. This means that high-fidelity detection is of vital importance.

\subsubsection{Site resolution}\label{sec06B02}
During the preparation of the eighty-one $\lvert \text{GHZ}^{(3)}\rangle$, up to three gate-measurement circuits may be needed as detailed in Sec.~\ref{sec02C2}. We use Circuit 1 as an example to discuss site resolution. Before the ancilla detection, the qubit state $\lvert0\rangle$ is first shelved to the ground state as discussed above Eq.~(\ref{s06-an02}). For each parallel detection of the ancilla, the imaging light will only be absorbed by a certain number $N_1$ of the eighty-one ancilla, while all the other ancilla stay intact. According to contents around Eq.~(\ref{p-x-yes}), $N_1$ is around $81/4$ during the step of Circuit 1-measurement. This means that during the detection of the eighty-one ancilla in the eighty-one blue triangles of Fig.~\ref{figure-4-step}(a4), one out of four ancilla in each row or column of the blue triangles will scatter photons. For $\mathbb{L}=5.5~\mu$m as discussed in Sec.~\ref{sec02C2}, the average distance between two ancilla that can scatter photons is over $4\mathbb{L}$, which enhances the site resolution. For the following two circuits as detailed in Sec.~\ref{sec03A}, the average number of ancilla that are transferred to the ground state is again $81/4$, and thus the site resolution stays the same as in Circuit 1. Note that even if two ancilla in two nearby blue triangles of Fig.~\ref{figure-4-step}(a4) scatter light, they are separated by $\sqrt{3}\mathbb{L}$, which is large enough to be distinguished when $\mathbb{L}=5.5~\mu$m.

As long as the sites can be distinguished, the detection-based GHZ state generation can proceed. In particular, there is no issue of crosstalk during the detection of the ancilla for the creation of large-scale GHZ state with a configuration as in Fig.~\ref{figure-ghz4}. As understood from the probability numbers in Fig.~\ref{figure-product-GHZ}, only part of the ancilla atoms will be in the ground state that can be imaged at each detection event. So, during Circuit 1-measurement, the anciila atoms in the ground state will scatter light and be imaged, while the scattered photons can not be absorbed by the data atoms for they are not in the ground state. In other words, if a photon emitted from ancilla A is absorbed by ancilla B, it will contribute to the fluorescence in ancilla B. This is not a problem because ancilla B is supposed to be detected to be in the ground state. During the detection of Circuit 2, the ancilla that have been detected during Circuit 1 will also scatter light which is not a problem since their locations are already known.

\section{Conclusions}\label{sec07}
In conclusion, we show that by using nuclear-spin qubits in alkaline-earth-like atoms, a phase gate between four nuclear-spin qubits can arise, where the gate imprints a $\pi$ phase to the input states $\lvert0000\rangle$ and $\lvert1111\rangle$. We show that by using this gate and several single-qubit operations, a three-qubit GHZ state can be generated in three data atoms by using up to three projective measurements of an ancilla atom. Repetition of this procedure can be used to glue three nearby smaller GHZ states to a larger GHZ state. By using an example of a two-layer atomic array, we outline a procedure to generate a 243-qubit GHZ state. It turns out that a detection fidelity around 0.999 is required, while a smaller~(larger) detection fidelity can help to generate GHZ state with several tens~(hundreds) of qubits. 


\section{acknowledgments}
The research leading to the results here has received funding from the National Natural Science Foundation of China under Grant No. 12074300 and the Quantum Science and Technology-National Science and Technology Major project Grant No. 2021ZD0302100. We acknowledge the anonymous referees
for their valuable comments, and the Beijing Super Cloud Center for providing HPC resources for the large-scale optimization calculation reported in Fig.~\ref{figure-gate}.

\appendix
\section{Quantum ferromagnetic gate in alkali-metal atoms}\label{app-A}
If qubits are defined by using hyperfine-Zeeman substates in alkali-metal atoms, the QFG gate can be realized by combining two controlled-controlled-controlled-Z~(CCCZ) gates. Many different protocols can realize such a gate.
A simple way to realize this gate requires two steps, each step with their own Rydberg lasers: \newline
{\bf Step 1}: by using Rydberg excitation of the qubit state $\lvert1\rangle$ assisted by single-qubit rotations, one can realize the canonical CCCZ gate~\cite{Isenhower2011},
\begin{eqnarray}
\lvert \alpha_1\beta_2\chi_{3}\zeta_4\rangle \mapsto \left\{\begin{array}{cc} -\lvert \alpha_1\beta_2\chi_{3}\zeta_4\rangle,&\text{ if }\alpha_1\beta_2\chi_{3}\zeta_4=1,\\\lvert \alpha_1\beta_2\chi_{3}\zeta_4\rangle,&\text{  if }\alpha_1\beta_2\chi_{3}\zeta_4=0, \end{array}\right.                                                                                                                                                                                                                              \label{4qfg-2}
\end{eqnarray}
where the analyses in Ref.~\cite{Pelegr2022} showed that with a gate duration $T\approx6.08\frac{2\pi}{\Omega_{\text{max}}}$, a CCCZ gate of fidelity 0.991 can be realized, and Ref.~\cite{PhysRevA.110.032619} analyzed that a CCCZ gate of fidelity 0.996 is realizable with a gate duration $T=12\frac{2\pi}{\Omega_{\text{max}}}$. Note that if Forster resonant Rydberg-Rydberg interaction is employed for such gates~\cite{M_Farouk_2023}, faster operations can be possible, with gate duration $\frac{3\pi}{\Omega}$~\cite{Yu_2022}. It was also shown that CCCZ gates with distant atoms can be realized with a longer gate duration~\cite{Petrosym2025}.
\newline
{\bf Setp 2}: by using another set of lasers which realize Rydberg excitation of the qubit state $\lvert0\rangle$, the same method for
generating Eq.~(\ref{4qfg-2}) can generate the CCCZ gate of the following form
\begin{eqnarray}
\lvert \alpha_1\beta_2\chi_{3}\zeta_4\rangle \mapsto \left\{\begin{array}{cc} -\lvert \alpha_1\beta_2\chi_{3}\zeta_4\rangle,&\text{ if }\alpha_1=\beta_2=\chi_{3}=\zeta_4=0,\\\lvert \alpha_1\beta_2\chi_{3}\zeta_4\rangle,&\text{  else, } \end{array}\right.                                                                                                                                                                                                                              \nonumber\label{4qfg-1}
\end{eqnarray}
then, the combination of the two steps above yields the QFG studied in Sec.~\ref{sec02}.
This means that by using alkali-metal atoms, the time required to realize the gate is $12\frac{2\pi}{\Omega_{\text{max}}}$ with the protocol in Ref.~\cite{Pelegr2022} or $24\frac{2\pi}{\Omega_{\text{max}}}$ with the one in Ref.~\cite{PhysRevA.110.032619}.

Though it looks like that using alkali-metal atoms is fine, the realization of the QFG with alkali-metal atoms needs two sets of Rydberg lasers, while with nuclear-spin qubits of alkaline-earth-like atoms, one set of Rydberg laser is sufficient.

\section{Details of the numerical simulation}\label{app-B}
The method to write the code for the optimization has been given in Appendix B of Ref.~\cite{ShiLu2024}. Here, we outline the essentials of the the numerical method for the QFG.

We first construct the Hamiltonian matrix used for the optimization. Because the inclusion of the Rydberg blockade of Eq.~(\ref{Ryd-H}) yields a matrix with a large dimension, while the optimization can only handle a small enough matrix, we ignore Rydberg blockade. In this case, the Hamiltonians relevant for the input states $\lvert 0_10_20_30_4\rangle$ and $\lvert 1_11_21_31_4\rangle$ are
\begin{eqnarray}
\hat{H}_{0000}&=&\left[\frac{2\Omega(t)}{2}\lvert \overline{w4}  \rangle \langle0_10_20_30_4\rvert +\text{H.c.} \right] + \Delta\lvert \overline{w4}  \rangle \langle \overline{w4}\rvert,\nonumber\\
\hat{H}_{1111}&=&\left[\frac{2\Omega(t)}{2}\lvert \overline{W4}  \rangle \langle 1_11_21_31_4\rvert +\text{H.c.} \right] - \Delta\lvert \overline{W4}  \rangle \langle \overline{W4}\rvert \label{H-0000}
\end{eqnarray}
where the factor 2 in the numerator of $\frac{2\Omega(t)}{2}$ is kept to highlight its origin from the collective-excitation enhancement due to the blockade, and
\begin{eqnarray}
\lvert \overline{w4}  \rangle &=& \frac{1}{2} ( \lvert   r_10_20_30_4 \rangle + \lvert   0_1r_20_30_4 \rangle +  \lvert   0_10_2r_30_4 \rangle +  \lvert   0_10_20_3r_4 \rangle),\nonumber\\
\lvert \overline{W4}  \rangle &=& \frac{1}{2} ( \lvert   R_11_21_31_4 \rangle + \lvert   1_1R_21_31_4 \rangle +  \lvert   1_11_2R_31_4 \rangle +  \lvert   1_11_21_3R_4 \rangle),\label{app-B-w4}
\end{eqnarray}
where the notation $\lvert \overline{w4(W4)}  \rangle$ is used because it is like a W state among the 4 atoms. One can see that either $\hat{H}_{0000}$ or $\hat{H}_{1111}$ can be represented by a $2\times2$ matrix. For the input states $\lvert 0_10_20_31_4\rangle$ and $\lvert 1_11_21_30_4\rangle$, the Hamiltonians are
\begin{eqnarray}
\hat{H}_{0001}&=&\left[\left(\frac{\sqrt{3}\Omega(t)}{2}\lvert \overline{w3}  \rangle \langle0_10_20_31_4\rvert +\frac{\Omega(t)}{2}\lvert  0_10_20_3R_4 \rangle \langle0_10_20_31_4\rvert\right)+\text{H.c.} \right] + \Delta(\lvert \overline{w3}  \rangle \langle \overline{w3}\rvert -\lvert  0_10_20_3R_4 \rangle \langle 0_10_20_3R_4 \rvert  ),\nonumber\\
\hat{H}_{1110}&=& \left[\left(\frac{\sqrt{3}\Omega(t)}{2}\lvert \overline{W3}  \rangle \langle1_11_21_30_4\rvert +\frac{\Omega(t)}{2}\lvert  1_11_21_3r_4 \rangle \langle1_11_21_30_4 \rvert\right)+\text{H.c.} \right] - \Delta(\lvert \overline{W3}  \rangle \langle \overline{W3}\rvert -\lvert  1_11_21_3r_4\rangle \langle 1_11_21_3r_4 \rvert  ),\nonumber\\\label{H-0001}
\end{eqnarray}
which shows that the dimension of $\hat{H}_{0001}$ or $\hat{H}_{1110}$ is three, where
\begin{eqnarray}
\lvert \overline{w3}  \rangle &=& \frac{1}{\sqrt{3}} ( \lvert   r_10_20_31_4 \rangle + \lvert   0_1r_20_31_4 \rangle +  \lvert   0_10_2r_31_4 \rangle  ),\nonumber\\
\lvert \overline{W3}  \rangle &=& \frac{1}{\sqrt{3}} ( \lvert   R_11_21_30_4 \rangle + \lvert   1_1R_21_30_4 \rangle +  \lvert   1_11_2R_30_4 \rangle  ).\label{app-B-w3}
\end{eqnarray}
For the input state $\lvert 0_10_21_31_4\rangle$ and $\lvert 1_11_20_30_4\rangle$, we have
\begin{eqnarray}
\hat{H}_{0011}&=&\left[\left(\frac{\sqrt{2}\Omega(t)}{2}\lvert \overline{w2}  \rangle \langle0_10_21_31_4\rvert +\frac{\sqrt{2}\Omega(t)}{2}\lvert  \overline{W2}  \rangle \langle0_10_21_31_4\rvert\right)+\text{H.c.} \right] + \Delta(\lvert \overline{w2}  \rangle \langle \overline{w2}\rvert -  \lvert \overline{W2}  \rangle \langle \overline{W2}\rvert ),\nonumber\\
\hat{H}_{1100}&=&  \left[\left(\frac{\sqrt{2}\Omega(t)}{2}\lvert \overline{w2}'  \rangle \langle1_11_20_30_4\rvert +\frac{\sqrt{2}\Omega(t)}{2}\lvert  \overline{W2}'  \rangle \langle 1_11_20_30_4\rvert\right)+\text{H.c.} \right] + \Delta(\lvert \overline{w2}'  \rangle \langle \overline{w2}'\rvert -  \lvert \overline{W2}'  \rangle \langle \overline{W2}'\rvert ),\label{H-0011}
\end{eqnarray}
where
\begin{eqnarray}
\lvert \overline{w2}  \rangle &=& \frac{1}{\sqrt{2}} ( \lvert   r_10_21_31_4 \rangle + \lvert   0_1r_21_31_4 \rangle     ),\nonumber\\
\lvert \overline{W2}  \rangle &=& \frac{1}{\sqrt{2}} ( \lvert  0_10_2R_31_4   \rangle + \lvert    0_10_21_3R_4 \rangle  ) , \nonumber\\
\lvert \overline{w2}'  \rangle &=& \frac{1}{\sqrt{2}} ( \lvert  1_11_2r_30_4  \rangle + \lvert    \rangle 1_11_20_3r_4 \rangle  ),\nonumber\\
\lvert \overline{W2}'  \rangle &=& \frac{1}{\sqrt{2}} ( \lvert   1_1R_20_30_4   \rangle + \lvert      R_11_20_30_4  \rangle   ) ,\label{app-B-w2}
\end{eqnarray}
and one can see that the dimension of $\hat{H}_{0011}$ or $\hat{H}_{1100}$ is three. A further examination on $\hat{H}_{0011}$ or $\hat{H}_{1100}$ reveal that when we write both of them in a matrix form, they have the same form though with different basis states. This means that in numerical simulation the results for $\lvert 0_10_21_31_4\rangle$ also apply to $\lvert 1_11_20_30_4\rangle$ when using appropriate basis states.

In the condition when the laser field on the four atoms have the same strength and the same time-dependent phase, the time dynamics for the input states $\{\lvert 0_10_20_31_4\rangle,\lvert 0_10_21_30_4\rangle, \lvert 0_11_20_30_4\rangle, \lvert 1_10_20_30_4\rangle\}$ will be similar, i.e., they can be studied by a Hamiltonian similar to $\hat{H}_{0001}$ in Eq.~(\ref{H-0001}). Similarly, the Hamiltonians for the input states $\{  \lvert 1_11_20_31_4\rangle, \lvert 1_10_21_31_4\rangle, \lvert 0_11_21_31_4\rangle \} $ can be studied by a Hamiltonian similar to $\hat{H}_{1110}$ in Eq.~(\ref{H-0001}). In this case, five types of Hamiltonians can be used for the eight different input eigenstates,
\begin{eqnarray}
\hat{H}_{0000} &:& \{ \lvert 0_10_20_30_4\rangle\},\nonumber\\
\hat{H}_{0001} &:& \{ \lvert 0_10_20_31_4\rangle, \lvert 0_10_21_30_4\rangle, \lvert 0_11_20_30_4\rangle, \lvert 1_10_20_30_4\rangle\},\nonumber\\
\hat{H}_{0011} &:& \{ \lvert 0_10_21_31_4\rangle, \lvert 0_11_20_31_4\rangle, \lvert 1_10_20_31_4\rangle, \lvert 0_11_21_30_4\rangle, \lvert 1_10_21_30_4\rangle, \lvert 1_11_20_30_4\rangle \},\nonumber\\
\hat{H}_{1110} &:& \{ \lvert 0_11_21_31_4\rangle, \lvert 1_10_21_31_4\rangle, \lvert 1_11_20_31_4\rangle, \lvert 1_11_21_30_4\rangle  \},\nonumber\\
\hat{H}_{1111} &:& \{ \lvert 1_11_21_31_4\rangle \},\label{5-type}
\end{eqnarray}
and the dimensions for the five types of Hamiltonians above are 2, 3, 3, 3, and 2, respectively. Because the numerical optimization of Ref.~\cite{ShiLu2024} needs to find a pulse profile that can induce the state transform of Eq.~(\ref{QFG01}) which can be converted to Eq.~(\ref{QFG02}) via Eq.~(\ref{phase-gate0}), the optimization shall simultaneously simulate the time dynamics with the five types of Hamiltonians in Eq.~(\ref{5-type}). For convenience, one can say that the matrix used in the optimization is
\begin{eqnarray}
\hat{\mathbb{H}}&=&\hat{H}_{0000} \oplus
\hat{H}_{0001} \oplus
\hat{H}_{0011} \oplus
\hat{H}_{1110}  \oplus
\hat{H}_{1111} ,
\end{eqnarray}
while in practice one can one by one calculate the propagation operators for each of the five Hamiltonian matrix at each time step.
The optimization start from a random value of $\{y_0,~y_1\}$ along with a random profile of arg$[\Omega(t)]$. We also need a total gate duration $t_{\text{\tiny g}}$ which is measured in unit of $2\pi/|\Omega(t)|$. Following Ref.~\cite{Jandura2022}, we would like to find the shortest possible pulse to realize the gate, but before the optimization we don't know what is it, nor do we know what is the optimal value of $\Delta$ that can render the smallest $t_{\text{\tiny g}}$. So, we use HPC to simultaneously simulate multiple optimization tasks with different sets of $\{\Delta, t_{\text{\tiny g}}\}$. The target is to maximize $F$ defined by~\cite{ShiLu2024}
\begin{eqnarray}
F&=&[|\text{Tr}(U_{\text{ideal}}^\dag U_{\text{real}} )|^2 + \text{Tr}(U_{\text{ideal}}^\dag U_{\text{real}}  U_{\text{real}}^\dag U_{\text{ideal}})]/272,\label{F0-definition}
\end{eqnarray}
or, equivalently, to find the smallest $-F$, where $U_{\text{ideal}}$ is the $16\times16$ diagonal matrix where only the first and last matrix element is $-1$ according to Eq.~(\ref{QFG02}) when the other diagonal matrix elements are 1. Here, $U_{\text{real}}$ is the $16\times16$ matrix simulated which incorporates the state transform for each input eigenstate. The matrix elements in $U_{\text{real}}$ are from the corresponding time evolution operators. For example, the first diagonal matrix element is equal to $e^{4iy_0}\langle 0_10_20_30_4\lvert\mathcal{T}$exp$[-it\hat{H}_{0000}]\lvert 0_10_20_30_4\rangle $ where the phase factor $4y_0$ is due to that the final QFG is realized by single-qubit phase gates in Eq.~(\ref{phase-gate0}).

For each set of $\{\Delta, t_{\text{\tiny g}}\}$, and in each iteration, we calculate the unitary time evolution, the derivative of the target function with respect to the change of arg$\Omega(t_i)$ at the time step $i$, and the values for $\{\text{arg}([\Omega(t_i)]\}$ are updated according to the simulated gradients. In the same way, the values for $\{y_0,~y_1\}$ must also be updated according to the simulated gradients for them. Because we don't know what are the optimal values for $\{y_0,~y_1\}$ with a certain set of $\{\Delta, t_{\text{\tiny g}}\}$, most initial values $\{y_0,~y_1\}$ yield useless results. Also, we seek after a small $t_{\text{\tiny g}}$, and the optimization process depends on $\Delta/|\Omega(t)|$, the increase of $F$ with growing iteration numbers slows down. This requires us to use parallel simulation of large sets of $\{\Delta, t_{\text{\tiny g}}\}$. Moreover, a large iteration number, e.g., $2\times10^5$, can be used to look for the optimal $t_{\text{\tiny g}}$ though one can also set a threshold value to stop. For example, we can stop the optimization if $1-F\leq 10^{-8}$. The total time steps can be, e.g., 100. For most initial sets of $\{\Delta, t_{\text{\tiny g}}\}$, the final $1-F$ is larger than $0.01$ at the end of the optimization. For certain values of $\{\Delta, t_{\text{\tiny g}}\}_{\text{good}}$ we have $F$ nearer to $1$, which means that we can vary $\{\Delta, t_{\text{\tiny g}}\}$ around $\{\Delta, t_{\text{\tiny g}}\}_{\text{good}}$ to do more simulations. To find the phase profile in Fig.~\ref{figure-gate}, more than four hundred sets of $\{\Delta, t_{\text{\tiny g}}\}$ were used and within one week the simulation finished where the CPU frequency was about 800~MHz. Note, however, it turns out the if we start from the $\{\Delta, t_{\text{\tiny g}}\}$ of Fig.~\ref{figure-gate}, only three hours of optimization can yield the desired pulse. This means that the most time-consuming part is to find the optimal combination of $\Delta$ and $t_{\text{\tiny g}}$.

\section{Finite blockade}\label{app-C}
There will be excitations with more than one Rydberg atoms in the QFG. However, the probability to have two-atom Rydberg state is already small, and a three-atom Rydberg state is excited from a two-atom Rydberg state. So we can ignore states with three or four Rydberg atoms. Again, there are still five classes of input states Eq.~(\ref{5-type}) that we should consider in this case. The Hamiltonian for $\lvert 0_10_20_30_4\rangle$ is
\begin{eqnarray}
\hat{H}_{0000}(V)&=& \left(\begin{array}{ccc}V+2\Delta & \frac{\sqrt{6}}{2}\Omega(t) & 0\\\frac{\sqrt{6}}{2}\Omega^\ast(t)&   \Delta& \Omega(t)\\0&\Omega^\ast(t)&0
                                      \end{array}
\right)\label{VH0000}
\end{eqnarray}
with basis of $\lvert \overline{w4r}  \rangle$, $\lvert \overline{w4}  \rangle$, and $\lvert 0_10_20_30_4\rangle$ where
\begin{eqnarray}
 \lvert \overline{w4r}  \rangle &=& \frac{1}{\sqrt{6}} ( \lvert   r_1r_20_30_4 \rangle + \lvert   0_1r_2r_30_4 \rangle +  \lvert   0_10_2r_3r_4 \rangle +  \lvert   r_10_20_3r_4 \rangle+  \lvert   0_1r_20_3r_4 \rangle+  \lvert   r_10_2r_30_4 \rangle),
\end{eqnarray}
and $\lvert \overline{w4}  \rangle$ is given in Eq.~(\ref{app-B-w4}). The Hamiltonian for $\lvert 0_10_20_31_4\rangle$ is
\begin{eqnarray}
\hat{H}_{0001}(V)&=& \left(\begin{array}{ccccc}V+2\Delta  &  0 & \Omega(t)& 0 &  0\\  0 & V &\frac{1}{2}\Omega(t)  & \frac{\sqrt{3}}{2}\Omega(t) &  0\\ \Omega^\ast(t)& \frac{1}{2}\Omega^\ast(t)&   \Delta& 0  & \frac{\sqrt{3}}{2} \Omega(t)\\0&\frac{\sqrt{3}}{2}\Omega^\ast(t)& 0&  -\Delta& \frac{1}{2} \Omega(t)\\0&0 &\frac{\sqrt{3}}{2} \Omega^\ast(t)&\frac{1}{2}\Omega^\ast(t)&0
                                      \end{array}
\right)\label{VH0001}
\end{eqnarray}
with basis of $\lvert \overline{w3rr}  \rangle,~\lvert \overline{w3rR}  \rangle$, $\lvert \overline{w3}  \rangle$, $\lvert 0_10_20_3R_4\rangle$, and $\lvert 0_10_20_31_4\rangle$, where
\begin{eqnarray}
 \lvert \overline{w3rr}  \rangle &=&  \frac{1}{\sqrt{3}} ( \lvert   r_1r_20_31_4 \rangle + \lvert   0_1r_2r_31_4 \rangle +  \lvert   r_10_2r_31_4 \rangle  ),\nonumber\\
 \lvert \overline{w3rR}  \rangle &=&  \frac{1}{\sqrt{3}} ( \lvert   r_10_20_3R_4 \rangle + \lvert   0_1r_20_3R_4 \rangle +  \lvert   0_10_2r_3R_4 \rangle  ),
\end{eqnarray}
and $\lvert \overline{w3}  \rangle$ is given in Eq.~(\ref{app-B-w3}). The Hamiltonian for $\lvert 0_10_21_31_4\rangle$ is
\begin{eqnarray}
\hat{H}_{0011}(V)&=& \left(\begin{array}{cccccc}V+2\Delta  &0  & 0 &  \frac{\sqrt{2}}{2}\Omega(t) &  0 &  0\\  0 & V-2\Delta  & 0&  0& \frac{\sqrt{2}}{2}\Omega(t) &  0\\ 0 & 0 & V & \frac{\sqrt{2}}{2}\Omega(t) &   \frac{\sqrt{2}}{2}\Omega(t)&  0\\ \frac{\sqrt{2}}{2}\Omega^\ast(t)& 0 &  \frac{\sqrt{2}}{2}\Omega^\ast(t)& \Delta& 0& \frac{\sqrt{2}}{2} \Omega(t)\\0&\frac{\sqrt{2}}{2}\Omega^\ast(t) & \frac{\sqrt{2}}{2}\Omega^\ast(t) & 0&  -\Delta& \frac{\sqrt{2}}{2} \Omega(t)\\0& 0& 0 &\frac{\sqrt{2}}{2} \Omega^\ast(t)&\frac{\sqrt{2}}{2}\Omega^\ast(t)&0
                                      \end{array}
\right) \label{VH0011}
\end{eqnarray}
with basis of $\lvert \overline{w2rr}  \rangle,~\lvert \overline{w2RR}  \rangle ,~\lvert \overline{w2rR}  \rangle$, $\lvert \overline{w2}  \rangle$, $\lvert \overline{W2}  \rangle$, and $\lvert 0_10_21_31_4\rangle$, where
\begin{eqnarray}
 \lvert \overline{w2rr}  \rangle &=&    \lvert   r_1r_21_31_4 \rangle  ,\nonumber\\
 \lvert \overline{w2RR}  \rangle &=&    \lvert   0_10_2R_3R_4 \rangle  ,\nonumber\\
 \lvert \overline{w2rR}  \rangle &=&  \frac{1}{2} (   \lvert   r_10_21_3R_4 \rangle + \lvert   r_10_2R_31_4 \rangle +  \lvert   0_1r_21_3R_4 \rangle + \lvert   0_1r_2R_31_4 \rangle ),
\end{eqnarray}
and $\lvert \overline{w2}  \rangle$ and $\lvert \overline{W2}  \rangle$ are given in Eq.~(\ref{app-B-w2}).

The Hamiltonian for the input $\lvert 1_11_21_31_4\rangle$ can be constructed similar to $\hat{H}_{0000}(V)$ with appropriate basis, and with $\Delta$ replaced by $-\Delta$. The Hamiltonian for the input $\lvert 0_11_21_31_4\rangle$ can be constructed similar to $\hat{H}_{0001}(V)$ with appropriate basis, and with $\pm\Delta$ replaced by $\mp\Delta$. By this, we have the Hamiltonians for all the five classes of input states Eq.~(\ref{5-type}).

It was shown in Ref.~\cite{Levine2019} that a finite Rydberg blockade can be treated as a renormalization of the energy of the Rydberg state. But one can see that the many-body enhanced Rabi frequency from single-Rydberg state to double-Rydberg state is different for different input states. This makes the renormalization of the energy of the Rydberg state state-dependent, which is sharp contrast to the case of Ref.~\cite{Levine2019} where only one many-body enhanced Rabi frequency appeared. The correct change of $\Delta$, if possible, should be $\frac{3\Omega^2}{2V}$ in Eq.~(\ref{VH0000}), $(\frac{5\Omega^2}{4V},~\frac{3\Omega^2}{4V})$ for $\pm\Delta$ in Eq.~(\ref{VH0001}), and $\frac{\Omega^2}{V}$ in Eq.~(\ref{VH0011}), respectively. So, we can take the average with respect to the all the input state, yielding a change of $\delta\approx \frac{\Omega^2}{V}$. By changing the frequency of the Rydberg laser by $\delta$, $\pm \Delta$ should become $\delta \pm \Delta$ in Eqs.~(\ref{VH0000}),~(\ref{VH0001}), and~(\ref{VH0011}).

\section{Position fluctuation of qubits}\label{app-D}
A finite blockade would be subjected to fluctuation. The Extended Data Fig.~2 of Ref.~\cite{Graham2022} showed that the longitudinal and transverse r.m.s. position fluctuations of the atoms in their traps were about $(\sigma_{z}, \sigma_{\perp})=(0.75,~0.15)~\mu$m. As an example, we assume a similar character of position fluctuations. We suppose that the data atoms are in $xy$ plane, and the quantization axis is along $\mathbf{z}$. Because of the larger position fluctuation along $\mathbf{z}$, the fluctuation of atomic separations between one data atom and the ancilla would be larger than the fluctuation of atomic separations between two data atoms. To estimate a lower bound of the fidelity in the presence of fluctuation of Rydberg interaction due to the atomic position fluctuation, we consider  the interaction between one data atom and the ancilla. We suppose that the center of the trap for the ancilla atom is at $(0,0,0)$, then the center of the trap for the data atom is at $(\mathbb{L}\sqrt{3}/3,0,\mathbb{L}\sqrt{6}/3)$. Because of the finiteness of the trap depths, the positions of the ancilla and data atoms are $(x_{\text{a}},y_{\text{a}},z_{\text{a}})$ and $(x_{\text{d}} + \mathbb{L}\sqrt{3}/3, ,y_{\text{d}} ,z_{\text{d}}+\mathbb{L}\sqrt{6}/3)$, respectively, where the distribution function
\begin{eqnarray}
\mathscr{G}(\zeta) &=& \frac{1}{\sqrt{2\pi} \sigma_\perp}e^{-\zeta^2/(2\sigma_\perp^2)}\nonumber
\end{eqnarray}
characterizes $\zeta\in\{x_{\text{a}},y_{\text{a}},x_{\text{d}},y_{\text{d}}\}$, and
\begin{eqnarray}
\mathscr{G}(\zeta) &=& \frac{1}{\sqrt{2\pi} \sigma_z}e^{-\zeta^2/(2\sigma_z^2)}\nonumber
\end{eqnarray}
is the distribution function for $\zeta=z_{\text{a}}$ or $z_{\text{d}}$. In the numerical simulation, the value of $V$ is calculated by using the actual distance of the qubits $\mathcal{L}=[( \mathbb{L}\sqrt{3}/3+ x_{\text{d}}-x_{\text{a}} )^2 +( y_{\text{d}}-y_{\text{a}})^2 +(\mathbb{L}\sqrt{6}/3+z_{\text{d}}-z_{\text{a}})^2 ]^{1/2}$. With $V$ replaced by $V (\mathbb{L}/\mathcal{L})^6$, the average fidelity can in principle be sampled with 
\begin{eqnarray}
\overline{\mathcal{F}}&=& \int F\mathscr{G}(x_{\text{a}} )\mathscr{G}(y_{\text{a}} )\mathscr{G}(z_{\text{a}} )\mathscr{G}(x_{\text{d}} )\mathscr{G}(y_{\text{d}} )\mathscr{G}(z_{\text{d}} ) d x_{\text{d}} d y_{\text{d}}d z_{\text{d}}d x_{\text{a}} d y_{\text{a}}d z_{\text{a}} ,\nonumber%
\end{eqnarray}
where $F$ is defined in Eq.~(\Ref{F0-definition}). In practice, one can approximate the sampling by taking several discrete values. For instance, by taking three $\zeta$ in each $\mathscr{G}(\zeta)$, the average $1-\overline{\mathcal{F}}$ is $1.84\times10^{-3}$ with $V=50\Omega$ and $\mathbb{L}=5.5~\mu$m, where the Rydberg laser frequency is shifted by $ \Omega^2/V$. Here, $\mathbb{L}=5.5~\mu$m is considered so as to have $V\approx2\pi\times1$~GHz as discussed in Sec.~\ref{sec02C2}.  Note that in Fig.~\ref{figure-Verror} the value of $1-\overline{\mathcal{F}}$ is $1.78\times10^{-3}$ with $V=50\Omega$, which means that the fluctuation of the atomic separation increases the blockade error by 3\%. This is due to the robustness of the blockade mechanism as observed in previous numerical simulation~\cite{ShiLu2024}.

\section{Gluing cycle without using single-qubit phase gate}\label{app-E}
We show that no single-qubit phase gate is needed in the elementary gluing cycle. Again, we consider a product state of a 4-qubit system as in Eq.~(\ref{s01}). Like in the main text, we study the three circuits, Circuit 1, Circuit 2, and Circuit 3, respectively, shown in Fig.~\ref{figure-product-GHZ}. Both Circuit 1 and Circuit 2 consist of three laser pulses and one measurement operation, while Circuit 3 consists of four laser pulses and one measurement operation.  \newline
{\bf Circuit 1-pulse 1}. For the three data atoms and the ancilla, we excite the atomic transition $\lvert 0\rangle\leftrightarrow\lvert 1\rangle$ with a Rabi frequency $ie^{iy}\Omega_{\text{c}}$, i.e., with a Hamiltonian 
\begin{eqnarray}
\hat{H}_j(\Omega_{\text{c}}, \frac{\pi}{2}+y)=ie^{iy}\frac{\Omega_{\text{c}}}{2}\lvert 1_{\text{j}}\rangle\langle 0_{\text{j}}\rvert+\text{H.c.},\label{app-Hamiltonian-atom-j}
\end{eqnarray}
where j$=1,~2,~3$, or a, and $y$ is given around Eq.~(\ref{QFG02}). Then, Eq.~(\ref{s01}) becomes,
\begin{eqnarray}
\left(\prod_{i=1}^3\frac{\lvert0_{\text{i}}\rangle +e^{iy}\lvert1_{\text{i}}\rangle  }{\sqrt{2}}\right)\otimes \frac{\lvert0_{\text{a}}\rangle +e^{iy}\lvert1_{\text{a}}\rangle  }{\sqrt{2}},\label{app-s02}
\end{eqnarray}
with a $\frac{\pi}{2}$ pulse. The part enclosed by $(\cdots)$ in Eq.~(\ref{app-s02}) can be expanded, leading to
\begin{eqnarray}
\frac{1}{2}\left(\frac{\lvert0_{1}0_{2}0_{3}\rangle +e^{3iy}\lvert1_{1}1_{2}1_{3}\rangle   }{\sqrt{2}}+ \sqrt{3}\lvert \text{O}_0\rangle \right)\otimes \frac{\lvert0_{\text{a}}\rangle +e^{iy}\lvert1_{\text{a}}\rangle  }{\sqrt{2}},\label{app-s04}
\end{eqnarray}
where
\begin{eqnarray}
\lvert\text{O}_0\rangle &=&  \frac{1}{\sqrt{6}}[e^{iy}(\lvert0_{1}0_{2}1_{3} \rangle + \lvert0_{1}1_{2}0_{3} \rangle + \lvert1_{1}0_{2}0_{3} \rangle) +e^{2iy} ( \lvert0_{1}1_{2}1_{3} \rangle + \lvert1_{1}0_{2}1_{3} \rangle + \lvert1_{1}1_{2}0_{3} \rangle)],\nonumber\\
\lvert\text{O}\rangle &=&  \frac{1}{\sqrt{6}}(\lvert0_{1}0_{2}1_{3} \rangle + \lvert0_{1}1_{2}0_{3} \rangle + \lvert1_{1}0_{2}0_{3} \rangle + \lvert0_{1}1_{2}1_{3} \rangle + \lvert1_{1}0_{2}1_{3} \rangle + \lvert1_{1}1_{2}0_{3} \rangle),\label{app-operatorO}
\end{eqnarray}
where $\lvert\text{O}\rangle$ will appear below. \newline
{\bf Circuit 1-pulse 2}. To illustrate the gluing of the states by QFG, we note that Eq.~(\ref{app-s04}) can be written as
\begin{eqnarray}
\frac{\sqrt{3}}{2}\lvert \text{O}_0\rangle \otimes \frac{\lvert0_{\text{a}}\rangle +e^{iy}\lvert1_{\text{a}}\rangle  }{\sqrt{2}} + \frac{1}{4}\left(    \lvert0_{1}0_{2}0_{3}\rangle\otimes \lvert0_{\text{a}}\rangle +e^{3iy}\lvert1_{1}1_{2}1_{3}\rangle\otimes \lvert0_{\text{a}}\rangle +  e^{iy} \lvert0_{1}0_{2}0_{3}\rangle\otimes \lvert1_{\text{a}}\rangle +e^{4iy}\lvert1_{1}1_{2}1_{3}\rangle\otimes \lvert1_{\text{a}}\rangle    \right) .\label{app-s04-vary1}
\end{eqnarray}
By using the 4-qubit QFG which induces state-dependent phase $x_1-x_5$ as in Eq.~(\ref{QFG01}), one can verify that the QFG in Eq.~(\ref{QFG02}) emerges. Then, Eq.~(\ref{app-s04-vary1}) becomes
\begin{eqnarray}
&&\frac{\sqrt{3}e^{4iy_0}}{2}\lvert \text{O}\rangle \otimes \frac{\lvert0_{\text{a}}\rangle +\lvert1_{\text{a}}\rangle  }{\sqrt{2}} + \frac{e^{4iy_0}}{4}\left(  -  \lvert0_{1}0_{2}0_{3}\rangle\otimes \lvert0_{\text{a}}\rangle +\lvert1_{1}1_{2}1_{3}\rangle\otimes \lvert0_{\text{a}}\rangle +   \lvert0_{1}0_{2}0_{3}\rangle\otimes \lvert1_{\text{a}}\rangle -\lvert1_{1}1_{2}1_{3}\rangle\otimes \lvert1_{\text{a}}\rangle    \right)\nonumber\\
&=&\frac{\sqrt{3}e^{4iy_0}}{2}\lvert \text{O}\rangle \otimes \frac{\lvert0_{\text{a}}\rangle +\lvert1_{\text{a}}\rangle  }{\sqrt{2}}-e^{4iy_0}
\frac{\lvert0_{1}0_{2}0_{3}\rangle -\lvert1_{1}1_{2}1_{3}\rangle   }{2^{3/2}}\otimes \frac{\lvert0_{\text{a}}\rangle -\lvert1_{\text{a}}\rangle  }{\sqrt{2}}\nonumber\\
&=&\frac{\sqrt{3}e^{4iy_0}}{2}\lvert \text{O}\rangle \otimes \frac{\lvert0_{\text{a}}\rangle +\lvert1_{\text{a}}\rangle  }{\sqrt{2}}-\frac{e^{4iy_0}}{2} \lvert \text{GHZ}^{(3)}_-\rangle
 \otimes \frac{\lvert0_{\text{a}}\rangle -\lvert1_{\text{a}}\rangle  }{\sqrt{2}}.
\label{app-s05}
\end{eqnarray}
In this step, the duration for realizing the QFG is less than $6\pi/\Omega$, which can be less than $1~\mu$s with a MHz-scale Rydberg Rabi frequency.
\newline
{\bf Circuit 1-pulse 3}. We excite the transition $\lvert 0\rangle\leftrightarrow\lvert 1\rangle$ of the ancillary atom with a Rabi frequency $i\Omega_{\text{c}}$ for a duration $\frac{\pi}{2\Omega_{\text{c}}}$, i.e., with a Hamiltonian $\hat{H}_{\text{a}}(\Omega_{\text{c}}, \frac{\pi}{2})$ as defined in Eq.~(\ref{app-Hamiltonian-atom-j}). This pulse changes Eq.~(\ref{app-s05}) to
\begin{eqnarray}
|\text{M}\rangle_{1}&=&\frac{\sqrt{3}e^{4iy_0}}{2}\lvert \text{O}\rangle \otimes  \lvert1_{\text{a}}\rangle -
\frac{e^{4iy_0}}{2}\frac{\lvert0_{1}0_{2}0_{3}\rangle -\lvert1_{1}1_{2}1_{3}\rangle   }{\sqrt{2}}\otimes  \lvert0_{\text{a}}\rangle\nonumber\\
&\equiv& \frac{\sqrt{3}e^{4iy_0}}{2}\lvert \text{O}\rangle \otimes  \lvert1_{\text{a}}\rangle -
\frac{e^{4iy_0}}{2}   \lvert \text{GHZ}^{(\text{3})}_-\rangle  \otimes\lvert0_{\text{a}}\rangle .\label{app-s06}
\end{eqnarray}
\newline
{\bf Circuit 1-measurement}. If the imaging via the $^1S_0\leftrightarrow~^1P_1$ transition is positive, i.e., an atom is detected, the state in Eq.~(\ref{app-s06}) collapses to
\begin{eqnarray}
\lvert \text{GHZ}^{(\text{3})}_-\rangle \otimes  \lvert {\text{Mixed state}}\rangle,\label{app-s06-an02}
\end{eqnarray}
where $\lvert {\text{Mixed state}}\rangle$ denotes a state of the ancilla without any coherence in the nuclear spin due to the imaging light. But if no atom is detected, it means that the atom is still in the state $|1\rangle$. In other words, the state in Eq.~(\ref{app-s06}) collapses to
\begin{eqnarray}
e^{4iy_0}\lvert \text{O}\rangle \otimes  \lvert1_{\text{a}}\rangle, \label{app-s06-an01}
\end{eqnarray}
which is still a pure state due to that no optical pumping occurs since the ancillary atom is still in the clock state.  \newline
{\bf Circuit 2-pulse 1}. Circuit 2 is shown in Fig.~\ref{figure-product-GHZ}(b). If the measurement result is $\lvert1\rangle$ in the last step, we have got Eq.~(\ref{app-s06-an01}), where the overall phase term $e^{4iy_0}$ is trivial and can be ignored. For the third data atom, we excite the transition $\lvert 0\rangle\leftrightarrow\lvert 1\rangle$ with a Rabi frequency $e^{\frac{3}{2}iy}\Omega_{\text{c}}$ for a $\pi$ pulse that induces
\begin{eqnarray}
&&  \lvert0_{\text{3}}\rangle \xrightarrow{e^{\frac{3}{2}iy}\Omega_{\text{c}},~ \pi \text{pulse}} -ie^{\frac{3}{2}iy}\lvert1_{\text{3}}\rangle,\nonumber\\
&&   \lvert1_{\text{3}}\rangle \xrightarrow{e^{\frac{3}{2}iy}\Omega_{\text{c}},~ \pi \text{pulse}} -ie^{-\frac{3}{2}iy}\lvert0_{\text{3}}\rangle,\label{app-rabi-pi}
\end{eqnarray}
and for the ancilla, we excite the transition $\lvert 0\rangle\leftrightarrow\lvert 1\rangle$ with a Rabi frequency $-ie^{iy}\Omega_{\text{c}}$ for a $\pi/2$ plus,
\begin{eqnarray}
&&  \lvert0_{\text{a}}\rangle \xrightarrow{-ie^{iy}\Omega_{\text{c}},~ \frac{\pi}{2} \text{pulse}}\frac{\lvert0_{\text{a}}\rangle -e^{iy}\lvert1_{\text{a}}\rangle  }{\sqrt{2}},\nonumber\\
&&   \lvert1_{\text{a}}\rangle \xrightarrow{-ie^{iy}\Omega_{\text{c}}, ~\frac{\pi}{2} \text{pulse}}\frac{e^{-iy}\lvert0_{\text{a}}\rangle +\lvert1_{\text{a}}\rangle  }{\sqrt{2}}.\label{app-C2-p1-ancilla}
\end{eqnarray}
Then,  Eq.~(\ref{app-s06-an01}) becomes,
\begin{eqnarray}
 &&  \frac{-i}{×\sqrt{6}} (e^{-\frac{3}{2}iy} \lvert0_{1}0_{2}0_{3} \rangle + e^{\frac{3}{2}iy}\lvert0_{1}1_{2}1_{3} \rangle + e^{\frac{3}{2}iy}\lvert1_{1}0_{2}1_{3} \rangle +e^{-\frac{3}{2}iy}\lvert0_{1}1_{2}0_{3} \rangle +e^{-\frac{3}{2}iy}\lvert1_{1}0_{2}0_{3} \rangle + e^{\frac{3}{2}iy}\lvert1_{1}1_{2}1_{3} \rangle)\otimes \frac{e^{-iy}\lvert0_{\text{a}}\rangle +\lvert1_{\text{a}}\rangle  }{\sqrt{2}}.\nonumber\\
 \label{app-an03}
\end{eqnarray}
 \newline
{\bf Circuit 2-pulse 2}. Before examining what will happen if we use QFG, we rewrite Eq.~(\ref{app-an03}) as
\begin{eqnarray}
 &&  \frac{-i}{×\sqrt{6}}\bigg [e^{-\frac{1}{2}iy} (e^{2iy}\lvert0_{1}1_{2}1_{3} \rangle + e^{2iy}\lvert1_{1}0_{2}1_{3} \rangle + e^{- iy}\lvert0_{1}1_{2}0_{3} \rangle + e^{- iy}\lvert1_{1}0_{2}0_{3} \rangle ) \otimes \frac{e^{-iy}\lvert0_{\text{a}}\rangle +\lvert1_{\text{a}}\rangle  }{\sqrt{2}} \nonumber\\
 &&+ e^{-\frac{3}{2}iy}( \lvert0_{1}0_{2}0_{3} \rangle + e^{3iy} \lvert1_{1}1_{2}1_{3} \rangle)\otimes \frac{e^{-iy}\lvert0_{\text{a}}\rangle + \lvert1_{\text{a}}\rangle  }{\sqrt{2}}\bigg].\label{app-an03-v1}
\end{eqnarray}
Using the the 4-qubit QFG on the three data atoms and the ancilla, one can show that Eq.~(\ref{app-an03-v1}) becomes
\begin{eqnarray}
&& \frac{-ie^{-\frac{5}{2}iy+4iy_0}}{×\sqrt{6}}\bigg [e^{2iy} (\lvert0_{1}1_{2}1_{3} \rangle + \lvert1_{1}0_{2}1_{3} \rangle + e^{- 2iy}\lvert0_{1}1_{2}0_{3} \rangle + e^{-2 iy}\lvert1_{1}0_{2}0_{3} \rangle ) \otimes \frac{\lvert0_{\text{a}}\rangle +\lvert1_{\text{a}}\rangle  }{\sqrt{2}}-\sqrt{2}\lvert \text{GHZ}^{(\text{3})}_-\rangle\nonumber\\&&\otimes \frac{\lvert0_{\text{a}}\rangle -\lvert1_{\text{a}}\rangle  }{\sqrt{2}}\bigg].
\label{app-an05}
\end{eqnarray}
\newline
{\bf Circuit 2-pulse 3}. In Eq.~(\ref{an05}), $ie^{-\frac{5}{2}iy+4iy_0}$ is from an overall phase $-\pi + \frac{5}{2}y+4y_0$ which is trivial, and can be ignored. Then, with the laser field as used in Circuit 1-pulse 3 which results in the state transform of Eqs.~(\ref{c1p3laser1}) and~(\ref{c1p3laser2}), Eq.~(\ref{app-an05}) becomes
\begin{eqnarray}
|\text{M}\rangle_{2}&=&  \frac{\sqrt{2}}{×\sqrt{3}}  \lvert \text{O}'\rangle \otimes \lvert1_{\text{a}}\rangle -\frac{1}{×\sqrt{3}} \lvert \text{GHZ}^{(\text{3})}_-\rangle\otimes \lvert0_{\text{a}}\rangle .\label{app-an05-2}
\end{eqnarray}
where
\begin{eqnarray}
\lvert \text{O}'\rangle &=& \frac{e^{2iy}}{2}(\lvert0_{1}1_{2}1_{3} \rangle + \lvert1_{1}0_{2}1_{3} \rangle + e^{- 2iy}\lvert0_{1}1_{2}0_{3} \rangle + e^{- 2iy}\lvert1_{1}0_{2}0_{3} \rangle ).
\end{eqnarray}
\newline
{\bf Circuit 2-measurement}. When we measure the state of the ancillary qubit and the result is $\lvert0\rangle$, the state in Eq.~(\ref{app-an05-2}) can collapse to Eq.~(\ref{app-s06-an02}) with a conditional probability $1/3$, and if the measurement result is $\lvert1\rangle$, the state in Eq.~(\ref{app-an05-2}) collapses to
\begin{eqnarray}
\lvert \text{O}'\rangle \otimes \lvert1_{\text{a}}\rangle\label{app-an06}
\end{eqnarray}
with a conditional probability $2/3$.
\newline
{\bf Circuit 3-pulse 1}. Circuit 3 is shown in Fig.~\ref{figure-product-GHZ}(c). If the measurement result is $\lvert1\rangle$ in the last step, we shall do as in Circuit 2-pulse 1, use a $\pi$ pulse to one data atom and a $\pi/2$ pulse to the ancilla, with the change here that the second data atom should be addressed in this pulse. Different from Circuit 2-pulse 1, here the Rabi frequency $e^{\frac{1}{2}iy}\Omega_{\text{c}}$ is used for the data atom 2,
\begin{eqnarray}
&&  \lvert0_{\text{2}}\rangle \xrightarrow{e^{\frac{1}{2}iy}\Omega_{\text{c}},~ \pi \text{ pulse}} -ie^{\frac{1}{2}iy}\lvert1_{\text{2}}\rangle,\nonumber\\
&&   \lvert1_{\text{2}}\rangle \xrightarrow{e^{\frac{1}{2}iy}\Omega_{\text{c}},~ \pi \text{ pulse}} -ie^{-\frac{1}{2}iy}\lvert0_{\text{2}}\rangle.\label{app-rabi-pi-C3p1}
\end{eqnarray}
Then, one can find that Eq.~(\ref{app-an06}) becomes,
\begin{eqnarray}
 &&    \frac{-i e^{-\frac{5}{2}iy}}{2}( e^{2iy}\lvert0_{1}0_{2}1_{3} \rangle + e^{3iy}\lvert1_{1}1_{2}1_{3} \rangle + \lvert0_{1}0_{2}0_{3} \rangle + e^{iy}\lvert1_{1}1_{2}0_{3} \rangle ) \otimes \frac{\lvert0_{\text{a}}\rangle +e^{iy}\lvert1_{\text{a}}\rangle  }{\sqrt{2}} .\label{app-an07}
\end{eqnarray} \newline
{\bf Circuit 3-pulse 2}. The 4-qubit QFG, as in Circuit 1-pulse 2, will induce state-dependent phase $x_1-x_5$ as in Eq.~(\ref{QFG01}), and one can show that Eq.~(\ref{app-an07}) becomes
\begin{eqnarray}
&& \frac{-ie^{-\frac{5}{2}iy+4iy_0
}}{2}\bigg [ (e^{iy} \lvert0_{1}0_{2}1_{3} \rangle +   e^{-iy}\lvert1_{1}1_{2}0_{3} \rangle )  \otimes   \frac{\lvert0_{\text{a}}\rangle +\lvert1_{\text{a}}\rangle  }{\sqrt{2}} - ( \lvert0_{1}0_{2}0_{3} \rangle -  \lvert1_{1}1_{2}1_{3} \rangle)\otimes \frac{\lvert0_{\text{a}}\rangle -\lvert1_{\text{a}}\rangle  }{\sqrt{2}}\bigg].\label{app-an08}
\end{eqnarray}
with analyses as used in Eqs.~(\ref{s04-vary1}) and~(\ref{s05}).
\newline
{\bf Circuit 3-pulse 3}. As in Circuit 1-pulse 3, we excite the transition $\lvert 0\rangle\leftrightarrow\lvert 1\rangle$ of the ancillary atom with a Rabi frequency $i\Omega_{\text{c}}$ for a $\frac{\pi}{2 }$ pulse. With the understanding shown in  Eqs.~(\ref{c1p3laser1}) and~(\ref{c1p3laser2}), one can find that Eq.~(\ref{app-an08}) becomes
\begin{eqnarray}
&& \frac{1}{2}\bigg [ (e^{iy} \lvert0_{1}0_{2}1_{3} \rangle +   e^{-iy}\lvert1_{1}1_{2}0_{3} \rangle )  \otimes   \lvert1_{\text{a}}\rangle   - ( \lvert0_{1}0_{2}0_{3} \rangle -  \lvert1_{1}1_{2}1_{3} \rangle)\otimes \lvert0_{\text{a}}\rangle \bigg],\label{app-an08-2}
\end{eqnarray}
where we have dropped the trivial overall phase $4y_0+\frac{5}{2}y-\frac{\pi}{2}  $ of Eq.~(\ref{app-an08}). For clarity, it can be written as
\begin{eqnarray}
|\text{M}\rangle_{3}&=& \frac{1}{\sqrt{2}} \frac{e^{iy}\lvert0_{1}0_{2}1_{3} \rangle +  e^{-iy} \lvert1_{1}1_{2}0_{3} \rangle }{\sqrt{2}} \otimes   \lvert1_{\text{a}}\rangle   -\frac{1}{\sqrt{2}}\lvert \text{GHZ}^{(3)}_-\rangle \otimes \lvert0_{\text{a}}\rangle \nonumber\\
&=& \frac{i}{\sqrt{2}} \hat{Z}_{3}(2y)\hat{X}_{3}(\pi)\lvert \text{GHZ}^{(3)}_+\rangle   \otimes   \lvert1_{\text{a}}\rangle   -\frac{1}{\sqrt{2}}\lvert \text{GHZ}^{(3)}_-\rangle \otimes \lvert0_{\text{a}}\rangle.\label{app-an08-3}
\end{eqnarray}
where $\hat{X}_{j}(\pi)$ and $ \hat{Z}_{j}(2y)$ denote the single-qubit rotation around x and z axes with an angle $\pi$ and $2y$, respectively.
\newline
{\bf Circuit 3-measurement}. If the measurement result is $\lvert0\rangle$ for the ancilla, the state in Eq.~(\ref{app-an08-3}) collapses to exactly Eq.~(\ref{app-s06-an02}), i.e., we have got the 3-qubit GHZ state, with a conditional probability $1/2$ according to the coefficients in the state components in Eq.~(\ref{an08-3}). If the measurement result is $\lvert1\rangle$, the state in Eq.~(\ref{app-an08-3}) collapses to
\begin{eqnarray}
 &&i\hat{Z}_{3}(2y) \hat{X}_{3}(\pi)\lvert \text{GHZ}^{(3)}_+\rangle \otimes \lvert1_{\text{a}}\rangle\label{app-an09}
\end{eqnarray}
with a conditional probability $1/2$.
\newline
{\bf Circuit 3-pulse 4}. If the measurement result is $\lvert1\rangle$ at the last step, we can use the transition $\lvert 0\rangle\leftrightarrow\lvert 1\rangle$ in data atom 3 with a Rabi frequency $-ie^{iy}\Omega_{\text{c}}$ for a $\pi$ pulse,
\begin{eqnarray}
&&  \lvert0_{\text{3}}\rangle \xrightarrow{-ie^{iy}\Omega_{\text{c}},~ \pi \text{ pulse}} -e^{ iy}\lvert1_{\text{3}}\rangle,\nonumber\\
&&   \lvert1_{\text{3}}\rangle \xrightarrow{-ie^{iy}\Omega_{\text{c}},~ \pi \text{ pulse}} e^{- iy}\lvert0_{\text{3}}\rangle.\label{app-rabi-pi-C3p4}
\end{eqnarray}
so that Eq.~(\ref{app-an09}) becomes $\lvert \text{GHZ}^{(3)}_-\rangle\otimes \lvert1_{\text{a}}\rangle$ with a probability $1/2$. So if Circuit 3-pulse 4 is required, the final state is pure in the ancilla.

\section*{DATA AVAILABILITY}
The data leading to the results shown in the manuscript can be requested by sending emails to the authors.

\section*{AUTHOR CONTRIBUTIONS}
Both authors contributed to the manuscript.
\newline
\newline
{\bf COMPETING INTERESTS}.--The authors declare no competing financial or non-financial interests.
\newline
\newline
%


\begin{thebibliography}{54}%
\makeatletter
\providecommand \@ifxundefined [1]{%
 \@ifx{#1\undefined}
}%
\providecommand \@ifnum [1]{%
 \ifnum #1\expandafter \@firstoftwo
 \else \expandafter \@secondoftwo
 \fi
}%
\providecommand \@ifx [1]{%
 \ifx #1\expandafter \@firstoftwo
 \else \expandafter \@secondoftwo
 \fi
}%
\providecommand \natexlab [1]{#1}%
\providecommand \enquote  [1]{``#1''}%
\providecommand \bibnamefont  [1]{#1}%
\providecommand \bibfnamefont [1]{#1}%
\providecommand \citenamefont [1]{#1}%
\providecommand \href@noop [0]{\@secondoftwo}%
\providecommand \href [0]{\begingroup \@sanitize@url \@href}%
\providecommand \@href[1]{\@@startlink{#1}\@@href}%
\providecommand \@@href[1]{\endgroup#1\@@endlink}%
\providecommand \@sanitize@url [0]{\catcode `\\12\catcode `\$12\catcode
  `\&12\catcode `\#12\catcode `\^12\catcode `\_12\catcode `\%12\relax}%
\providecommand \@@startlink[1]{}%
\providecommand \@@endlink[0]{}%
\providecommand \url  [0]{\begingroup\@sanitize@url \@url }%
\providecommand \@url [1]{\endgroup\@href {#1}{\urlprefix }}%
\providecommand \urlprefix  [0]{URL }%
\providecommand \Eprint [0]{\href }%
\providecommand \doibase [0]{http://dx.doi.org/}%
\providecommand \selectlanguage [0]{\@gobble}%
\providecommand \bibinfo  [0]{\@secondoftwo}%
\providecommand \bibfield  [0]{\@secondoftwo}%
\providecommand \translation [1]{[#1]}%
\providecommand \BibitemOpen [0]{}%
\providecommand \bibitemStop [0]{}%
\providecommand \bibitemNoStop [0]{.\EOS\space}%
\providecommand \EOS [0]{\spacefactor3000\relax}%
\providecommand \BibitemShut  [1]{\csname bibitem#1\endcsname}%
\let\auto@bib@innerbib\@empty
\bibitem [{\citenamefont {Greenberger}\ \emph {et~al.}(1989)\citenamefont
  {Greenberger}, \citenamefont {Horne},\ and\ \citenamefont
  {Zeilinger}}]{GHZ1989}%
  \BibitemOpen
  \bibfield  {author} {\bibinfo {author} {\bibfnamefont {D.~M.}\ \bibnamefont
  {Greenberger}}, \bibinfo {author} {\bibfnamefont {M.}~\bibnamefont {Horne}},
  \ and\ \bibinfo {author} {\bibfnamefont {A.}~\bibnamefont {Zeilinger}},\
  }\href@noop {} {\emph {\bibinfo {title} {{Bell’s Theorem, Quantum Theory,
  and Conceptions of the Universe}}}},\ edited by\ \bibinfo {editor}
  {\bibfnamefont {M.}~\bibnamefont {Kafatos}}\ (\bibinfo  {publisher}
  {Kluwer},\ \bibinfo {address} {Dordrecht},\ \bibinfo {year}
  {1989})\BibitemShut {NoStop}%
\bibitem [{\citenamefont {Hillery}\ \emph {et~al.}(1999)\citenamefont
  {Hillery}, \citenamefont {Bu\ifmmode~\check{z}\else \v{z}\fi{}ek},\ and\
  \citenamefont {Berthiaume}}]{Hillery1999}%
  \BibitemOpen
  \bibfield  {author} {\bibinfo {author} {\bibfnamefont {M.}~\bibnamefont
  {Hillery}}, \bibinfo {author} {\bibfnamefont {V.}~\bibnamefont
  {Bu\ifmmode~\check{z}\else \v{z}\fi{}ek}}, \ and\ \bibinfo {author}
  {\bibfnamefont {A.}~\bibnamefont {Berthiaume}},\ }\bibfield  {title}
  {\bibinfo {title} {Quantum secret sharing},\ }\href {\doibase
  10.1103/PhysRevA.59.1829} {\bibfield  {journal} {\bibinfo  {journal} {Phys.
  Rev. A}\ }\textbf {\bibinfo {volume} {59}},\ \bibinfo {pages} {1829}
  (\bibinfo {year} {1999})}\BibitemShut {NoStop}%
\bibitem [{\citenamefont {Proietti}\ \emph {et~al.}(2021)\citenamefont
  {Proietti}, \citenamefont {Ho}, \citenamefont {Grasselli}, \citenamefont
  {Barrow}, \citenamefont {Malik},\ and\ \citenamefont
  {Fedrizzi}}]{proietti2021}%
  \BibitemOpen
  \bibfield  {author} {\bibinfo {author} {\bibfnamefont {M.}~\bibnamefont
  {Proietti}}, \bibinfo {author} {\bibfnamefont {J.}~\bibnamefont {Ho}},
  \bibinfo {author} {\bibfnamefont {F.}~\bibnamefont {Grasselli}}, \bibinfo
  {author} {\bibfnamefont {P.}~\bibnamefont {Barrow}}, \bibinfo {author}
  {\bibfnamefont {M.}~\bibnamefont {Malik}}, \ and\ \bibinfo {author}
  {\bibfnamefont {A.}~\bibnamefont {Fedrizzi}},\ }\bibfield  {title} {\bibinfo
  {title} {Experimental quantum conference key agreement},\ }\href {\doibase
  10.1126/sciadv.abe0395} {\bibfield  {journal} {\bibinfo  {journal} {Science
  Advances}\ }\textbf {\bibinfo {volume} {7}},\ \bibinfo {pages} {eabe0395}
  (\bibinfo {year} {2021})}\BibitemShut {NoStop}%
\bibitem [{\citenamefont {Moses}\ \emph {et~al.}(2023)\citenamefont {Moses},
  \citenamefont {Baldwin}, \citenamefont {Allman}, \citenamefont {Ancona},
  \citenamefont {Ascarrunz}, \citenamefont {Barnes}, \citenamefont
  {Bartolotta}, \citenamefont {Bjork}, \citenamefont {Blanchard}, \citenamefont
  {Bohn}, \citenamefont {Bohnet}, \citenamefont {Brown}, \citenamefont
  {Burdick}, \citenamefont {Burton}, \citenamefont {Campbell}, \citenamefont
  {Campora}, \citenamefont {Carron}, \citenamefont {Chambers}, \citenamefont
  {Chan}, \citenamefont {Chen}, \citenamefont {Chernoguzov}, \citenamefont
  {Chertkov}, \citenamefont {Colina}, \citenamefont {Curtis}, \citenamefont
  {Daniel}, \citenamefont {DeCross}, \citenamefont {Deen}, \citenamefont
  {Delaney}, \citenamefont {Dreiling}, \citenamefont {Ertsgaard}, \citenamefont
  {Esposito}, \citenamefont {Estey}, \citenamefont {Fabrikant}, \citenamefont
  {Figgatt}, \citenamefont {Foltz}, \citenamefont {Foss-Feig}, \citenamefont
  {Francois}, \citenamefont {Gaebler}, \citenamefont {Gatterman}, \citenamefont
  {Gilbreth}, \citenamefont {Giles}, \citenamefont {Glynn}, \citenamefont
  {Hall}, \citenamefont {Hankin}, \citenamefont {Hansen}, \citenamefont
  {Hayes}, \citenamefont {Higashi}, \citenamefont {Hoffman}, \citenamefont
  {Horning}, \citenamefont {Hout}, \citenamefont {Jacobs}, \citenamefont
  {Johansen}, \citenamefont {Jones}, \citenamefont {Karcz}, \citenamefont
  {Klein}, \citenamefont {Lauria}, \citenamefont {Lee}, \citenamefont {Liefer},
  \citenamefont {Lu}, \citenamefont {Lucchetti}, \citenamefont {Lytle},
  \citenamefont {Malm}, \citenamefont {Matheny}, \citenamefont {Mathewson},
  \citenamefont {Mayer}, \citenamefont {Miller}, \citenamefont {Mills},
  \citenamefont {Neyenhuis}, \citenamefont {Nugent}, \citenamefont {Olson},
  \citenamefont {Parks}, \citenamefont {Price}, \citenamefont {Price},
  \citenamefont {Pugh}, \citenamefont {Ransford}, \citenamefont {Reed},
  \citenamefont {Roman}, \citenamefont {Rowe}, \citenamefont {Ryan-Anderson},
  \citenamefont {Sanders}, \citenamefont {Sedlacek}, \citenamefont {Shevchuk},
  \citenamefont {Siegfried}, \citenamefont {Skripka}, \citenamefont {Spaun},
  \citenamefont {Sprenkle}, \citenamefont {Stutz}, \citenamefont {Swallows},
  \citenamefont {Tobey}, \citenamefont {Tran}, \citenamefont {Tran},
  \citenamefont {Vogt}, \citenamefont {Volin}, \citenamefont {Walker},
  \citenamefont {Zolot},\ and\ \citenamefont {Pino}}]{Moses2023}%
  \BibitemOpen
  \bibfield  {author} {\bibinfo {author} {\bibfnamefont {S.}~\bibnamefont
  {Moses}}, \bibinfo {author} {\bibfnamefont {C.}~\bibnamefont {Baldwin}},
  \bibinfo {author} {\bibfnamefont {M.}~\bibnamefont {Allman}}, \bibinfo
  {author} {\bibfnamefont {R.}~\bibnamefont {Ancona}}, \bibinfo {author}
  {\bibfnamefont {L.}~\bibnamefont {Ascarrunz}}, \bibinfo {author}
  {\bibfnamefont {C.}~\bibnamefont {Barnes}}, \bibinfo {author} {\bibfnamefont
  {J.}~\bibnamefont {Bartolotta}}, \bibinfo {author} {\bibfnamefont
  {B.}~\bibnamefont {Bjork}}, \bibinfo {author} {\bibfnamefont
  {P.}~\bibnamefont {Blanchard}}, \bibinfo {author} {\bibfnamefont
  {M.}~\bibnamefont {Bohn}}, \bibinfo {author} {\bibfnamefont {J.}~\bibnamefont
  {Bohnet}}, \bibinfo {author} {\bibfnamefont {N.}~\bibnamefont {Brown}},
  \bibinfo {author} {\bibfnamefont {N.}~\bibnamefont {Burdick}}, \bibinfo
  {author} {\bibfnamefont {W.}~\bibnamefont {Burton}}, \bibinfo {author}
  {\bibfnamefont {S.}~\bibnamefont {Campbell}}, \bibinfo {author}
  {\bibfnamefont {J.}~\bibnamefont {Campora}}, \bibinfo {author} {\bibfnamefont
  {C.}~\bibnamefont {Carron}}, \bibinfo {author} {\bibfnamefont
  {J.}~\bibnamefont {Chambers}}, \bibinfo {author} {\bibfnamefont
  {J.}~\bibnamefont {Chan}}, \bibinfo {author} {\bibfnamefont {Y.}~\bibnamefont
  {Chen}}, \bibinfo {author} {\bibfnamefont {A.}~\bibnamefont {Chernoguzov}},
  \bibinfo {author} {\bibfnamefont {E.}~\bibnamefont {Chertkov}}, \bibinfo
  {author} {\bibfnamefont {J.}~\bibnamefont {Colina}}, \bibinfo {author}
  {\bibfnamefont {J.}~\bibnamefont {Curtis}}, \bibinfo {author} {\bibfnamefont
  {R.}~\bibnamefont {Daniel}}, \bibinfo {author} {\bibfnamefont
  {M.}~\bibnamefont {DeCross}}, \bibinfo {author} {\bibfnamefont
  {D.}~\bibnamefont {Deen}}, \bibinfo {author} {\bibfnamefont {C.}~\bibnamefont
  {Delaney}}, \bibinfo {author} {\bibfnamefont {J.}~\bibnamefont {Dreiling}},
  \bibinfo {author} {\bibfnamefont {C.}~\bibnamefont {Ertsgaard}}, \bibinfo
  {author} {\bibfnamefont {J.}~\bibnamefont {Esposito}}, \bibinfo {author}
  {\bibfnamefont {B.}~\bibnamefont {Estey}}, \bibinfo {author} {\bibfnamefont
  {M.}~\bibnamefont {Fabrikant}}, \bibinfo {author} {\bibfnamefont
  {C.}~\bibnamefont {Figgatt}}, \bibinfo {author} {\bibfnamefont
  {C.}~\bibnamefont {Foltz}}, \bibinfo {author} {\bibfnamefont
  {M.}~\bibnamefont {Foss-Feig}}, \bibinfo {author} {\bibfnamefont
  {D.}~\bibnamefont {Francois}}, \bibinfo {author} {\bibfnamefont
  {J.}~\bibnamefont {Gaebler}}, \bibinfo {author} {\bibfnamefont
  {T.}~\bibnamefont {Gatterman}}, \bibinfo {author} {\bibfnamefont
  {C.}~\bibnamefont {Gilbreth}}, \bibinfo {author} {\bibfnamefont
  {J.}~\bibnamefont {Giles}}, \bibinfo {author} {\bibfnamefont
  {E.}~\bibnamefont {Glynn}}, \bibinfo {author} {\bibfnamefont
  {A.}~\bibnamefont {Hall}}, \bibinfo {author} {\bibfnamefont {A.}~\bibnamefont
  {Hankin}}, \bibinfo {author} {\bibfnamefont {A.}~\bibnamefont {Hansen}},
  \bibinfo {author} {\bibfnamefont {D.}~\bibnamefont {Hayes}}, \bibinfo
  {author} {\bibfnamefont {B.}~\bibnamefont {Higashi}}, \bibinfo {author}
  {\bibfnamefont {I.}~\bibnamefont {Hoffman}}, \bibinfo {author} {\bibfnamefont
  {B.}~\bibnamefont {Horning}}, \bibinfo {author} {\bibfnamefont
  {J.}~\bibnamefont {Hout}}, \bibinfo {author} {\bibfnamefont {R.}~\bibnamefont
  {Jacobs}}, \bibinfo {author} {\bibfnamefont {J.}~\bibnamefont {Johansen}},
  \bibinfo {author} {\bibfnamefont {L.}~\bibnamefont {Jones}}, \bibinfo
  {author} {\bibfnamefont {J.}~\bibnamefont {Karcz}}, \bibinfo {author}
  {\bibfnamefont {T.}~\bibnamefont {Klein}}, \bibinfo {author} {\bibfnamefont
  {P.}~\bibnamefont {Lauria}}, \bibinfo {author} {\bibfnamefont
  {P.}~\bibnamefont {Lee}}, \bibinfo {author} {\bibfnamefont {D.}~\bibnamefont
  {Liefer}}, \bibinfo {author} {\bibfnamefont {S.}~\bibnamefont {Lu}}, \bibinfo
  {author} {\bibfnamefont {D.}~\bibnamefont {Lucchetti}}, \bibinfo {author}
  {\bibfnamefont {C.}~\bibnamefont {Lytle}}, \bibinfo {author} {\bibfnamefont
  {A.}~\bibnamefont {Malm}}, \bibinfo {author} {\bibfnamefont {M.}~\bibnamefont
  {Matheny}}, \bibinfo {author} {\bibfnamefont {B.}~\bibnamefont {Mathewson}},
  \bibinfo {author} {\bibfnamefont {K.}~\bibnamefont {Mayer}}, \bibinfo
  {author} {\bibfnamefont {D.}~\bibnamefont {Miller}}, \bibinfo {author}
  {\bibfnamefont {M.}~\bibnamefont {Mills}}, \bibinfo {author} {\bibfnamefont
  {B.}~\bibnamefont {Neyenhuis}}, \bibinfo {author} {\bibfnamefont
  {L.}~\bibnamefont {Nugent}}, \bibinfo {author} {\bibfnamefont
  {S.}~\bibnamefont {Olson}}, \bibinfo {author} {\bibfnamefont
  {J.}~\bibnamefont {Parks}}, \bibinfo {author} {\bibfnamefont
  {G.}~\bibnamefont {Price}}, \bibinfo {author} {\bibfnamefont
  {Z.}~\bibnamefont {Price}}, \bibinfo {author} {\bibfnamefont
  {M.}~\bibnamefont {Pugh}}, \bibinfo {author} {\bibfnamefont {A.}~\bibnamefont
  {Ransford}}, \bibinfo {author} {\bibfnamefont {A.}~\bibnamefont {Reed}},
  \bibinfo {author} {\bibfnamefont {C.}~\bibnamefont {Roman}}, \bibinfo
  {author} {\bibfnamefont {M.}~\bibnamefont {Rowe}}, \bibinfo {author}
  {\bibfnamefont {C.}~\bibnamefont {Ryan-Anderson}}, \bibinfo {author}
  {\bibfnamefont {S.}~\bibnamefont {Sanders}}, \bibinfo {author} {\bibfnamefont
  {J.}~\bibnamefont {Sedlacek}}, \bibinfo {author} {\bibfnamefont
  {P.}~\bibnamefont {Shevchuk}}, \bibinfo {author} {\bibfnamefont
  {P.}~\bibnamefont {Siegfried}}, \bibinfo {author} {\bibfnamefont
  {T.}~\bibnamefont {Skripka}}, \bibinfo {author} {\bibfnamefont
  {B.}~\bibnamefont {Spaun}}, \bibinfo {author} {\bibfnamefont
  {R.}~\bibnamefont {Sprenkle}}, \bibinfo {author} {\bibfnamefont
  {R.}~\bibnamefont {Stutz}}, \bibinfo {author} {\bibfnamefont
  {M.}~\bibnamefont {Swallows}}, \bibinfo {author} {\bibfnamefont
  {R.}~\bibnamefont {Tobey}}, \bibinfo {author} {\bibfnamefont
  {A.}~\bibnamefont {Tran}}, \bibinfo {author} {\bibfnamefont {T.}~\bibnamefont
  {Tran}}, \bibinfo {author} {\bibfnamefont {E.}~\bibnamefont {Vogt}}, \bibinfo
  {author} {\bibfnamefont {C.}~\bibnamefont {Volin}}, \bibinfo {author}
  {\bibfnamefont {J.}~\bibnamefont {Walker}}, \bibinfo {author} {\bibfnamefont
  {A.}~\bibnamefont {Zolot}}, \ and\ \bibinfo {author} {\bibfnamefont
  {J.}~\bibnamefont {Pino}},\ }\bibfield  {title} {\bibinfo {title} {A
  race-track trapped-ion quantum processor},\ }\href {\doibase
  10.1103/physrevx.13.041052} {\bibfield  {journal} {\bibinfo  {journal} {Phys.
  Rev. X}\ }\textbf {\bibinfo {volume} {13}},\ \bibinfo {pages} {041052}
  (\bibinfo {year} {2023})}\BibitemShut {NoStop}%
\bibitem [{\citenamefont {Bao}\ \emph {et~al.}(2024)\citenamefont {Bao},
  \citenamefont {Xu}, \citenamefont {Song}, \citenamefont {Wang}, \citenamefont
  {Xiang}, \citenamefont {Zhu}, \citenamefont {Chen}, \citenamefont {Jin},
  \citenamefont {Zhu}, \citenamefont {Gao}, \citenamefont {Wu}, \citenamefont
  {Zhang}, \citenamefont {Wang}, \citenamefont {Zou}, \citenamefont {Tan},
  \citenamefont {Zhang}, \citenamefont {Cui}, \citenamefont {Shen},
  \citenamefont {Zhong}, \citenamefont {Li}, \citenamefont {Deng},
  \citenamefont {Zhang}, \citenamefont {Dong}, \citenamefont {Zhang},
  \citenamefont {Liu}, \citenamefont {Zhao}, \citenamefont {Hao}, \citenamefont
  {Li}, \citenamefont {Wang}, \citenamefont {Song}, \citenamefont {Guo},
  \citenamefont {Huang},\ and\ \citenamefont {Wang}}]{Bao2024}%
  \BibitemOpen
  \bibfield  {author} {\bibinfo {author} {\bibfnamefont {Z.}~\bibnamefont
  {Bao}}, \bibinfo {author} {\bibfnamefont {S.}~\bibnamefont {Xu}}, \bibinfo
  {author} {\bibfnamefont {Z.}~\bibnamefont {Song}}, \bibinfo {author}
  {\bibfnamefont {K.}~\bibnamefont {Wang}}, \bibinfo {author} {\bibfnamefont
  {L.}~\bibnamefont {Xiang}}, \bibinfo {author} {\bibfnamefont
  {Z.}~\bibnamefont {Zhu}}, \bibinfo {author} {\bibfnamefont {J.}~\bibnamefont
  {Chen}}, \bibinfo {author} {\bibfnamefont {F.}~\bibnamefont {Jin}}, \bibinfo
  {author} {\bibfnamefont {X.}~\bibnamefont {Zhu}}, \bibinfo {author}
  {\bibfnamefont {Y.}~\bibnamefont {Gao}}, \bibinfo {author} {\bibfnamefont
  {Y.}~\bibnamefont {Wu}}, \bibinfo {author} {\bibfnamefont {C.}~\bibnamefont
  {Zhang}}, \bibinfo {author} {\bibfnamefont {N.}~\bibnamefont {Wang}},
  \bibinfo {author} {\bibfnamefont {Y.}~\bibnamefont {Zou}}, \bibinfo {author}
  {\bibfnamefont {Z.}~\bibnamefont {Tan}}, \bibinfo {author} {\bibfnamefont
  {A.}~\bibnamefont {Zhang}}, \bibinfo {author} {\bibfnamefont
  {Z.}~\bibnamefont {Cui}}, \bibinfo {author} {\bibfnamefont {F.}~\bibnamefont
  {Shen}}, \bibinfo {author} {\bibfnamefont {J.}~\bibnamefont {Zhong}},
  \bibinfo {author} {\bibfnamefont {T.}~\bibnamefont {Li}}, \bibinfo {author}
  {\bibfnamefont {J.}~\bibnamefont {Deng}}, \bibinfo {author} {\bibfnamefont
  {X.}~\bibnamefont {Zhang}}, \bibinfo {author} {\bibfnamefont
  {H.}~\bibnamefont {Dong}}, \bibinfo {author} {\bibfnamefont {P.}~\bibnamefont
  {Zhang}}, \bibinfo {author} {\bibfnamefont {Y.-R.}\ \bibnamefont {Liu}},
  \bibinfo {author} {\bibfnamefont {L.}~\bibnamefont {Zhao}}, \bibinfo {author}
  {\bibfnamefont {J.}~\bibnamefont {Hao}}, \bibinfo {author} {\bibfnamefont
  {H.}~\bibnamefont {Li}}, \bibinfo {author} {\bibfnamefont {Z.}~\bibnamefont
  {Wang}}, \bibinfo {author} {\bibfnamefont {C.}~\bibnamefont {Song}}, \bibinfo
  {author} {\bibfnamefont {Q.}~\bibnamefont {Guo}}, \bibinfo {author}
  {\bibfnamefont {B.}~\bibnamefont {Huang}}, \ and\ \bibinfo {author}
  {\bibfnamefont {H.}~\bibnamefont {Wang}},\ }\bibfield  {title} {\bibinfo
  {title} {Creating and controlling global greenberger-horne-zeilinger
  entanglement on quantum processors},\ }\href {\doibase
  10.1038/s41467-024-53140-5} {\bibfield  {journal} {\bibinfo  {journal} {Nat.
  Commun.}\ }\textbf {\bibinfo {volume} {15}},\ \bibinfo {pages} {8823}
  (\bibinfo {year} {2024})}\BibitemShut {NoStop}%
\bibitem [{\citenamefont {Zhao}\ \emph {et~al.}(2021)\citenamefont {Zhao},
  \citenamefont {Zhang}, \citenamefont {Chen}, \citenamefont {Wang},\ and\
  \citenamefont {Hu}}]{zhao2021}%
  \BibitemOpen
  \bibfield  {author} {\bibinfo {author} {\bibfnamefont {Y.}~\bibnamefont
  {Zhao}}, \bibinfo {author} {\bibfnamefont {R.}~\bibnamefont {Zhang}},
  \bibinfo {author} {\bibfnamefont {W.}~\bibnamefont {Chen}}, \bibinfo {author}
  {\bibfnamefont {X.-B.}\ \bibnamefont {Wang}}, \ and\ \bibinfo {author}
  {\bibfnamefont {J.}~\bibnamefont {Hu}},\ }\bibfield  {title} {\bibinfo
  {title} {Creation of greenberger-horne-zeilinger states with thousands of
  atoms by entanglement amplification},\ }\href {\doibase
  10.1038/s41534-021-00364-8} {\bibfield  {journal} {\bibinfo  {journal} {npj
  Quantum Information}\ }\textbf {\bibinfo {volume} {7}},\ \bibinfo {pages}
  {24} (\bibinfo {year} {2021})}\BibitemShut {NoStop}%
\bibitem [{\citenamefont {Gallagher}(2005)}]{Gallagh2005}%
  \BibitemOpen
  \bibfield  {author} {\bibinfo {author} {\bibfnamefont {T.~F.}\ \bibnamefont
  {Gallagher}},\ }\href@noop {} {\emph {\bibinfo {title} {{Rydberg Atoms}}}}\
  (\bibinfo  {publisher} {Cambridge Univ. Press},\ \bibinfo {year}
  {2005})\BibitemShut {NoStop}%
\bibitem [{\citenamefont {Omran}\ \emph {et~al.}(2019)\citenamefont {Omran},
  \citenamefont {Levine}, \citenamefont {Keesling}, \citenamefont {Semeghini},
  \citenamefont {Wang}, \citenamefont {Ebadi}, \citenamefont {Bernien},
  \citenamefont {Zibrov}, \citenamefont {Pichler}, \citenamefont {Choi},
  \citenamefont {Cui}, \citenamefont {Rossignolo}, \citenamefont {Rembold},
  \citenamefont {Montangero}, \citenamefont {Calarco}, \citenamefont {Endres},
  \citenamefont {Greiner}, \citenamefont {Vuletić},\ and\ \citenamefont
  {Lukin}}]{Omran2019}%
  \BibitemOpen
  \bibfield  {author} {\bibinfo {author} {\bibfnamefont {A.}~\bibnamefont
  {Omran}}, \bibinfo {author} {\bibfnamefont {H.}~\bibnamefont {Levine}},
  \bibinfo {author} {\bibfnamefont {A.}~\bibnamefont {Keesling}}, \bibinfo
  {author} {\bibfnamefont {G.}~\bibnamefont {Semeghini}}, \bibinfo {author}
  {\bibfnamefont {T.~T.}\ \bibnamefont {Wang}}, \bibinfo {author}
  {\bibfnamefont {S.}~\bibnamefont {Ebadi}}, \bibinfo {author} {\bibfnamefont
  {H.}~\bibnamefont {Bernien}}, \bibinfo {author} {\bibfnamefont {A.~S.}\
  \bibnamefont {Zibrov}}, \bibinfo {author} {\bibfnamefont {H.}~\bibnamefont
  {Pichler}}, \bibinfo {author} {\bibfnamefont {S.}~\bibnamefont {Choi}},
  \bibinfo {author} {\bibfnamefont {J.}~\bibnamefont {Cui}}, \bibinfo {author}
  {\bibfnamefont {M.}~\bibnamefont {Rossignolo}}, \bibinfo {author}
  {\bibfnamefont {P.}~\bibnamefont {Rembold}}, \bibinfo {author} {\bibfnamefont
  {S.}~\bibnamefont {Montangero}}, \bibinfo {author} {\bibfnamefont
  {T.}~\bibnamefont {Calarco}}, \bibinfo {author} {\bibfnamefont
  {M.}~\bibnamefont {Endres}}, \bibinfo {author} {\bibfnamefont
  {M.}~\bibnamefont {Greiner}}, \bibinfo {author} {\bibfnamefont
  {V.}~\bibnamefont {Vuletić}}, \ and\ \bibinfo {author} {\bibfnamefont
  {M.~D.}\ \bibnamefont {Lukin}},\ }\bibfield  {title} {\bibinfo {title}
  {{Generation and manipulation of Schrödinger cat states in Rydberg atom
  arrays}},\ }\href {\doibase 10.1126/science.aax9743} {\bibfield  {journal}
  {\bibinfo  {journal} {Science}\ }\textbf {\bibinfo {volume} {365}},\ \bibinfo
  {pages} {570–574} (\bibinfo {year} {2019})}\BibitemShut {NoStop}%
\bibitem [{\citenamefont {Senoo}\ \emph {et~al.}(2025)\citenamefont {Senoo},
  \citenamefont {Baumgärtner}, \citenamefont {Lis}, \citenamefont {Vaidya},
  \citenamefont {Zeng}, \citenamefont {Giudici}, \citenamefont {Pichler},\ and\
  \citenamefont {Kaufman}}]{senoo2025}%
  \BibitemOpen
  \bibfield  {author} {\bibinfo {author} {\bibfnamefont {A.}~\bibnamefont
  {Senoo}}, \bibinfo {author} {\bibfnamefont {A.}~\bibnamefont {Baumgärtner}},
  \bibinfo {author} {\bibfnamefont {J.~W.}\ \bibnamefont {Lis}}, \bibinfo
  {author} {\bibfnamefont {G.~M.}\ \bibnamefont {Vaidya}}, \bibinfo {author}
  {\bibfnamefont {Z.}~\bibnamefont {Zeng}}, \bibinfo {author} {\bibfnamefont
  {G.}~\bibnamefont {Giudici}}, \bibinfo {author} {\bibfnamefont
  {H.}~\bibnamefont {Pichler}}, \ and\ \bibinfo {author} {\bibfnamefont
  {A.~M.}\ \bibnamefont {Kaufman}},\ }\href {\doibase
  10.48550/arXiv.2506.13632} {\bibinfo {title} {High-fidelity entanglement and
  coherent multi-qubit mapping in an atom array},\ } (\bibinfo {year} {2025}),\
  \bibinfo {note} {arXiv:2506.13632 [quant-ph]}\BibitemShut {NoStop}%
\bibitem [{\citenamefont {Saffman}\ \emph {et~al.}(2010)\citenamefont
  {Saffman}, \citenamefont {Walker},\ and\ \citenamefont
  {{Mølmer}}}]{Saffman2010}%
  \BibitemOpen
  \bibfield  {author} {\bibinfo {author} {\bibfnamefont {M.}~\bibnamefont
  {Saffman}}, \bibinfo {author} {\bibfnamefont {T.~G.}\ \bibnamefont {Walker}},
  \ and\ \bibinfo {author} {\bibfnamefont {K.}~\bibnamefont {{Mølmer}}},\
  }\bibfield  {title} {\bibinfo {title} {{Quantum information with Rydberg
  atoms}},\ }\href@noop {} {\bibfield  {journal} {\bibinfo  {journal} {Rev.
  Mod. Phys.}\ }\textbf {\bibinfo {volume} {82}},\ \bibinfo {pages} {2313}
  (\bibinfo {year} {2010})}\BibitemShut {NoStop}%
\bibitem [{\citenamefont {Ma}\ \emph {et~al.}(2023)\citenamefont {Ma},
  \citenamefont {Liu}, \citenamefont {Peng}, \citenamefont {Zhang},
  \citenamefont {Jandura}, \citenamefont {Burgers}, \citenamefont {Pupillo},
  \citenamefont {Puri},\ and\ \citenamefont {Thompson}}]{Ma2023}%
  \BibitemOpen
  \bibfield  {author} {\bibinfo {author} {\bibfnamefont {S.}~\bibnamefont
  {Ma}}, \bibinfo {author} {\bibfnamefont {G.}~\bibnamefont {Liu}}, \bibinfo
  {author} {\bibfnamefont {P.}~\bibnamefont {Peng}}, \bibinfo {author}
  {\bibfnamefont {B.}~\bibnamefont {Zhang}}, \bibinfo {author} {\bibfnamefont
  {S.}~\bibnamefont {Jandura}}, \bibinfo {author} {\bibfnamefont {A.~P.}\
  \bibnamefont {Burgers}}, \bibinfo {author} {\bibfnamefont {G.}~\bibnamefont
  {Pupillo}}, \bibinfo {author} {\bibfnamefont {S.}~\bibnamefont {Puri}}, \
  and\ \bibinfo {author} {\bibfnamefont {J.~D.}\ \bibnamefont {Thompson}},\
  }\bibfield  {title} {\bibinfo {title} {{High-fidelity gates with mid-circuit
  erasure conversion in a metastable neutral atom qubit}},\ }\href
  {https://www.nature.com/articles/s41586-023-06438-1} {\bibfield  {journal}
  {\bibinfo  {journal} {Nature}\ }\textbf {\bibinfo {volume} {622}},\ \bibinfo
  {pages} {279} (\bibinfo {year} {2023})}\BibitemShut {NoStop}%
\bibitem [{\citenamefont {Manetsch}\ \emph {et~al.}(2025)\citenamefont
  {Manetsch}, \citenamefont {Nomura}, \citenamefont {Bataille}, \citenamefont
  {Lv}, \citenamefont {Leung},\ and\ \citenamefont {Endres}}]{6100atoms}%
  \BibitemOpen
  \bibfield  {author} {\bibinfo {author} {\bibfnamefont {H.~J.}\ \bibnamefont
  {Manetsch}}, \bibinfo {author} {\bibfnamefont {G.}~\bibnamefont {Nomura}},
  \bibinfo {author} {\bibfnamefont {E.}~\bibnamefont {Bataille}}, \bibinfo
  {author} {\bibfnamefont {X.}~\bibnamefont {Lv}}, \bibinfo {author}
  {\bibfnamefont {K.~H.}\ \bibnamefont {Leung}}, \ and\ \bibinfo {author}
  {\bibfnamefont {M.}~\bibnamefont {Endres}},\ }\bibfield  {title} {\bibinfo
  {title} {A tweezer array with 6100 highly coherent atomic qubits},\ }\href
  {\doibase 10.1038/s41586-025-09641-4} {\bibfield  {journal} {\bibinfo
  {journal} {Nature}\ }\textbf {\bibinfo {volume} {647}},\ \bibinfo {pages}
  {60–67} (\bibinfo {year} {2025})}\BibitemShut {NoStop}%
\bibitem [{\citenamefont {Chiu}\ \emph {et~al.}(2025)\citenamefont {Chiu},
  \citenamefont {Trapp}, \citenamefont {Guo}, \citenamefont {Abobeih},
  \citenamefont {Stewart}, \citenamefont {Hollerith}, \citenamefont
  {Stroganov}, \citenamefont {Kalinowski}, \citenamefont {Geim}, \citenamefont
  {Evered}, \citenamefont {Li}, \citenamefont {Lyu}, \citenamefont {Peters},
  \citenamefont {Bluvstein}, \citenamefont {Wang}, \citenamefont {Greiner},
  \citenamefont {Vuletić},\ and\ \citenamefont {Lukin}}]{Chiu2025}%
  \BibitemOpen
  \bibfield  {author} {\bibinfo {author} {\bibfnamefont {N.-C.}\ \bibnamefont
  {Chiu}}, \bibinfo {author} {\bibfnamefont {E.~C.}\ \bibnamefont {Trapp}},
  \bibinfo {author} {\bibfnamefont {J.}~\bibnamefont {Guo}}, \bibinfo {author}
  {\bibfnamefont {M.~H.}\ \bibnamefont {Abobeih}}, \bibinfo {author}
  {\bibfnamefont {L.~M.}\ \bibnamefont {Stewart}}, \bibinfo {author}
  {\bibfnamefont {S.}~\bibnamefont {Hollerith}}, \bibinfo {author}
  {\bibfnamefont {P.~L.}\ \bibnamefont {Stroganov}}, \bibinfo {author}
  {\bibfnamefont {M.}~\bibnamefont {Kalinowski}}, \bibinfo {author}
  {\bibfnamefont {A.~A.}\ \bibnamefont {Geim}}, \bibinfo {author}
  {\bibfnamefont {S.~J.}\ \bibnamefont {Evered}}, \bibinfo {author}
  {\bibfnamefont {S.~H.}\ \bibnamefont {Li}}, \bibinfo {author} {\bibfnamefont
  {X.}~\bibnamefont {Lyu}}, \bibinfo {author} {\bibfnamefont {L.~M.}\
  \bibnamefont {Peters}}, \bibinfo {author} {\bibfnamefont {D.}~\bibnamefont
  {Bluvstein}}, \bibinfo {author} {\bibfnamefont {T.~T.}\ \bibnamefont {Wang}},
  \bibinfo {author} {\bibfnamefont {M.}~\bibnamefont {Greiner}}, \bibinfo
  {author} {\bibfnamefont {V.}~\bibnamefont {Vuletić}}, \ and\ \bibinfo
  {author} {\bibfnamefont {M.~D.}\ \bibnamefont {Lukin}},\ }\bibfield  {title}
  {\bibinfo {title} {Continuous operation of a coherent 3,000-qubit system},\
  }\href {\doibase 10.1038/s41586-025-09596-6} {\bibfield  {journal} {\bibinfo
  {journal} {Nature}\ }\textbf {\bibinfo {volume} {646}},\ \bibinfo {pages}
  {1075} (\bibinfo {year} {2025})}\BibitemShut {NoStop}%
\bibitem [{\citenamefont {Shi}\ and\ \citenamefont {Lu}(2024)}]{ShiLu2024}%
  \BibitemOpen
  \bibfield  {author} {\bibinfo {author} {\bibfnamefont {X.-F.}\ \bibnamefont
  {Shi}}\ and\ \bibinfo {author} {\bibfnamefont {Y.}~\bibnamefont {Lu}},\
  }\bibfield  {title} {\bibinfo {title} {Fast nuclear-spin entangling gates
  compatible with large-scale atomic arrays},\ }\href {\doibase
  10.1103/PhysRevA.110.012610} {\bibfield  {journal} {\bibinfo  {journal}
  {Phys. Rev. A}\ }\textbf {\bibinfo {volume} {110}},\ \bibinfo {pages}
  {012610} (\bibinfo {year} {2024})}\BibitemShut {NoStop}%
\bibitem [{\citenamefont {Shi}(2021{\natexlab{a}})}]{Shi2021}%
  \BibitemOpen
  \bibfield  {author} {\bibinfo {author} {\bibfnamefont {X.-F.}\ \bibnamefont
  {Shi}},\ }\bibfield  {title} {\bibinfo {title} {{Rydberg quantum computation
  with nuclear spins in two-electron neutral atoms}},\ }\href@noop {}
  {\bibfield  {journal} {\bibinfo  {journal} {Front. Phys.}\ }\textbf {\bibinfo
  {volume} {16}},\ \bibinfo {pages} {52501} (\bibinfo {year}
  {2021}{\natexlab{a}})}\BibitemShut {NoStop}%
\bibitem [{\citenamefont {Shi}(2021{\natexlab{b}})}]{Shi2021pra}%
  \BibitemOpen
  \bibfield  {author} {\bibinfo {author} {\bibfnamefont {X.-F.}\ \bibnamefont
  {Shi}},\ }\bibfield  {title} {\bibinfo {title} {{Hyperentanglement of
  divalent neutral atoms by Rydberg blockade}},\ }\href@noop {} {\bibfield
  {journal} {\bibinfo  {journal} {Phys. Rev. A}\ }\textbf {\bibinfo {volume}
  {104}},\ \bibinfo {pages} {042422} (\bibinfo {year}
  {2021}{\natexlab{b}})}\BibitemShut {NoStop}%
\bibitem [{\citenamefont {Shi}(2024)}]{Shi2024}%
  \BibitemOpen
  \bibfield  {author} {\bibinfo {author} {\bibfnamefont {X.-F.}\ \bibnamefont
  {Shi}},\ }\bibfield  {title} {\bibinfo {title} {{Fast nuclear-spin gates and
  electrons-nuclei entanglement of neutral atoms in weak magnetic fields}},\
  }\href@noop {} {\bibfield  {journal} {\bibinfo  {journal} {Front. Phys.}\
  }\textbf {\bibinfo {volume} {19}},\ \bibinfo {pages} {22203} (\bibinfo {year}
  {2024})}\BibitemShut {NoStop}%
\bibitem [{\citenamefont {Chen}\ \emph {et~al.}(2022)\citenamefont {Chen},
  \citenamefont {Li}, \citenamefont {Huie}, \citenamefont {Zhao}, \citenamefont
  {Vetter}, \citenamefont {Greene},\ and\ \citenamefont {Covey}}]{Chen2022}%
  \BibitemOpen
  \bibfield  {author} {\bibinfo {author} {\bibfnamefont {N.}~\bibnamefont
  {Chen}}, \bibinfo {author} {\bibfnamefont {L.}~\bibnamefont {Li}}, \bibinfo
  {author} {\bibfnamefont {W.}~\bibnamefont {Huie}}, \bibinfo {author}
  {\bibfnamefont {M.}~\bibnamefont {Zhao}}, \bibinfo {author} {\bibfnamefont
  {I.}~\bibnamefont {Vetter}}, \bibinfo {author} {\bibfnamefont {C.~H.}\
  \bibnamefont {Greene}}, \ and\ \bibinfo {author} {\bibfnamefont {J.~P.}\
  \bibnamefont {Covey}},\ }\bibfield  {title} {\bibinfo {title} {{Analyzing the
  Rydberg-based optical-metastable-ground architecture for $^{171}$Yb nuclear
  spins}},\ }\href {\doibase 10.1103/PhysRevA.105.052438} {\bibfield  {journal}
  {\bibinfo  {journal} {Phys. Rev. A}\ }\textbf {\bibinfo {volume} {105}},\
  \bibinfo {pages} {052438} (\bibinfo {year} {2022})}\BibitemShut {NoStop}%
\bibitem [{\citenamefont {Muniz}\ \emph {et~al.}(2025)\citenamefont {Muniz},
  \citenamefont {Stone}, \citenamefont {Stack}, \citenamefont {Jaffe},
  \citenamefont {Kindem}, \citenamefont {Wadleigh}, \citenamefont
  {Zalys-Geller}, \citenamefont {Zhang}, \citenamefont {Chen}, \citenamefont
  {Norcia}, \citenamefont {Epstein}, \citenamefont {Halperin}, \citenamefont
  {Hummel}, \citenamefont {Wilkason}, \citenamefont {Li}, \citenamefont
  {Barnes}, \citenamefont {Battaglino}, \citenamefont {Bohdanowicz},
  \citenamefont {Booth}, \citenamefont {Brown}, \citenamefont {Brown},
  \citenamefont {Cairncross}, \citenamefont {Cassella}, \citenamefont {Coxe},
  \citenamefont {Crow}, \citenamefont {Feldkamp}, \citenamefont {Griger},
  \citenamefont {Heinz}, \citenamefont {Jones}, \citenamefont {Kim},
  \citenamefont {King}, \citenamefont {Kotru}, \citenamefont {Lauigan},
  \citenamefont {Marjanovic}, \citenamefont {Megidish}, \citenamefont
  {Meredith}, \citenamefont {McDonald}, \citenamefont {Morshead}, \citenamefont
  {Narayanaswami}, \citenamefont {Nishiguchi}, \citenamefont {Paule},
  \citenamefont {Pawlak}, \citenamefont {Pudenz}, \citenamefont {Pérez},
  \citenamefont {Ryou}, \citenamefont {Simon}, \citenamefont {Smull},
  \citenamefont {Urbanek}, \citenamefont {van~de Veerdonk}, \citenamefont
  {Vendeiro}, \citenamefont {Wu}, \citenamefont {Xie},\ and\ \citenamefont
  {Bloom}}]{Muniz2025}%
  \BibitemOpen
  \bibfield  {author} {\bibinfo {author} {\bibfnamefont {J.~A.}\ \bibnamefont
  {Muniz}}, \bibinfo {author} {\bibfnamefont {M.}~\bibnamefont {Stone}},
  \bibinfo {author} {\bibfnamefont {D.~T.}\ \bibnamefont {Stack}}, \bibinfo
  {author} {\bibfnamefont {M.}~\bibnamefont {Jaffe}}, \bibinfo {author}
  {\bibfnamefont {J.~M.}\ \bibnamefont {Kindem}}, \bibinfo {author}
  {\bibfnamefont {L.}~\bibnamefont {Wadleigh}}, \bibinfo {author}
  {\bibfnamefont {E.}~\bibnamefont {Zalys-Geller}}, \bibinfo {author}
  {\bibfnamefont {X.}~\bibnamefont {Zhang}}, \bibinfo {author} {\bibfnamefont
  {C.-A.}\ \bibnamefont {Chen}}, \bibinfo {author} {\bibfnamefont {M.~A.}\
  \bibnamefont {Norcia}}, \bibinfo {author} {\bibfnamefont {J.}~\bibnamefont
  {Epstein}}, \bibinfo {author} {\bibfnamefont {E.}~\bibnamefont {Halperin}},
  \bibinfo {author} {\bibfnamefont {F.}~\bibnamefont {Hummel}}, \bibinfo
  {author} {\bibfnamefont {T.}~\bibnamefont {Wilkason}}, \bibinfo {author}
  {\bibfnamefont {M.}~\bibnamefont {Li}}, \bibinfo {author} {\bibfnamefont
  {K.}~\bibnamefont {Barnes}}, \bibinfo {author} {\bibfnamefont
  {P.}~\bibnamefont {Battaglino}}, \bibinfo {author} {\bibfnamefont {T.~C.}\
  \bibnamefont {Bohdanowicz}}, \bibinfo {author} {\bibfnamefont
  {G.}~\bibnamefont {Booth}}, \bibinfo {author} {\bibfnamefont
  {A.}~\bibnamefont {Brown}}, \bibinfo {author} {\bibfnamefont {M.~O.}\
  \bibnamefont {Brown}}, \bibinfo {author} {\bibfnamefont {W.~B.}\ \bibnamefont
  {Cairncross}}, \bibinfo {author} {\bibfnamefont {K.}~\bibnamefont
  {Cassella}}, \bibinfo {author} {\bibfnamefont {R.}~\bibnamefont {Coxe}},
  \bibinfo {author} {\bibfnamefont {D.}~\bibnamefont {Crow}}, \bibinfo {author}
  {\bibfnamefont {M.}~\bibnamefont {Feldkamp}}, \bibinfo {author}
  {\bibfnamefont {C.}~\bibnamefont {Griger}}, \bibinfo {author} {\bibfnamefont
  {A.}~\bibnamefont {Heinz}}, \bibinfo {author} {\bibfnamefont {A.~M.~W.}\
  \bibnamefont {Jones}}, \bibinfo {author} {\bibfnamefont {H.}~\bibnamefont
  {Kim}}, \bibinfo {author} {\bibfnamefont {J.}~\bibnamefont {King}}, \bibinfo
  {author} {\bibfnamefont {K.}~\bibnamefont {Kotru}}, \bibinfo {author}
  {\bibfnamefont {J.}~\bibnamefont {Lauigan}}, \bibinfo {author} {\bibfnamefont
  {J.}~\bibnamefont {Marjanovic}}, \bibinfo {author} {\bibfnamefont
  {E.}~\bibnamefont {Megidish}}, \bibinfo {author} {\bibfnamefont
  {M.}~\bibnamefont {Meredith}}, \bibinfo {author} {\bibfnamefont
  {M.}~\bibnamefont {McDonald}}, \bibinfo {author} {\bibfnamefont
  {R.}~\bibnamefont {Morshead}}, \bibinfo {author} {\bibfnamefont
  {S.}~\bibnamefont {Narayanaswami}}, \bibinfo {author} {\bibfnamefont
  {C.}~\bibnamefont {Nishiguchi}}, \bibinfo {author} {\bibfnamefont
  {T.}~\bibnamefont {Paule}}, \bibinfo {author} {\bibfnamefont {K.~A.}\
  \bibnamefont {Pawlak}}, \bibinfo {author} {\bibfnamefont {K.~L.}\
  \bibnamefont {Pudenz}}, \bibinfo {author} {\bibfnamefont {D.~R.}\
  \bibnamefont {Pérez}}, \bibinfo {author} {\bibfnamefont {A.}~\bibnamefont
  {Ryou}}, \bibinfo {author} {\bibfnamefont {J.}~\bibnamefont {Simon}},
  \bibinfo {author} {\bibfnamefont {A.}~\bibnamefont {Smull}}, \bibinfo
  {author} {\bibfnamefont {M.}~\bibnamefont {Urbanek}}, \bibinfo {author}
  {\bibfnamefont {R.~J.~M.}\ \bibnamefont {van~de Veerdonk}}, \bibinfo {author}
  {\bibfnamefont {Z.}~\bibnamefont {Vendeiro}}, \bibinfo {author}
  {\bibfnamefont {T.-Y.}\ \bibnamefont {Wu}}, \bibinfo {author} {\bibfnamefont
  {X.}~\bibnamefont {Xie}}, \ and\ \bibinfo {author} {\bibfnamefont {B.~J.}\
  \bibnamefont {Bloom}},\ }\bibfield  {title} {\bibinfo {title} {High-fidelity
  universal gates in the $^{171}$yb ground-state nuclear-spin qubit},\ }\href
  {\doibase 10.1103/prxquantum.6.020334} {\bibfield  {journal} {\bibinfo
  {journal} {PRX Quantum}\ }\textbf {\bibinfo {volume} {6}},\ \bibinfo {pages}
  {020334} (\bibinfo {year} {2025})}\BibitemShut {NoStop}%
\bibitem [{\citenamefont {Shi}(2017)}]{Shi2017}%
  \BibitemOpen
  \bibfield  {author} {\bibinfo {author} {\bibfnamefont {X.-F.}\ \bibnamefont
  {Shi}},\ }\bibfield  {title} {\bibinfo {title} {{Rydberg Quantum Gates Free
  from Blockade Error}},\ }\href@noop {} {\bibfield  {journal} {\bibinfo
  {journal} {Phys. Rev. Appl.}\ }\textbf {\bibinfo {volume} {7}},\ \bibinfo
  {pages} {064017} (\bibinfo {year} {2017})}\BibitemShut {NoStop}%
\bibitem [{\citenamefont {Khaneja}\ \emph {et~al.}(2005)\citenamefont
  {Khaneja}, \citenamefont {Reiss}, \citenamefont {Kehlet}, \citenamefont
  {Schulte-Herbrüggen},\ and\ \citenamefont {Glaser}}]{Khaneja2005}%
  \BibitemOpen
  \bibfield  {author} {\bibinfo {author} {\bibfnamefont {N.}~\bibnamefont
  {Khaneja}}, \bibinfo {author} {\bibfnamefont {T.}~\bibnamefont {Reiss}},
  \bibinfo {author} {\bibfnamefont {C.}~\bibnamefont {Kehlet}}, \bibinfo
  {author} {\bibfnamefont {T.}~\bibnamefont {Schulte-Herbrüggen}}, \ and\
  \bibinfo {author} {\bibfnamefont {S.~J.}\ \bibnamefont {Glaser}},\ }\bibfield
   {title} {\bibinfo {title} {{Optimal control of coupled spin dynamics: Design
  of NMR pulse sequences by gradient ascent algorithms}},\ }\href {\doibase
  10.1016/j.jmr.2004.11.004} {\bibfield  {journal} {\bibinfo  {journal}
  {Journal of Magnetic Resonance}\ }\textbf {\bibinfo {volume} {172}},\
  \bibinfo {pages} {296} (\bibinfo {year} {2005})}\BibitemShut {NoStop}%
\bibitem [{\citenamefont {Evered}\ \emph {et~al.}(2023)\citenamefont {Evered},
  \citenamefont {Bluvstein}, \citenamefont {Kalinowski}, \citenamefont {Ebadi},
  \citenamefont {Manovitz}, \citenamefont {Zhou}, \citenamefont {Li},
  \citenamefont {Geim}, \citenamefont {Wang}, \citenamefont {Maskara},
  \citenamefont {Levine}, \citenamefont {Semeghini}, \citenamefont {Greiner},
  \citenamefont {Vuletic},\ and\ \citenamefont {Lukin}}]{Evered2023}%
  \BibitemOpen
  \bibfield  {author} {\bibinfo {author} {\bibfnamefont {S.~J.}\ \bibnamefont
  {Evered}}, \bibinfo {author} {\bibfnamefont {D.}~\bibnamefont {Bluvstein}},
  \bibinfo {author} {\bibfnamefont {M.}~\bibnamefont {Kalinowski}}, \bibinfo
  {author} {\bibfnamefont {S.}~\bibnamefont {Ebadi}}, \bibinfo {author}
  {\bibfnamefont {T.}~\bibnamefont {Manovitz}}, \bibinfo {author}
  {\bibfnamefont {H.}~\bibnamefont {Zhou}}, \bibinfo {author} {\bibfnamefont
  {S.~H.}\ \bibnamefont {Li}}, \bibinfo {author} {\bibfnamefont {A.~A.}\
  \bibnamefont {Geim}}, \bibinfo {author} {\bibfnamefont {T.~T.}\ \bibnamefont
  {Wang}}, \bibinfo {author} {\bibfnamefont {N.}~\bibnamefont {Maskara}},
  \bibinfo {author} {\bibfnamefont {H.}~\bibnamefont {Levine}}, \bibinfo
  {author} {\bibfnamefont {G.}~\bibnamefont {Semeghini}}, \bibinfo {author}
  {\bibfnamefont {M.}~\bibnamefont {Greiner}}, \bibinfo {author} {\bibfnamefont
  {V.}~\bibnamefont {Vuletic}}, \ and\ \bibinfo {author} {\bibfnamefont
  {M.~D.}\ \bibnamefont {Lukin}},\ }\bibfield  {title} {\bibinfo {title}
  {{High-fidelity parallel entangling gates on a neutral-atom quantum
  computer}},\ }\href {https://www.nature.com/articles/s41586-023-06481-y}
  {\bibfield  {journal} {\bibinfo  {journal} {Nature}\ }\textbf {\bibinfo
  {volume} {622}},\ \bibinfo {pages} {268} (\bibinfo {year}
  {2023})}\BibitemShut {NoStop}%
\bibitem [{\citenamefont {Finkelstein}\ \emph {et~al.}(2024)\citenamefont
  {Finkelstein}, \citenamefont {Tsai}, \citenamefont {Sun}, \citenamefont
  {Scholl}, \citenamefont {Direkci}, \citenamefont {Gefen}, \citenamefont
  {Choi}, \citenamefont {Shaw},\ and\ \citenamefont
  {Endres}}]{Finkelstein2024}%
  \BibitemOpen
  \bibfield  {author} {\bibinfo {author} {\bibfnamefont {R.}~\bibnamefont
  {Finkelstein}}, \bibinfo {author} {\bibfnamefont {R.~B.-S.}\ \bibnamefont
  {Tsai}}, \bibinfo {author} {\bibfnamefont {X.}~\bibnamefont {Sun}}, \bibinfo
  {author} {\bibfnamefont {P.}~\bibnamefont {Scholl}}, \bibinfo {author}
  {\bibfnamefont {S.}~\bibnamefont {Direkci}}, \bibinfo {author} {\bibfnamefont
  {T.}~\bibnamefont {Gefen}}, \bibinfo {author} {\bibfnamefont
  {J.}~\bibnamefont {Choi}}, \bibinfo {author} {\bibfnamefont {A.~L.}\
  \bibnamefont {Shaw}}, \ and\ \bibinfo {author} {\bibfnamefont
  {M.}~\bibnamefont {Endres}},\ }\bibfield  {title} {\bibinfo {title}
  {Universal quantum operations and ancilla-based read-out for tweezer
  clocks},\ }\href {\doibase 10.1038/s41586-024-08005-8} {\bibfield  {journal}
  {\bibinfo  {journal} {Nature}\ }\textbf {\bibinfo {volume} {634}},\ \bibinfo
  {pages} {321} (\bibinfo {year} {2024})}\BibitemShut {NoStop}%
\bibitem [{\citenamefont {Tsai}\ \emph {et~al.}(2025)\citenamefont {Tsai},
  \citenamefont {Sun}, \citenamefont {Shaw}, \citenamefont {Finkelstein},\ and\
  \citenamefont {Endres}}]{tsai2024fid}%
  \BibitemOpen
  \bibfield  {author} {\bibinfo {author} {\bibfnamefont {R.~B.-S.}\
  \bibnamefont {Tsai}}, \bibinfo {author} {\bibfnamefont {X.}~\bibnamefont
  {Sun}}, \bibinfo {author} {\bibfnamefont {A.~L.}\ \bibnamefont {Shaw}},
  \bibinfo {author} {\bibfnamefont {R.}~\bibnamefont {Finkelstein}}, \ and\
  \bibinfo {author} {\bibfnamefont {M.}~\bibnamefont {Endres}},\ }\bibfield
  {title} {\bibinfo {title} {Benchmarking and fidelity response theory of
  high-fidelity rydberg entangling gates},\ }\href@noop {} {\bibfield
  {journal} {\bibinfo  {journal} {PRX Quantum}\ }\textbf {\bibinfo {volume}
  {6}},\ \bibinfo {pages} {010331} (\bibinfo {year} {2025})}\BibitemShut
  {NoStop}%
\bibitem [{\citenamefont {Jandura}\ and\ \citenamefont
  {Pupillo}(2022)}]{Jandura2022}%
  \BibitemOpen
  \bibfield  {author} {\bibinfo {author} {\bibfnamefont {S.}~\bibnamefont
  {Jandura}}\ and\ \bibinfo {author} {\bibfnamefont {G.}~\bibnamefont
  {Pupillo}},\ }\bibfield  {title} {\bibinfo {title} {{Time-Optimal Two- And
  Three-Qubit Gates for Rydberg Atoms}},\ }\href {\doibase
  10.22331/Q-2022-05-13-712} {\bibfield  {journal} {\bibinfo  {journal}
  {Quantum}\ }\textbf {\bibinfo {volume} {6}},\ \bibinfo {pages} {38} (\bibinfo
  {year} {2022})}\BibitemShut {NoStop}%
\bibitem [{\citenamefont {Beterov}\ \emph {et~al.}(2009)\citenamefont
  {Beterov}, \citenamefont {Ryabtsev}, \citenamefont {Tretyakov},\ and\
  \citenamefont {Entin}}]{Beterov2009}%
  \BibitemOpen
  \bibfield  {author} {\bibinfo {author} {\bibfnamefont {I.}~\bibnamefont
  {Beterov}}, \bibinfo {author} {\bibfnamefont {I.}~\bibnamefont {Ryabtsev}},
  \bibinfo {author} {\bibfnamefont {D.}~\bibnamefont {Tretyakov}}, \ and\
  \bibinfo {author} {\bibfnamefont {V.}~\bibnamefont {Entin}},\ }\bibfield
  {title} {\bibinfo {title} {{Quasiclassical calculations of
  blackbody-radiation-induced depopulation rates and effective lifetimes of
  Rydberg nS, nP, and nD alkali-metal atoms with n$\leq$80}},\ }\href {\doibase
  10.1103/PhysRevA.79.052504} {\bibfield  {journal} {\bibinfo  {journal} {Phys.
  Rev. A}\ }\textbf {\bibinfo {volume} {79}},\ \bibinfo {pages} {052504}
  (\bibinfo {year} {2009})}\BibitemShut {NoStop}%
\bibitem [{\citenamefont {Levine}\ \emph {et~al.}(2019)\citenamefont {Levine},
  \citenamefont {Keesling}, \citenamefont {Semeghini}, \citenamefont {Omran},
  \citenamefont {Wang}, \citenamefont {Ebadi}, \citenamefont {Bernien},
  \citenamefont {Greiner}, \citenamefont {Vuletić}, \citenamefont {Pichler},\
  and\ \citenamefont {Lukin}}]{Levine2019}%
  \BibitemOpen
  \bibfield  {author} {\bibinfo {author} {\bibfnamefont {H.}~\bibnamefont
  {Levine}}, \bibinfo {author} {\bibfnamefont {A.}~\bibnamefont {Keesling}},
  \bibinfo {author} {\bibfnamefont {G.}~\bibnamefont {Semeghini}}, \bibinfo
  {author} {\bibfnamefont {A.}~\bibnamefont {Omran}}, \bibinfo {author}
  {\bibfnamefont {T.~T.}\ \bibnamefont {Wang}}, \bibinfo {author}
  {\bibfnamefont {S.}~\bibnamefont {Ebadi}}, \bibinfo {author} {\bibfnamefont
  {H.}~\bibnamefont {Bernien}}, \bibinfo {author} {\bibfnamefont
  {M.}~\bibnamefont {Greiner}}, \bibinfo {author} {\bibfnamefont
  {V.}~\bibnamefont {Vuletić}}, \bibinfo {author} {\bibfnamefont
  {H.}~\bibnamefont {Pichler}}, \ and\ \bibinfo {author} {\bibfnamefont
  {M.~D.}\ \bibnamefont {Lukin}},\ }\bibfield  {title} {\bibinfo {title}
  {{Parallel implementation of high-fidelity multi-qubit gates with neutral
  atoms}},\ }\href {\doibase 10.1103/PhysRevLett.123.170503} {\bibfield
  {journal} {\bibinfo  {journal} {Phys. Rev. Lett.}\ }\textbf {\bibinfo
  {volume} {123}},\ \bibinfo {pages} {170503} (\bibinfo {year}
  {2019})}\BibitemShut {NoStop}%
\bibitem [{\citenamefont {Madjarov}\ \emph {et~al.}(2020)\citenamefont
  {Madjarov}, \citenamefont {Covey}, \citenamefont {Shaw}, \citenamefont
  {Choi}, \citenamefont {Kale}, \citenamefont {Cooper}, \citenamefont
  {Pichler}, \citenamefont {Schkolnik}, \citenamefont {Williams},\ and\
  \citenamefont {Endres}}]{Madjarov2020}%
  \BibitemOpen
  \bibfield  {author} {\bibinfo {author} {\bibfnamefont {I.~S.}\ \bibnamefont
  {Madjarov}}, \bibinfo {author} {\bibfnamefont {J.~P.}\ \bibnamefont {Covey}},
  \bibinfo {author} {\bibfnamefont {A.~L.}\ \bibnamefont {Shaw}}, \bibinfo
  {author} {\bibfnamefont {J.}~\bibnamefont {Choi}}, \bibinfo {author}
  {\bibfnamefont {A.}~\bibnamefont {Kale}}, \bibinfo {author} {\bibfnamefont
  {A.}~\bibnamefont {Cooper}}, \bibinfo {author} {\bibfnamefont
  {H.}~\bibnamefont {Pichler}}, \bibinfo {author} {\bibfnamefont
  {V.}~\bibnamefont {Schkolnik}}, \bibinfo {author} {\bibfnamefont {J.~R.}\
  \bibnamefont {Williams}}, \ and\ \bibinfo {author} {\bibfnamefont
  {M.}~\bibnamefont {Endres}},\ }\bibfield  {title} {\bibinfo {title}
  {{High-fidelity entanglement and detection of alkaline-earth Rydberg
  atoms}},\ }\href {http://dx.doi.org/10.1038/s41567-020-0903-z} {\bibfield
  {journal} {\bibinfo  {journal} {Nat. Phys.}\ }\textbf {\bibinfo {volume}
  {16}},\ \bibinfo {pages} {857} (\bibinfo {year} {2020})}\BibitemShut
  {NoStop}%
\bibitem [{\citenamefont {Shi}(2022)}]{Shi2021qst}%
  \BibitemOpen
  \bibfield  {author} {\bibinfo {author} {\bibfnamefont {X.-F.}\ \bibnamefont
  {Shi}},\ }\bibfield  {title} {\bibinfo {title} {{Quantum logic and
  entanglement by neutral Rydberg atoms: methods and fidelity}},\ }\href
  {https://doi.org/10.1088/2058-9565/ac18b8} {\bibfield  {journal} {\bibinfo
  {journal} {Quantum Sci. Technol.}\ }\textbf {\bibinfo {volume} {7}},\
  \bibinfo {pages} {023002} (\bibinfo {year} {2022})}\BibitemShut {NoStop}%
\bibitem [{\citenamefont {Shi}(2025)}]{Shi2025pra}%
  \BibitemOpen
  \bibfield  {author} {\bibinfo {author} {\bibfnamefont {X.-F.}\ \bibnamefont
  {Shi}},\ }\bibfield  {title} {\bibinfo {title} {{Barium-based Rydberg atom
  quantum technologies with long Rydberg coherence}},\ }\href {\doibase
  https://doi.org/10.1103/bbv3-d4ch} {\bibfield  {journal} {\bibinfo  {journal}
  {Phys. Rev. A}\ }\textbf {\bibinfo {volume} {112}},\ \bibinfo {pages}
  {042401} (\bibinfo {year} {2025})}\BibitemShut {NoStop}%
\bibitem [{\citenamefont {Saffman}\ \emph {et~al.}(2020)\citenamefont
  {Saffman}, \citenamefont {Beterov}, \citenamefont {Dalal}, \citenamefont
  {Paez},\ and\ \citenamefont {Sanders}}]{Saffman2020}%
  \BibitemOpen
  \bibfield  {author} {\bibinfo {author} {\bibfnamefont {M.}~\bibnamefont
  {Saffman}}, \bibinfo {author} {\bibfnamefont {I.~I.}\ \bibnamefont
  {Beterov}}, \bibinfo {author} {\bibfnamefont {A.}~\bibnamefont {Dalal}},
  \bibinfo {author} {\bibfnamefont {E.~J.}\ \bibnamefont {Paez}}, \ and\
  \bibinfo {author} {\bibfnamefont {B.~C.}\ \bibnamefont {Sanders}},\
  }\bibfield  {title} {\bibinfo {title} {{Symmetric Rydberg controlled- Z gates
  with adiabatic pulses control target}},\ }\href {\doibase
  10.1103/PhysRevA.101.062309} {\bibfield  {journal} {\bibinfo  {journal}
  {Phys. Rev. A}\ }\textbf {\bibinfo {volume} {101}},\ \bibinfo {pages}
  {062309} (\bibinfo {year} {2020})}\BibitemShut {NoStop}%
\bibitem [{\citenamefont {Graham}\ \emph {et~al.}(2022)\citenamefont {Graham},
  \citenamefont {Song}, \citenamefont {Scott}, \citenamefont {Poole},
  \citenamefont {Phuttitarn}, \citenamefont {Jooya}, \citenamefont {Eichler},
  \citenamefont {Jiang}, \citenamefont {Marra}, \citenamefont {Grinkemeyer},
  \citenamefont {Kwon}, \citenamefont {Ebert}, \citenamefont {Cherek},
  \citenamefont {Lichtman}, \citenamefont {Gillette}, \citenamefont {Gilbert},
  \citenamefont {Bowman}, \citenamefont {Ballance}, \citenamefont {Campbell},
  \citenamefont {Dahl}, \citenamefont {Crawford}, \citenamefont {Blunt},
  \citenamefont {Rogers}, \citenamefont {Noel},\ and\ \citenamefont
  {Saffman}}]{Graham2022}%
  \BibitemOpen
  \bibfield  {author} {\bibinfo {author} {\bibfnamefont {T.~M.}\ \bibnamefont
  {Graham}}, \bibinfo {author} {\bibfnamefont {Y.}~\bibnamefont {Song}},
  \bibinfo {author} {\bibfnamefont {J.}~\bibnamefont {Scott}}, \bibinfo
  {author} {\bibfnamefont {C.}~\bibnamefont {Poole}}, \bibinfo {author}
  {\bibfnamefont {L.}~\bibnamefont {Phuttitarn}}, \bibinfo {author}
  {\bibfnamefont {K.}~\bibnamefont {Jooya}}, \bibinfo {author} {\bibfnamefont
  {P.}~\bibnamefont {Eichler}}, \bibinfo {author} {\bibfnamefont
  {X.}~\bibnamefont {Jiang}}, \bibinfo {author} {\bibfnamefont
  {A.}~\bibnamefont {Marra}}, \bibinfo {author} {\bibfnamefont
  {B.}~\bibnamefont {Grinkemeyer}}, \bibinfo {author} {\bibfnamefont
  {M.}~\bibnamefont {Kwon}}, \bibinfo {author} {\bibfnamefont {M.}~\bibnamefont
  {Ebert}}, \bibinfo {author} {\bibfnamefont {J.}~\bibnamefont {Cherek}},
  \bibinfo {author} {\bibfnamefont {M.~T.}\ \bibnamefont {Lichtman}}, \bibinfo
  {author} {\bibfnamefont {M.}~\bibnamefont {Gillette}}, \bibinfo {author}
  {\bibfnamefont {J.}~\bibnamefont {Gilbert}}, \bibinfo {author} {\bibfnamefont
  {D.}~\bibnamefont {Bowman}}, \bibinfo {author} {\bibfnamefont
  {T.}~\bibnamefont {Ballance}}, \bibinfo {author} {\bibfnamefont
  {C.}~\bibnamefont {Campbell}}, \bibinfo {author} {\bibfnamefont {E.~D.}\
  \bibnamefont {Dahl}}, \bibinfo {author} {\bibfnamefont {O.}~\bibnamefont
  {Crawford}}, \bibinfo {author} {\bibfnamefont {N.~S.}\ \bibnamefont {Blunt}},
  \bibinfo {author} {\bibfnamefont {B.}~\bibnamefont {Rogers}}, \bibinfo
  {author} {\bibfnamefont {T.}~\bibnamefont {Noel}}, \ and\ \bibinfo {author}
  {\bibfnamefont {M.}~\bibnamefont {Saffman}},\ }\bibfield  {title} {\bibinfo
  {title} {{Multi-qubit entanglement and algorithms on a neutral-atom quantum
  computer}},\ }\href {\doibase 10.1038/s41586-022-04603-6} {\bibfield
  {journal} {\bibinfo  {journal} {Nature}\ }\textbf {\bibinfo {volume} {604}},\
  \bibinfo {pages} {457} (\bibinfo {year} {2022})}\BibitemShut {NoStop}%
\bibitem [{\citenamefont {Peper}\ \emph {et~al.}(2025)\citenamefont {Peper},
  \citenamefont {Li}, \citenamefont {Knapp}, \citenamefont {Bileska},
  \citenamefont {Ma}, \citenamefont {Liu}, \citenamefont {Peng}, \citenamefont
  {Zhang}, \citenamefont {Horvath}, \citenamefont {Burgers},\ and\
  \citenamefont {Thompson}}]{Peper2025}%
  \BibitemOpen
  \bibfield  {author} {\bibinfo {author} {\bibfnamefont {M.}~\bibnamefont
  {Peper}}, \bibinfo {author} {\bibfnamefont {Y.}~\bibnamefont {Li}}, \bibinfo
  {author} {\bibfnamefont {D.~Y.}\ \bibnamefont {Knapp}}, \bibinfo {author}
  {\bibfnamefont {M.}~\bibnamefont {Bileska}}, \bibinfo {author} {\bibfnamefont
  {S.}~\bibnamefont {Ma}}, \bibinfo {author} {\bibfnamefont {G.}~\bibnamefont
  {Liu}}, \bibinfo {author} {\bibfnamefont {P.}~\bibnamefont {Peng}}, \bibinfo
  {author} {\bibfnamefont {B.}~\bibnamefont {Zhang}}, \bibinfo {author}
  {\bibfnamefont {S.~P.}\ \bibnamefont {Horvath}}, \bibinfo {author}
  {\bibfnamefont {A.~P.}\ \bibnamefont {Burgers}}, \ and\ \bibinfo {author}
  {\bibfnamefont {J.~D.}\ \bibnamefont {Thompson}},\ }\bibfield  {title}
  {\bibinfo {title} {Spectroscopy and modeling of $^{171}$yb rydberg states for
  high-fidelity two-qubit gates},\ }\href {\doibase 10.1103/physrevx.15.011009}
  {\bibfield  {journal} {\bibinfo  {journal} {Phys. Rev. X}\ }\textbf {\bibinfo
  {volume} {15}},\ \bibinfo {pages} {011009} (\bibinfo {year}
  {2025})}\BibitemShut {NoStop}%
\bibitem [{\citenamefont {Booth}\ \emph {et~al.}(2018)\citenamefont {Booth},
  \citenamefont {Isaacs},\ and\ \citenamefont {Saffman}}]{Booth2017}%
  \BibitemOpen
  \bibfield  {author} {\bibinfo {author} {\bibfnamefont {D.~W.}\ \bibnamefont
  {Booth}}, \bibinfo {author} {\bibfnamefont {J.}~\bibnamefont {Isaacs}}, \
  and\ \bibinfo {author} {\bibfnamefont {M.}~\bibnamefont {Saffman}},\
  }\bibfield  {title} {\bibinfo {title} {{Reducing the sensitivity of Rydberg
  atoms to dc electric fields using two-frequency ac field dressing}},\
  }\href@noop {} {\bibfield  {journal} {\bibinfo  {journal} {Phys. Rev. A}\
  }\textbf {\bibinfo {volume} {97}},\ \bibinfo {pages} {012515} (\bibinfo
  {year} {2018})}\BibitemShut {NoStop}%
\bibitem [{\citenamefont {Bohorquez}\ \emph {et~al.}(2023)\citenamefont
  {Bohorquez}, \citenamefont {Chinnarasu}, \citenamefont {Isaacs},
  \citenamefont {Booth}, \citenamefont {Beck}, \citenamefont {McDermott},\ and\
  \citenamefont {Saffman}}]{Bohorquez_2023}%
  \BibitemOpen
  \bibfield  {author} {\bibinfo {author} {\bibfnamefont {J.~C.}\ \bibnamefont
  {Bohorquez}}, \bibinfo {author} {\bibfnamefont {R.}~\bibnamefont
  {Chinnarasu}}, \bibinfo {author} {\bibfnamefont {J.}~\bibnamefont {Isaacs}},
  \bibinfo {author} {\bibfnamefont {D.}~\bibnamefont {Booth}}, \bibinfo
  {author} {\bibfnamefont {M.}~\bibnamefont {Beck}}, \bibinfo {author}
  {\bibfnamefont {R.}~\bibnamefont {McDermott}}, \ and\ \bibinfo {author}
  {\bibfnamefont {M.}~\bibnamefont {Saffman}},\ }\bibfield  {title} {\bibinfo
  {title} {Reducing rydberg-state dc polarizability by microwave dressing},\
  }\href {\doibase 10.1103/physreva.108.022805} {\bibfield  {journal} {\bibinfo
   {journal} {Physical Review A}\ }\textbf {\bibinfo {volume} {108}},\ \bibinfo
  {pages} {022805} (\bibinfo {year} {2023})}\BibitemShut {NoStop}%
\bibitem [{\citenamefont {Jiang}\ \emph {et~al.}(2023)\citenamefont {Jiang},
  \citenamefont {Scott}, \citenamefont {Friesen},\ and\ \citenamefont
  {Saffman}}]{Jiang_2023}%
  \BibitemOpen
  \bibfield  {author} {\bibinfo {author} {\bibfnamefont {X.}~\bibnamefont
  {Jiang}}, \bibinfo {author} {\bibfnamefont {J.}~\bibnamefont {Scott}},
  \bibinfo {author} {\bibfnamefont {M.}~\bibnamefont {Friesen}}, \ and\
  \bibinfo {author} {\bibfnamefont {M.}~\bibnamefont {Saffman}},\ }\bibfield
  {title} {\bibinfo {title} {Sensitivity of quantum gate fidelity to laser
  phase and intensity noise},\ }\href {\doibase 10.1103/physreva.107.042611}
  {\bibfield  {journal} {\bibinfo  {journal} {Physical Review A}\ }\textbf
  {\bibinfo {volume} {107}},\ \bibinfo {pages} {042611} (\bibinfo {year}
  {2023})}\BibitemShut {NoStop}%
\bibitem [{\citenamefont {Shi}(2019)}]{Shi2018Accuv1}%
  \BibitemOpen
  \bibfield  {author} {\bibinfo {author} {\bibfnamefont {X.-F.}\ \bibnamefont
  {Shi}},\ }\bibfield  {title} {\bibinfo {title} {{Fast, Accurate, and
  Realizable Two-Qubit Entangling Gates by Quantum Interference in Detuned Rabi
  Cycles of Rydberg Atoms}},\ }\href@noop {} {\bibfield  {journal} {\bibinfo
  {journal} {Phys. Rev. Appl.}\ }\textbf {\bibinfo {volume} {11}},\ \bibinfo
  {pages} {044035} (\bibinfo {year} {2019})}\BibitemShut {NoStop}%
\bibitem [{\citenamefont {Jenkins}\ \emph {et~al.}(2022)\citenamefont
  {Jenkins}, \citenamefont {Lis}, \citenamefont {Senoo}, \citenamefont
  {McGrew},\ and\ \citenamefont {Kaufman}}]{Jenkins2022}%
  \BibitemOpen
  \bibfield  {author} {\bibinfo {author} {\bibfnamefont {A.}~\bibnamefont
  {Jenkins}}, \bibinfo {author} {\bibfnamefont {J.~W.}\ \bibnamefont {Lis}},
  \bibinfo {author} {\bibfnamefont {A.}~\bibnamefont {Senoo}}, \bibinfo
  {author} {\bibfnamefont {W.~F.}\ \bibnamefont {McGrew}}, \ and\ \bibinfo
  {author} {\bibfnamefont {A.~M.}\ \bibnamefont {Kaufman}},\ }\bibfield
  {title} {\bibinfo {title} {{Ytterbium nuclear-spin qubits in an optical
  tweezer array}},\ }\href {\doibase 10.1103/PhysRevX.12.021027} {\bibfield
  {journal} {\bibinfo  {journal} {Phys. Rev. X}\ }\textbf {\bibinfo {volume}
  {12}},\ \bibinfo {pages} {021027} (\bibinfo {year} {2022})}\BibitemShut
  {NoStop}%
\bibitem [{\citenamefont {Shi}(2023)}]{PhysRevA.107.023102}%
  \BibitemOpen
  \bibfield  {author} {\bibinfo {author} {\bibfnamefont {X.-F.}\ \bibnamefont
  {Shi}},\ }\bibfield  {title} {\bibinfo {title} {{Coherence-preserving cooling
  of nuclear-spin qubits in a weak magnetic field}},\ }\href {\doibase
  10.1103/PhysRevA.107.023102} {\bibfield  {journal} {\bibinfo  {journal}
  {Phys. Rev. A}\ }\textbf {\bibinfo {volume} {107}},\ \bibinfo {pages}
  {023102} (\bibinfo {year} {2023})}\BibitemShut {NoStop}%
\bibitem [{\citenamefont {Jia}\ \emph {et~al.}(2024)\citenamefont {Jia},
  \citenamefont {Huie}, \citenamefont {Li}, \citenamefont {Sun}, \citenamefont
  {Hu}, \citenamefont {Aakash}, \citenamefont {Kogan}, \citenamefont {Karve},
  \citenamefont {Lee},\ and\ \citenamefont {Covey}}]{Jia2024}%
  \BibitemOpen
  \bibfield  {author} {\bibinfo {author} {\bibfnamefont {Z.}~\bibnamefont
  {Jia}}, \bibinfo {author} {\bibfnamefont {W.}~\bibnamefont {Huie}}, \bibinfo
  {author} {\bibfnamefont {L.}~\bibnamefont {Li}}, \bibinfo {author}
  {\bibfnamefont {W.~K.~C.}\ \bibnamefont {Sun}}, \bibinfo {author}
  {\bibfnamefont {X.}~\bibnamefont {Hu}}, \bibinfo {author} {\bibnamefont
  {Aakash}}, \bibinfo {author} {\bibfnamefont {H.}~\bibnamefont {Kogan}},
  \bibinfo {author} {\bibfnamefont {A.}~\bibnamefont {Karve}}, \bibinfo
  {author} {\bibfnamefont {J.~Y.}\ \bibnamefont {Lee}}, \ and\ \bibinfo
  {author} {\bibfnamefont {J.~P.}\ \bibnamefont {Covey}},\ }\bibfield  {title}
  {\bibinfo {title} {An architecture for two-qubit encoding in neutral
  ytterbium-171 atoms},\ }\href {\doibase 10.1038/s41534-024-00898-7}
  {\bibfield  {journal} {\bibinfo  {journal} {npj Quantum Information}\
  }\textbf {\bibinfo {volume} {10}},\ \bibinfo {pages} {106} (\bibinfo {year}
  {2024})}\BibitemShut {NoStop}%
\bibitem [{\citenamefont {Barredo}\ \emph {et~al.}(2018)\citenamefont
  {Barredo}, \citenamefont {Lienhard}, \citenamefont {de~Léséleuc},
  \citenamefont {Lahaye},\ and\ \citenamefont {Browaeys}}]{Barredo_2018}%
  \BibitemOpen
  \bibfield  {author} {\bibinfo {author} {\bibfnamefont {D.}~\bibnamefont
  {Barredo}}, \bibinfo {author} {\bibfnamefont {V.}~\bibnamefont {Lienhard}},
  \bibinfo {author} {\bibfnamefont {S.}~\bibnamefont {de~Léséleuc}}, \bibinfo
  {author} {\bibfnamefont {T.}~\bibnamefont {Lahaye}}, \ and\ \bibinfo {author}
  {\bibfnamefont {A.}~\bibnamefont {Browaeys}},\ }\bibfield  {title} {\bibinfo
  {title} {Synthetic three-dimensional atomic structures assembled atom by
  atom},\ }\href {\doibase 10.1038/s41586-018-0450-2} {\bibfield  {journal}
  {\bibinfo  {journal} {Nature}\ }\textbf {\bibinfo {volume} {561}},\ \bibinfo
  {pages} {79} (\bibinfo {year} {2018})}\BibitemShut {NoStop}%
\bibitem [{\citenamefont {Shao}\ \emph {et~al.}(2017)\citenamefont {Shao},
  \citenamefont {Wu}, \citenamefont {Yi},\ and\ \citenamefont
  {Long}}]{PhysRevA.96.062315}%
  \BibitemOpen
  \bibfield  {author} {\bibinfo {author} {\bibfnamefont {X.~Q.}\ \bibnamefont
  {Shao}}, \bibinfo {author} {\bibfnamefont {J.~H.}\ \bibnamefont {Wu}},
  \bibinfo {author} {\bibfnamefont {X.~X.}\ \bibnamefont {Yi}}, \ and\ \bibinfo
  {author} {\bibfnamefont {G.-L.}\ \bibnamefont {Long}},\ }\bibfield  {title}
  {\bibinfo {title} {Dissipative preparation of steady
  greenberger-horne-zeilinger states for rydberg atoms with quantum zeno
  dynamics},\ }\href {\doibase 10.1103/PhysRevA.96.062315} {\bibfield
  {journal} {\bibinfo  {journal} {Phys. Rev. A}\ }\textbf {\bibinfo {volume}
  {96}},\ \bibinfo {pages} {062315} (\bibinfo {year} {2017})}\BibitemShut
  {NoStop}%
\bibitem [{\citenamefont {Sauerwein}\ \emph {et~al.}(2023)\citenamefont
  {Sauerwein}, \citenamefont {Orsi}, \citenamefont {Uhrich}, \citenamefont
  {Bandyopadhyay}, \citenamefont {Mattiotti}, \citenamefont {Cantat-Moltrecht},
  \citenamefont {Pupillo}, \citenamefont {Hauke},\ and\ \citenamefont
  {Brantut}}]{Sauerwein_2023}%
  \BibitemOpen
  \bibfield  {author} {\bibinfo {author} {\bibfnamefont {N.}~\bibnamefont
  {Sauerwein}}, \bibinfo {author} {\bibfnamefont {F.}~\bibnamefont {Orsi}},
  \bibinfo {author} {\bibfnamefont {P.}~\bibnamefont {Uhrich}}, \bibinfo
  {author} {\bibfnamefont {S.}~\bibnamefont {Bandyopadhyay}}, \bibinfo {author}
  {\bibfnamefont {F.}~\bibnamefont {Mattiotti}}, \bibinfo {author}
  {\bibfnamefont {T.}~\bibnamefont {Cantat-Moltrecht}}, \bibinfo {author}
  {\bibfnamefont {G.}~\bibnamefont {Pupillo}}, \bibinfo {author} {\bibfnamefont
  {P.}~\bibnamefont {Hauke}}, \ and\ \bibinfo {author} {\bibfnamefont {J.-P.}\
  \bibnamefont {Brantut}},\ }\bibfield  {title} {\bibinfo {title} {Engineering
  random spin models with atoms in a high-finesse cavity},\ }\href {\doibase
  10.1038/s41567-023-02033-3} {\bibfield  {journal} {\bibinfo  {journal}
  {Nature Physics}\ }\textbf {\bibinfo {volume} {19}},\ \bibinfo {pages} {1128}
  (\bibinfo {year} {2023})}\BibitemShut {NoStop}%
\bibitem [{\citenamefont {Liu}\ \emph {et~al.}(2023)\citenamefont {Liu},
  \citenamefont {Wang}, \citenamefont {Yang}, \citenamefont {Wang},
  \citenamefont {Fan}, \citenamefont {Guan}, \citenamefont {Li}, \citenamefont
  {Zhang},\ and\ \citenamefont {Zhang}}]{Liu_2023}%
  \BibitemOpen
  \bibfield  {author} {\bibinfo {author} {\bibfnamefont {Y.}~\bibnamefont
  {Liu}}, \bibinfo {author} {\bibfnamefont {Z.}~\bibnamefont {Wang}}, \bibinfo
  {author} {\bibfnamefont {P.}~\bibnamefont {Yang}}, \bibinfo {author}
  {\bibfnamefont {Q.}~\bibnamefont {Wang}}, \bibinfo {author} {\bibfnamefont
  {Q.}~\bibnamefont {Fan}}, \bibinfo {author} {\bibfnamefont {S.}~\bibnamefont
  {Guan}}, \bibinfo {author} {\bibfnamefont {G.}~\bibnamefont {Li}}, \bibinfo
  {author} {\bibfnamefont {P.}~\bibnamefont {Zhang}}, \ and\ \bibinfo {author}
  {\bibfnamefont {T.}~\bibnamefont {Zhang}},\ }\bibfield  {title} {\bibinfo
  {title} {Realization of strong coupling between deterministic single-atom
  arrays and a high-finesse miniature optical cavity},\ }\href {\doibase
  10.1103/physrevlett.130.173601} {\bibfield  {journal} {\bibinfo  {journal}
  {Physical Review Letters}\ }\textbf {\bibinfo {volume} {130}},\ \bibinfo
  {pages} {173601} (\bibinfo {year} {2023})}\BibitemShut {NoStop}%
\bibitem [{\citenamefont {Endres}\ \emph {et~al.}(2016)\citenamefont {Endres},
  \citenamefont {Bernien}, \citenamefont {Keesling}, \citenamefont {Levine},
  \citenamefont {Anschuetz}, \citenamefont {Krajenbrink}, \citenamefont
  {Senko}, \citenamefont {Vuletic}, \citenamefont {Greiner},\ and\
  \citenamefont {Lukin}}]{Endres2016}%
  \BibitemOpen
  \bibfield  {author} {\bibinfo {author} {\bibfnamefont {M.}~\bibnamefont
  {Endres}}, \bibinfo {author} {\bibfnamefont {H.}~\bibnamefont {Bernien}},
  \bibinfo {author} {\bibfnamefont {A.}~\bibnamefont {Keesling}}, \bibinfo
  {author} {\bibfnamefont {H.}~\bibnamefont {Levine}}, \bibinfo {author}
  {\bibfnamefont {E.~R.}\ \bibnamefont {Anschuetz}}, \bibinfo {author}
  {\bibfnamefont {A.}~\bibnamefont {Krajenbrink}}, \bibinfo {author}
  {\bibfnamefont {C.}~\bibnamefont {Senko}}, \bibinfo {author} {\bibfnamefont
  {V.}~\bibnamefont {Vuletic}}, \bibinfo {author} {\bibfnamefont
  {M.}~\bibnamefont {Greiner}}, \ and\ \bibinfo {author} {\bibfnamefont
  {M.~D.}\ \bibnamefont {Lukin}},\ }\bibfield  {title} {\bibinfo {title}
  {Atom-by-atom assembly of defect-free one-dimensional cold atom arrays},\
  }\href {\doibase 10.1126/science.aah3752} {\bibfield  {journal} {\bibinfo
  {journal} {Science}\ }\textbf {\bibinfo {volume} {354}},\ \bibinfo {pages}
  {1024} (\bibinfo {year} {2016})}\BibitemShut {NoStop}%
\bibitem [{\citenamefont {Kim}\ \emph {et~al.}(2016)\citenamefont {Kim},
  \citenamefont {Lee}, \citenamefont {Lee}, \citenamefont {Jo}, \citenamefont
  {Song},\ and\ \citenamefont {Ahn}}]{Kim2016}%
  \BibitemOpen
  \bibfield  {author} {\bibinfo {author} {\bibfnamefont {H.}~\bibnamefont
  {Kim}}, \bibinfo {author} {\bibfnamefont {W.}~\bibnamefont {Lee}}, \bibinfo
  {author} {\bibfnamefont {H.-g.}\ \bibnamefont {Lee}}, \bibinfo {author}
  {\bibfnamefont {H.}~\bibnamefont {Jo}}, \bibinfo {author} {\bibfnamefont
  {Y.}~\bibnamefont {Song}}, \ and\ \bibinfo {author} {\bibfnamefont
  {J.}~\bibnamefont {Ahn}},\ }\bibfield  {title} {\bibinfo {title} {In situ
  single-atom array synthesis using dynamic holographic optical tweezers},\
  }\href {\doibase 10.1038/ncomms13317} {\bibfield  {journal} {\bibinfo
  {journal} {Nat. Commun.}\ }\textbf {\bibinfo {volume} {7}},\ \bibinfo {pages}
  {13317} (\bibinfo {year} {2016})}\BibitemShut {NoStop}%
\bibitem [{\citenamefont {Barredo}\ \emph {et~al.}(2016)\citenamefont
  {Barredo}, \citenamefont {de~Leseleuc}, \citenamefont {Lienhard},
  \citenamefont {Lahaye},\ and\ \citenamefont {Browaeys}}]{Barredo_2016}%
  \BibitemOpen
  \bibfield  {author} {\bibinfo {author} {\bibfnamefont {D.}~\bibnamefont
  {Barredo}}, \bibinfo {author} {\bibfnamefont {S.}~\bibnamefont
  {de~Leseleuc}}, \bibinfo {author} {\bibfnamefont {V.}~\bibnamefont
  {Lienhard}}, \bibinfo {author} {\bibfnamefont {T.}~\bibnamefont {Lahaye}}, \
  and\ \bibinfo {author} {\bibfnamefont {A.}~\bibnamefont {Browaeys}},\
  }\bibfield  {title} {\bibinfo {title} {An atom-by-atom assembler of
  defect-free arbitrary two-dimensional atomic arrays},\ }\href {\doibase
  10.1126/science.aah3778} {\bibfield  {journal} {\bibinfo  {journal}
  {Science}\ }\textbf {\bibinfo {volume} {354}},\ \bibinfo {pages} {1021}
  (\bibinfo {year} {2016})}\BibitemShut {NoStop}%
\bibitem [{\citenamefont {Lin}\ \emph {et~al.}(2025)\citenamefont {Lin},
  \citenamefont {Zhong}, \citenamefont {Li}, \citenamefont {Zhao},
  \citenamefont {Zheng}, \citenamefont {Hu}, \citenamefont {Wu}, \citenamefont
  {Wu}, \citenamefont {Ma}, \citenamefont {Gao}, \citenamefont {Zhu},
  \citenamefont {Su}, \citenamefont {Ouyang}, \citenamefont {Zhang},
  \citenamefont {Rui}, \citenamefont {Chen}, \citenamefont {Lu},\ and\
  \citenamefont {Pan}}]{Lin2024}%
  \BibitemOpen
  \bibfield  {author} {\bibinfo {author} {\bibfnamefont {R.}~\bibnamefont
  {Lin}}, \bibinfo {author} {\bibfnamefont {H.-S.}\ \bibnamefont {Zhong}},
  \bibinfo {author} {\bibfnamefont {Y.}~\bibnamefont {Li}}, \bibinfo {author}
  {\bibfnamefont {Z.-R.}\ \bibnamefont {Zhao}}, \bibinfo {author}
  {\bibfnamefont {L.-T.}\ \bibnamefont {Zheng}}, \bibinfo {author}
  {\bibfnamefont {T.-R.}\ \bibnamefont {Hu}}, \bibinfo {author} {\bibfnamefont
  {H.-M.}\ \bibnamefont {Wu}}, \bibinfo {author} {\bibfnamefont
  {Z.}~\bibnamefont {Wu}}, \bibinfo {author} {\bibfnamefont {W.-J.}\
  \bibnamefont {Ma}}, \bibinfo {author} {\bibfnamefont {Y.}~\bibnamefont
  {Gao}}, \bibinfo {author} {\bibfnamefont {Y.-K.}\ \bibnamefont {Zhu}},
  \bibinfo {author} {\bibfnamefont {Z.-F.}\ \bibnamefont {Su}}, \bibinfo
  {author} {\bibfnamefont {W.-L.}\ \bibnamefont {Ouyang}}, \bibinfo {author}
  {\bibfnamefont {Y.-C.}\ \bibnamefont {Zhang}}, \bibinfo {author}
  {\bibfnamefont {J.}~\bibnamefont {Rui}}, \bibinfo {author} {\bibfnamefont
  {M.-C.}\ \bibnamefont {Chen}}, \bibinfo {author} {\bibfnamefont {C.-Y.}\
  \bibnamefont {Lu}}, \ and\ \bibinfo {author} {\bibfnamefont {J.-W.}\
  \bibnamefont {Pan}},\ }\bibfield  {title} {\bibinfo {title} {Ai-enabled
  parallel assembly of thousands of defect-free neutral atom arrays},\
  }\href@noop {} {\bibfield  {journal} {\bibinfo  {journal} {Phys. Rev. Lett.}\
  }\textbf {\bibinfo {volume} {135}},\ \bibinfo {pages} {060602} (\bibinfo
  {year} {2025})}\BibitemShut {NoStop}%
\bibitem [{\citenamefont {Isenhower}\ \emph {et~al.}(2011)\citenamefont
  {Isenhower}, \citenamefont {Saffman},\ and\ \citenamefont
  {Mølmer}}]{Isenhower2011}%
  \BibitemOpen
  \bibfield  {author} {\bibinfo {author} {\bibfnamefont {L.}~\bibnamefont
  {Isenhower}}, \bibinfo {author} {\bibfnamefont {M.}~\bibnamefont {Saffman}},
  \ and\ \bibinfo {author} {\bibfnamefont {K.}~\bibnamefont {Mølmer}},\
  }\bibfield  {title} {\bibinfo {title} {{Multibit $C_k$NOT quantum gates via
  Rydberg blockade}},\ }\href {\doibase 10.1007/s11128-011-0292-4} {\bibfield
  {journal} {\bibinfo  {journal} {Quant. Inf. Proc.}\ }\textbf {\bibinfo
  {volume} {10}},\ \bibinfo {pages} {755} (\bibinfo {year} {2011})}\BibitemShut
  {NoStop}%
\bibitem [{\citenamefont {Pelegrí}\ \emph {et~al.}(2022)\citenamefont
  {Pelegrí}, \citenamefont {Daley},\ and\ \citenamefont
  {Pritchard}}]{Pelegr2022}%
  \BibitemOpen
  \bibfield  {author} {\bibinfo {author} {\bibfnamefont {G.}~\bibnamefont
  {Pelegrí}}, \bibinfo {author} {\bibfnamefont {A.~J.}\ \bibnamefont {Daley}},
  \ and\ \bibinfo {author} {\bibfnamefont {J.~D.}\ \bibnamefont {Pritchard}},\
  }\bibfield  {title} {\bibinfo {title} {High-fidelity multiqubit rydberg gates
  via two-photon adiabatic rapid passage},\ }\href {\doibase
  10.1088/2058-9565/ac823a} {\bibfield  {journal} {\bibinfo  {journal} {Quantum
  Science and Technology}\ }\textbf {\bibinfo {volume} {7}},\ \bibinfo {pages}
  {045020} (\bibinfo {year} {2022})}\BibitemShut {NoStop}%
\bibitem [{\citenamefont {Xue}\ \emph {et~al.}(2024)\citenamefont {Xue},
  \citenamefont {Xu}, \citenamefont {Li},\ and\ \citenamefont
  {Li}}]{PhysRevA.110.032619}%
  \BibitemOpen
  \bibfield  {author} {\bibinfo {author} {\bibfnamefont {M.}~\bibnamefont
  {Xue}}, \bibinfo {author} {\bibfnamefont {S.}~\bibnamefont {Xu}}, \bibinfo
  {author} {\bibfnamefont {X.}~\bibnamefont {Li}}, \ and\ \bibinfo {author}
  {\bibfnamefont {X.}~\bibnamefont {Li}},\ }\bibfield  {title} {\bibinfo
  {title} {High-fidelity and robust controlled-$z$ gates implemented with
  rydberg atoms via echoing rapid adiabatic passage},\ }\href {\doibase
  10.1103/PhysRevA.110.032619} {\bibfield  {journal} {\bibinfo  {journal}
  {Phys. Rev. A}\ }\textbf {\bibinfo {volume} {110}},\ \bibinfo {pages}
  {032619} (\bibinfo {year} {2024})}\BibitemShut {NoStop}%
\bibitem [{\citenamefont {M.~Farouk}\ \emph {et~al.}(2023)\citenamefont
  {M.~Farouk}, \citenamefont {Beterov}, \citenamefont {Xu }, \citenamefont
  {Bergamini },\ and\ \citenamefont {Ryabtsev }}]{M_Farouk_2023}%
  \BibitemOpen
  \bibfield  {author} {\bibinfo {author} {\bibfnamefont {A.}~\bibnamefont
  {M.~Farouk}}, \bibinfo {author} {\bibfnamefont {I.~I.}\ \bibnamefont
  {Beterov}}, \bibinfo {author} {\bibfnamefont {P.}~\bibnamefont {Xu }},
  \bibinfo {author} {\bibfnamefont {S.}~\bibnamefont {Bergamini }}, \ and\
  \bibinfo {author} {\bibfnamefont {I.~I.}\ \bibnamefont {Ryabtsev }},\
  }\bibfield  {title} {\bibinfo {title} {Parallel implementation of cnotn and
  c2not2 gates via homonuclear and heteronuclear förster interactions of
  rydberg atoms},\ }\href {\doibase 10.3390/photonics10111280} {\bibfield
  {journal} {\bibinfo  {journal} {Photonics}\ }\textbf {\bibinfo {volume}
  {10}},\ \bibinfo {pages} {1280} (\bibinfo {year} {2023})}\BibitemShut
  {NoStop}%
\bibitem [{\citenamefont {Yu}\ \emph {et~al.}(2022)\citenamefont {Yu},
  \citenamefont {Wang}, \citenamefont {Liu}, \citenamefont {Su}, \citenamefont
  {Qian},\ and\ \citenamefont {Zhang}}]{Yu_2022}%
  \BibitemOpen
  \bibfield  {author} {\bibinfo {author} {\bibfnamefont {D.}~\bibnamefont
  {Yu}}, \bibinfo {author} {\bibfnamefont {H.}~\bibnamefont {Wang}}, \bibinfo
  {author} {\bibfnamefont {J.-M.}\ \bibnamefont {Liu}}, \bibinfo {author}
  {\bibfnamefont {S.-L.}\ \bibnamefont {Su}}, \bibinfo {author} {\bibfnamefont
  {J.}~\bibnamefont {Qian}}, \ and\ \bibinfo {author} {\bibfnamefont
  {W.}~\bibnamefont {Zhang}},\ }\bibfield  {title} {\bibinfo {title}
  {Multiqubit toffoli gates and optimal geometry with rydberg atoms},\
  }\href@noop {} {\bibfield  {journal} {\bibinfo  {journal} {Phys. Rev. Appl.}\
  }\textbf {\bibinfo {volume} {18}},\ \bibinfo {pages} {034072} (\bibinfo
  {year} {2022})}\BibitemShut {NoStop}%
\bibitem [{\citenamefont {Delakouras}\ \emph {et~al.}(2025)\citenamefont
  {Delakouras}, \citenamefont {Doultsinos},\ and\ \citenamefont
  {Petrosyan}}]{Petrosym2025}%
  \BibitemOpen
  \bibfield  {author} {\bibinfo {author} {\bibfnamefont {A.}~\bibnamefont
  {Delakouras}}, \bibinfo {author} {\bibfnamefont {G.}~\bibnamefont
  {Doultsinos}}, \ and\ \bibinfo {author} {\bibfnamefont {D.}~\bibnamefont
  {Petrosyan}},\ }\href {\doibase 10.48550/arXiv.2507.16602} {\bibinfo {title}
  {Multi-qubit {Rydberg} gates between distant atoms},\ } (\bibinfo {year}
  {2025}),\ \bibinfo {note} {arXiv:2507.16602 [quant-ph]}\BibitemShut {NoStop}%
\end{thebibliography}
\end{document}